\documentclass[fleqn,usenatbib]{mnras}

\usepackage{newtxtext,newtxmath}

\usepackage[T1]{fontenc}

\DeclareRobustCommand{\VAN}[3]{#2}
\let\VANthebibliography\thebibliography
\def\thebibliography{\DeclareRobustCommand{\VAN}[3]{##3}\VANthebibliography}

\usepackage{graphicx}	
\usepackage{amsmath}	

\usepackage{amssymb}	
\usepackage{hyperref}
\usepackage{multirow}
\usepackage{rotating}
\usepackage[flushleft]{threeparttable}
\usepackage{subfloat}
\usepackage{xcolor}
\usepackage{longtable, makecell}

\usepackage{siunitx}
\usepackage{tablefootnote}
\usepackage{footnote}
\usepackage{float}
\usepackage{placeins}
\usepackage{afterpage}
\usepackage{perpage}
\usepackage{pdflscape}
\usepackage{threeparttablex}

\newcommand{\hii}{H~\textsc{ii}}
\newcommand{\msun}{$\rm M_\odot$}

\newcommand{\kms}{km~s$^{-1}$}
\newcommand{\jybeam}{Jy beam$^{-1}$}
\newcommand{\mjybeam}{mJy beam$^{-1}$}
\newcommand{\hmole}{H$_2$}

\newcommand{\cm}{cm$^{-2}$}
\newcommand{\pressure}{K~cm$^{-3}$}

\newcommand{\hcop}{HCO$^+$}
\newcommand{\hcopone}{HCO$^+$~$J=1-0$}
\newcommand{\htcop}{H$^{13}$CO$^+$}
\newcommand{\htcopone}{H$^{13}$CO$^+$~$J=1-0$}

\newcommand{\hcn}{HCN}
\newcommand{\hcnone}{HCN~$J=1-0$}
\newcommand{\htcn}{H$^{13}$CN}
\newcommand{\htcnone}{H$^{13}$CN~$J=1-0$}

\newcommand{\cs}{CS}

\newcommand{\sio}{SiO}

\newcommand{\hctn}{HC$_3$N}

\newcommand{\hfourtyalpha}{H40$\alpha$}
\newcommand{\chtoh}{CH$_3$OH}

\newcommand{\cch}{CCH}
\newcommand{\cchline}{CCH~$N_{J,F}=1_{3/2,2}-0_{1/2,1}$}

\newcommand{\cchhyperfine}{CCH~$N_{J,F}=1_{3/2,1}-0_{1/2,0}$}

\newcommand{\so}{SO}

\newcommand{\nht}{NH$_3$}

\newcommand{\mclump}{$M_{\rm clp}$}

\newcommand{\RGC}{$R_{\rm GC}$}
\newcommand{\mvir}{$M_{\rm vir}$}

\newcommand{\tkin}{$T_{\rm kin}$}
\newcommand{\tex}{$T_{\rm ex}$}
\newcommand{\tbg}{$T_{\rm bg}$}
\newcommand{\tr}{$T_{\rm r}$}
\newcommand{\te}{$T_{\rm e}$}

\newcommand{\co}{CO}
\newcommand{\tco}{$^{13}$CO}
\newcommand{\ceo}{C$^{18}$O}

\newcommand{\cothr}{$^{12}$CO $J=3-2$}

\newcommand{\tcoone}{$^{13}$CO $J=1-0$}

\newcommand{\ceoone}{C$^{18}$O~$J=1-0$}

\newcommand{\degree}{$^{\circ}$}

\newcommand{\nhtcd}{$N_{\rm H_{2}}$}
\newcommand{\dustt}{$T_{\rm dust}$}
\newcommand{\nhtnd}{$n_{\rm H_{2}}$}

\newcommand{\ncchcd}{$N_{\rm CCH}$}
\newcommand{\nnhtcd}{$N_{\rm NH_{3}}$}

\newcommand{\vlsr}{$\rm v_{lsr}$}
\newcommand{\vinfall}{$\rm v_{infall}$}

\newcommand{\mjth}{$M_{\rm J}^{\rm th}$}
\newcommand{\lambdajth}{$\lambda_{\rm J}^{\rm th}$}

\newcommand{\mjthclump}{$M_{\rm J, clp}^{\rm th}$}

\newcommand{\mjtotclump}{$M_{\rm J, clp}^{\rm tot}$}
\newcommand{\lambdajtotclump}{$\lambda_{\rm J, clp}^{\rm tot}$}

\newcommand{\mjflowclump}{$M_{\rm J, clp}^{\rm com, flow}$}
\newcommand{\lambdajflowclump}{$\lambda_{\rm J, clp}^{\rm com, flow}$}

\newcommand{\Pex}{$P_{\rm ex}$}

\newcommand{\av}{$A_{\rm v}$}

\title[ATOMS clump-scale massive bright-rimmed cloud]{ATOMS: ALMA Three-millimeter Observations of Massive Star-forming regions -XIII. Ongoing triggered star formation within clump-fed scenario found in the massive ($\sim1500$~\msun) clump}

\author[Zhang et al.]{
Siju Zhang,$^{1}$\thanks{E-mail: sijuzhangastro@gmail.com}
Ke Wang,$^{1}$\thanks{E-mail: kwang.astro@pku.edu.cn}
Tie Liu,$^{2,3}$
Annie Zavagno,$^{4,5}$
Mika Juvela,$^{6}$
Hongli Liu,$^{7}$
Anandmayee Tej,$^{8}$
\newauthor
Amelia M. Stutz,$^{9}$
Shanghuo Li,$^{10}$ 
Leonardo Bronfman,$^{11}$
Qizhou Zhang,$^{12}$
Paul F. Goldsmith,$^{13}$
\newauthor
Chang Won Lee,$^{14,15}$
Enrique Vázquez-Semadeni,$^{16}$
Ken'ichi Tatematsu,$^{17,18}$
Wenyu Jiao,$^{1,19}$
Fengwei Xu,$^{1,19}$
\newauthor
Chao Wang,$^{1,19}$ 
Jian-Wen Zhou$^{20,21}$
\\
Affiliations are listed at the end of the paper}

\date{Accepted 2022 December 28. Received 2022 December 28; in original form 2022 June 15}

\pubyear{2022}

\begin{document}
\label{firstpage}
\pagerange{\pageref{firstpage}--\pageref{lastpage}}
\maketitle

\begin{abstract}
 Whether ionization feedback triggers the formation of massive stars is highly debated. Using ALMA 3~mm observations with a spatial resolution of $\sim0.05$ pc and a mass sensitivity of 1.1~\msun~beam$^{-1}$ at 20~K, we investigate the star formation and gas flow structures within the ionizing feedback-driven structure, a clump-scale massive ($\gtrsim1500$~\msun) bright-rimmed cloud (BRC) associated with IRAS 18290$-$0924. This BRC is bound only if external compression from ionized gas is considered. A small-scale ($\lesssim1$~pc) age sequence along the direction of ionizing radiation is revealed for the embedded cores and protostars, which suggests triggered star formation via radiation-driven implosion (RDI). Furthermore, filamentary gas structures converge towards the cores located in the BRC's center, indicating that these filaments are fueling mass towards cores. The local core-scale mass infall rate derived from \htcopone\ blue profile is of the same order of magnitude as the filamentary mass inflow rate, approximately 1~\msun~kyr$^{-1}$. A photodissociation region (PDR) covering the irradiated clump surface is detected in several molecules, such as \cch, \hcop, and \cs\ whereas the spatial distribution stratification of these molecules is indistinct. \cch\ spectra of the PDR possibly indicate a photoevaporation flow leaving the clump surface with a projected velocity of $\sim2$~\kms. Our new observations show that RDI accompanied by a clump-fed process is operating in this massive BRC. Whether this combined process works in other massive BRCs is worth exploring with dedicated surveys. 
 

\end{abstract}




\section{Introduction} \label{INTRODUCTION}
How massive stars gain their mass during formation is a longstanding problem \citep{Bonnell01, Mckee03, Zinnecker07, Krumholz09}. \citet{Motte18} highlighted the key roles of global hierarchical collapse and clump-fed process in high-mass star formation (HMSF). \citet{Padoan20} proposed that the large-scale and converging inertial flows in supersonic-turbulence environment are assembled to form clumps and cores as a first step toward formation of massive stars. The intricate interplay between birthplaces and environment in these scenarios shows that the environmental factors beyond the core scale ($\sim0.1$~pc) are crucial to understand HMSF.

One of the widely discussed environmental factors is the feedback from external ionized (\hii) regions. The close associations between  \hii\ regions and other HMSF sites have been well discussed in a number of works, for example by \citet{Thompson12} for UC\hii\ regions, \citet{Kendrew16} for massive cold clumps,  \citet{Palmeirim17} for young stellar objects (YSOs), and \citet{Zhang20,Zhang21} for high-mass starless clumps. However, whether ionization feedback induces or suppresses formation of the next-generation stars is highly debatable and is potentially linked to the key questions of star formation efficiency (SFE) of the Milky Way \citep{Geen17,Fukushima20,Gonzalez20}.
The mechanical and radiative feedback of ionized regions could induce star formation mainly via two mechanisms: (1) Collect and collapse: A massive molecular shell is collected between an ionization front (IF) and a shock front (SF), and then this shell fragments and collapses to form stars \citep{Whitworth94, Deharveng05, Zavagno07,LiuHL15,LiuHL16,LiuHL17}. (2) Radiation-driven implosion \citep[RDI,][]{Bertoldi89}: The surface of a pre-existing molecular clump is ionized by the UV radiation from nearby OB stars, plus the pre-existing ionized gas of the \hii\ region that is jammed on the clump surface, forming an over-pressure ionized boundary layer (IBL). This IBL drives a photoevaporation flow leaving clump surface and a shock penetrating the clump interior, compressing the clump and inducing its collapse for future star formation \citep{Lefloch94}.

\begin{figure}
     \centering
   \includegraphics[width=0.48\textwidth]{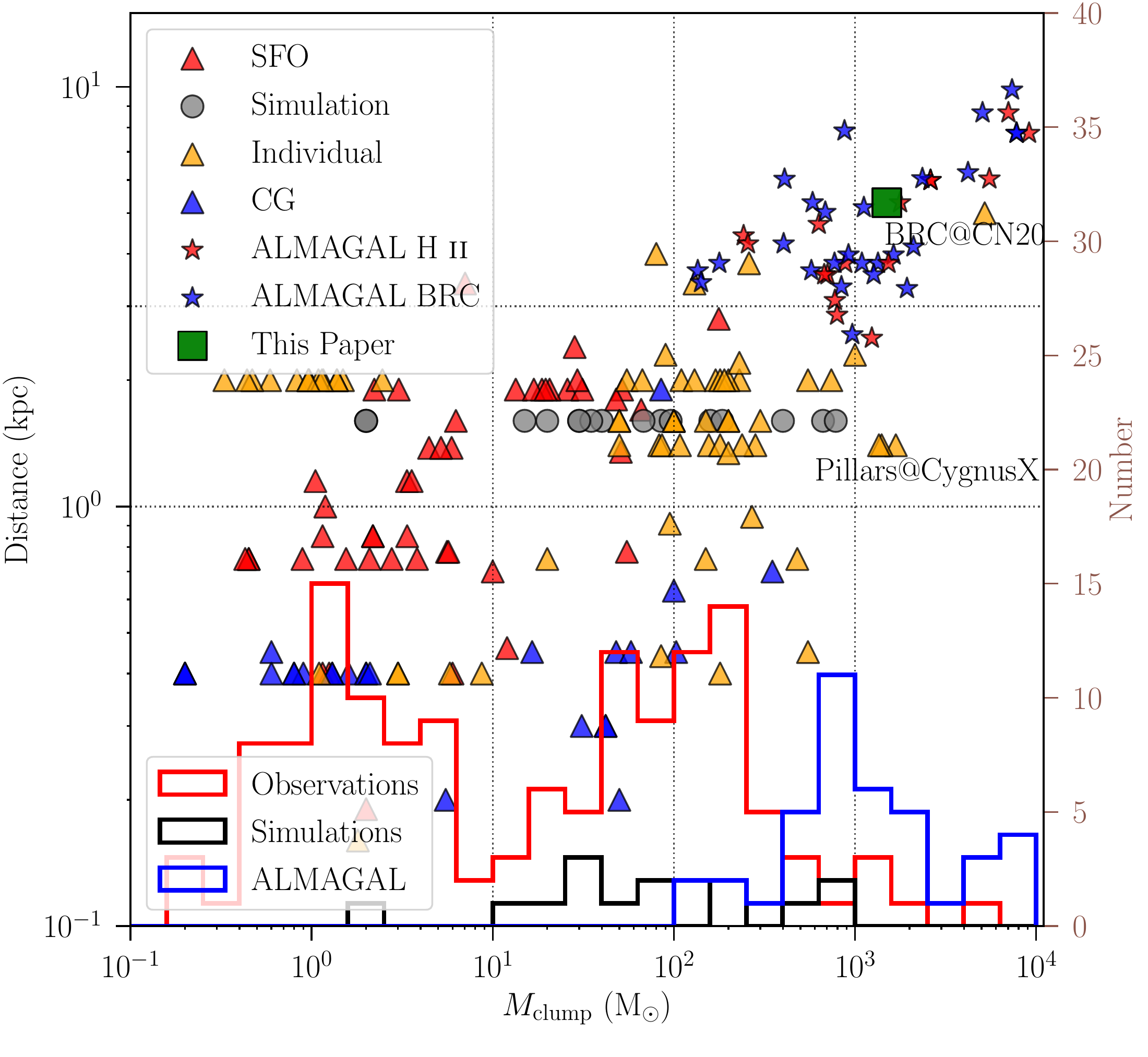}
       \caption{Summary of the published studies about RDI. The meanings of ``SFO'', ``CG'', ``individual'', ``ALMAGAL \hii'', and ``ALMAGAL BRC'' can be found in Sect~\ref{INTRODUCTION}. ``BRC@CN20'' \citep{Orgeta16} and ``Pillars@CygnusX'' \citep{Schneider16} highlight the two single-dish observations which have the clump mass of $> 1000$~\msun. The histograms show the number distributions of different types of studies. Dashed grids mark the mass of 10, $10^2$, and $10^3$~\msun\ and the distances of 1 and 3~kpc. The mass and distance of ALMAGAL sources are taken from \citet{Urquhart18}. The mass for other sources is mainly derived from either continuum emission of dust or line emission of molecules, e.g. \tco\ and \ceo. Therefore, the uncertainty of the mass estimation for all sources in the figure is difficult to unify, and thus should be cautioned. The distance for simulation work is assumed to be the averaged distance of the entire sample because many simulations do not set the distance. The detailed information for the whole sample in this figure is in Appendix~\ref{App:Figure1Information}.}
        \label{StatisticsMap}
    \end{figure}

The bright-rimmed clouds (BRCs), with the bright rim generally tracing the IBL that can be detected in free-free  and/or recombination line emission, are candidate sites of the RDI mechanism being at work. BRCs were first systematically searched by Sugitani, Fukui and Ogura \citep[so called ``SFO'' BRCs,][]{Sugitani91, Sugitani94} in the surrounding of the optically visible \hii\ regions \citep[e.g.][]{Sharpless59} with a size larger than 60\arcmin, which biases their selected BRCs to the near distances and thus to the low- or intermediate-mass regime. Figure~\ref{StatisticsMap} shows the mass distribution for BRCs and other objects that may relate RDI, such as cometary globules (CGs) and pillars, collected by us from a number of observational and modelling works (200+ sources in 50+ papers, see Appendix Table~\ref{SourceInformation}). In the figure, the ``SFO'', ``CG'', and ``individual'' represent the BRCs catalogued by Sugitani, Fukui and Ogura, the CGs catalogued by \citet{Maheswar08}, and other individual case studies not included in the two catalogues, respectively. Although it is impractical to list all related works, we see that not only ``SFO'' BRCs are mainly in low- to intermediate-mass regime, but also for other potentially RDI-driven objects in many follow-up works.

This observed regime may be due to both observational biases and physical natures. On the one hand, nearby bright \hii\ regions attract more attention in the previous selection of BRCs, leading to the bias toward the low- to intermediate-mass BRCs. On the other hand, a large fraction of clump mass are being peeled off by photoionization, dissociation, and evaporation. In addition, the turbulent nature of the massive clumps could also assist in forming less massive BRCs. Turbulent HMSF regions prefer to form low-mass pillar-like structures when strong UV radiation penetrates low-density regions in the turbulent clouds \citep{Gritschneder09a,Gritschneder09b}.  

Actually, massive BRCs with a mass more than few thousand solar masses can be found in our Galaxy, such as those selected by us from the ALMAGAL survey in Fig.~\ref{StatisticsMap}. The ALMAGAL (ALMA Evolutionary study of High Mass Protocluster Formation in the Galaxy, Project ID: 2019.1.00195.L) targets at using ALMA to survey nearly 1000 dense ($\gtrsim 0.1$~g~\cm) and massive ($\gtrsim 500$~\msun) Hi-GAL clumps \citep{Elia17} within distance $\lesssim7.5$~kpc. A check on $4\arcmin\times4\arcmin$ environment of 860 ALMAGAL clumps using the \textit{Spitzer} 24/8.0/4.5~\micron\ RGB image reveals that a number of ALMAGAL clumps are close to the extended 24~\micron\ emission region. The two example maps are shown in Appendix Figs.~\ref{108768-fig} and \ref{109047-fig}. The extended 24~\micron\ emission of hot dust strongly correlates with the radio free-free emission of \hii\ region, as revealed by \citet{Ingallinera14} and \citet{Makai17}. Moreover, this correlation forms the basis that mid-infrared images such as \textit{Spitzer} 24/8.0/4.5 \micron\ RGB images can be used for identification of bubble-like \hii\ regions \citep{Churchwell06, Anderson14}.   Some of \hii-region-nearby ALMAGAL clumps present a convex shape pointing to the inner of extended 24~\micron\ emission region and thus show a BRC morphology (``ALMAGAL BRC'' in Fig.~\ref{StatisticsMap}). It is worth noting that these massive BRCs are not studied yet regarding RDI. Some ALMAGAL clumps are simply located on the edge of \hii\ regions and do not present a prominent convex shape (so called ``ALMAGAL \hii'' in Fig.~\ref{StatisticsMap}).

The existence of ALMAGAL BRCs in Fig.~\ref{StatisticsMap} raises the questions on whether and how RDI works in massive BRCs, which have not been well explored yet by existing observations. 
The few BRCs \citep{Orgeta16, Schneider16} with clump mass $\rm M_{\rm clp} \gtrsim~1000~M_{\odot}$ which belong to ``individual'' type in Fig.~\ref{StatisticsMap} have been only studied with single-dish radio telescopes; hence, because of their far distances of several kpcs, the small scales of $\lesssim0.1$~pc representative of the core scale remain inaccessible in this mass regime. In this work, we present a comprehensive, pilot study of the clump-scale massive ($\sim1500$~\msun) bright-rimmed cloud IRAS 18290$-$0924 (I18290 hereafter). Using ALMA high-resolution data, we demonstrate that the radiation-driven implosion is at work in a
framework of clump-fed scenario for this massive BRC by investigating in great detail star formation therein and its gas kinematics. The results of this pilot work open a new window for studying details of the radiation-driven implosion in massive BRCs.

\section{Main Data}
\label{MAINDATA}
I18290 was observed by the ALMA project ATOMS (ALMA Three-millimeter Observations of Massive Star-forming regions, Project ID: 2019.1.00685.S; PI: Tie Liu, \citealt{LiuT20ATOMSI}), which aims to survey 146 massive star-forming clumps \citep{Faundez04} at ALMA Band 3 using single-point mapping. The combined main array and ACA 7~m data result in a continuum sensitivity of 0.08~\mjybeam\ and a spatial resolution of $\sim2$\arcsec\ ($\sim0.05$~pc at $D = 5.34$~kpc, see the next section), with an imaged field radius of 44\arcsec\ (down to 20\% primary beam response). More details about the ALMA data can be found in Appendix~\ref{App:ALMADataDetails} and ATOMS papers \citep[e.g.][]{LiuT20ATOMSII,LiuHL21ATOMSIII,LiuHL22ATOMSV,Liu22b}. Besides, a series of supplementary data, including single-dish CO maps, the Very Large Array (VLA) 20~cm continuum maps and \nht\ cube, the \textit{Herschel} column density and dust temperature maps, and the \textit{Spitzer} images are described in Appendix \ref{App:SupplementaryDataDetails}.

\section{BRC environment and global status} \label{ENVIRONMENTSANDSTATUS}
\subsection{Environment}
\begin{figure}
     \centering
   \includegraphics[width=0.49\textwidth]{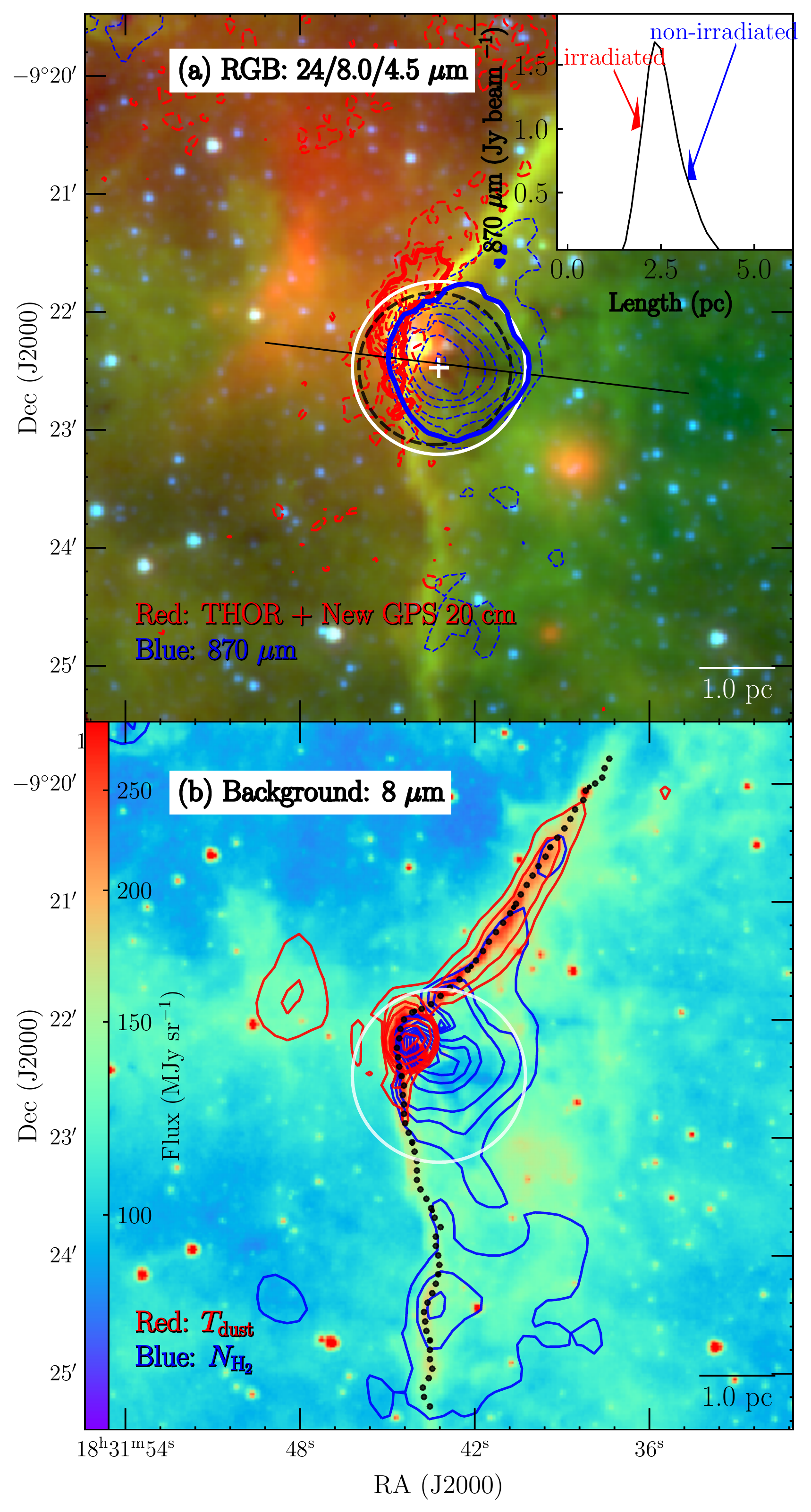}
 \caption{Environment of I18290. \textit{Panel a:} \textit{Spitzer} 24/8.0/4.5~\micron\ RGB image overlaid with the contours of 20~cm (red dashed) and ATLASGAL 870~\micron\ (blue) continuum. The levels of 20~cm continuum are [4, 5, 6, 7, 8, 9, 10, 12, 14, 16]~$\times \sigma_{\rm 20~cm}$, where noise of the 20~cm image $\sigma_{\rm 20~cm} = 0.45$~\mjybeam. The red bold contour of 20 cm emission (level: 2.38~\mjybeam\ = 40\% of the continuum peak) highlights the region considered in the IBL calculations. The levels of 870~\micron\ emission are [0.2, 0.3, 0.4, 0.5, 0.7, 0.9, 1.3, 1.8]~\jybeam. The white and black circles indicate the ATOMS-imaged field and the clump effective radius $r_{\rm clp}$ given by \citet{Urquhart18}, respectively. The blue bold contour of 870~\micron\ continuum with a level of 0.3~\mjybeam\ indicates a region with a size roughly equivalent to the circle size of $r_{\rm clp}$.  \textit{The small panel} shows the 870~\micron\ intensity profile along the black line. \textit{Panel b:} The 8~\micron\ emission overlaid with the contours of the \textit{Herschel} dust temperature \dustt\ (red) and column density \nhtcd\ (blue), with levels of [21.3, 21.6, 21.9, 22.2, 23.2, 24.2, 25.2]~K and [2.4, 2.7, 3, 4, 5, 6, 7, 8, 9, 10]~$\times 10^{22}$~\cm, respectively. The dotted line marks the rim spine extracted by \texttt{Filfinder} from 8 \micron\ image.}
        \label{GlobalEnviron}
    \end{figure}

 The environment of I18290 is shown in Fig.~\ref{GlobalEnviron}. A rim bright at 8~\micron, mainly tracing the mid-infrared emission of the aromatic infrared bands of polycyclic aromatic hydrocarbons (PAHs) in photodissociation region (PDR), extends from the ATLASGAL 870~\micron\ central clump \citep{Urquhart18} to the north and south. The 8-\micron\ rim extracted by the algorithm \texttt{Filfinder} \citep{Koch15} has a length of $\sim10$~pc, a beam-deconvolved FWHM width of $\sim0.36$~pc, and a mean column density \nhtcd\ from \textit{Herschel} maps of $\sim2\times10^{22}$~\cm. The total mass of the 10-pc rim excluding the central clump derived from the \textit{Herschel} \nhtcd\ map is $\sim1100$~\msun. The details of rim extraction and mass calculations can be found in Appendix~\ref{App:BrightRimIdentification}. An IBL covering nearly half of the clump surface is traced by 20~cm free-free emission shown in Fig.~\ref{GlobalEnviron}. 
 
 I18290 is located in the giant molecular complex (GMC) G23.3$-$0.3 which harbours several \hii\ regions and supernova remnants (SNRs) at a distance around 4 to 5~kpc \citep{Su14,Messineo14}. The rim points to the geometric center of an OB cluster with more than 10 OB stars \citep[regions REG7 and GLIMPSE09 in Fig.~8 of][]{Messineo14}, with a projected separation of $\sim30$\arcmin\ (45~pc). Therefore, we propose that these OB stars are potential exciting sources for the rim. 

\subsection{Clump under external compression}  \label{COMPRESSION}
The kinematic distance and corresponding SED-resulted clump mass \mclump\ derived by \citet{Urquhart18} are 5.34~kpc and 1420~\msun, respectively, consistent with other studies \citep{Lu14,Mege21}. Our independent calculations using \tcoone\ transition give $M_{\rm clp} \simeq1450$~\msun\ with an error $\sim30$\% (details in Appendix~\ref{COLUMNDENSITYANDMASS-CLUMPMASS}). Considering errors from abundance, excitation temperature \tex, and beam size, the \tco-derived \mclump\ is in good agreement with the continuum-derived \mclump. The total mass of the clump and the rim is $\sim2500$~\msun.

Virial status is key to understanding the state of equilibrium for the clump, especially if the clump is deeply influenced by the external compression from IBL. The compression is confirmed by the remarkably steeper radial profile of cold dust emission for the side facing the ionizing radiation shown in \textit{Panel a} of Fig.~\ref{GlobalEnviron}, similar to the RCW 120 bubble shell revealed by \citet{Zavagno20}. The power-law fittings to the ATLASGAL 870~\micron\ intensity profiles \citep{Schuller09} of I18290 yield the power-law indexes of $-1.34\pm0.3$ and $-0.55\pm0.1$ for the irradiated and non-irradiated sides, respectively. 

We first estimate the external pressure \Pex\ exerted by the ionized region which is composed of two parts: (1) the ram pressure from the sonic ionized flow in the form of $P_{\rm ram} = \rho_{\rm i} {c_{\rm i}}^2$ and (2) the bulk pressure of the IBL via $P_{\rm IBL} = \rho_{\rm i} {c_{\rm i}}^2$ \citep{Lefloch94,Morgan04,Haworth12}, where $c_{\rm i} = \sqrt{2.2 k_{\rm B} T_{\rm e} / \mu_{\rm i} m_{\rm H} }$ is the sound speed of ionized gas at the electron temperature $T_{\rm e}$, $k_{\rm B}$ is the Boltzmann constant, $\mu_{\rm i}=1.4$, $m_{\rm H}$ is the mass of the atomic hydrogen, and the factor of 2.2 appears owing to that there are 2.2 free particles per H nucleus \citep[0.1 He per H, and 1.1 electrons per H,][]{Krumholz17}. The total external pressure can be expressed as below:
\begin{equation} \label{ExternalPressureEquation}
P_{\rm ex} = P_{\rm IBL} +  P_{\rm ram} = 2\rho_{\rm i} {c_{\rm i}}^2 = 2n_{\rm e}m_{\rm H}{c_{\rm i}}^2.
\end{equation}

Using an electron number density $n_{\rm e}$ of 240~cm$^{-3}$ derived from the 20~cm free-free emission and an electron temperature \te\ of 7180~K derived from the  empirical relation between \te\ and Galactic distance \RGC, we have $P_{\rm ex}/k_{\rm B}\simeq 5.7\times10^6$~K~cm$^{-3}$. Here the details of calculating $n_{\rm e}$ and \te\ can be found in Appendix \ref{App:IBLCalculations}.

We then estimate the clump inner pressure $P_{\rm in} \simeq \rho_{\rm in} \sigma_{\rm 1D}^2$, where $\rho_{\rm in}$ and $\sigma_{\rm 1D}$ are the clump mass density and the deconvolved 1D velocity dispersion, respectively. Considering that \tco\ traces intermediate-density gas compared to \co\ and \ceo, and that \tco\ higher transitions trace warmer and denser gas compared to $J = 1 - 0$ (e.g. ${\rm E}_{u} = 5.5$ and 33.2~K for $J = 1 - 0$ and $J = 3 - 2$, respectively), we use FOREST Unbiased Galactic plane Imaging survey with the \textit{Nobeyama} 45~m telescope \citep[FUGIN,][]{Umemoto17} \tcoone\ to derive $\sigma_{\rm 1D}^2 = \sigma_{\rm sp}^2 - \sigma_{\rm ch}^2$. The $\sigma_{\rm ch}$ and $\sigma_{\rm sp}$ are the velocity dispersions for the channel and the fitted Gaussian of clump-averaged spectra, respectively. The corresponding spectra are presented in Appendix Fig.~\ref{SingleDishSpectraFigure}. The clump radius $r_{\rm clp} \simeq 1$~pc is from the effective radius given by \citet{Urquhart18}, which corresponds a size approximately equivalent to the region with the ATLASGAL 870~\micron\ emission above 0.3~\mjybeam\ (indicated by the blue bold contour in \textit{Panel a} of Fig.~\ref{GlobalEnviron}). We have $P_{\rm in}/k_{\rm B} = 3.4\pm1.0\times10^6$~K~cm$^{-3}$ with $\sigma_{{\rm sp, ^{13}CO}~J=1-0}$ = 1.66~\kms.

The derived \Pex\ is in a range similar to those of the low-mass BRCs surveyed by \citet{Thompson04}. In the most massive BRC that RDI has been discussed, BRC@CN20 ($M_{\rm clp}\simeq5.2\times10^3$~\msun), \citet{Orgeta16} derived $P_{\rm in}/k_{\rm B}$ = $1.2\pm0.5\times10^8$~\pressure\ and $P_{\rm ex}/k_{\rm B} = 5.4\pm2.5\times10^6$~\pressure, illustrating that BRC@CN20 IBL compression is too weak compared to the clump inner pressure and therefore RDI is unable to operate. 

Differing from BRC@CN20, I18290 has an over-pressure IBL compressing the clump. The virial status of I18290 can be significantly changed by this compression. First, we take into account two simplified cases for the external pressure, \Pex, from ionized gas: (1) \textit{The IBL enshrouds the whole clump surface}. The corresponding virial equation without magnetic fields and rotation is expressed as \citep{Lequeux05,Bodenheimer11}:
\begin{equation} \label{VrialEquation}
F = 2U + \Omega - 4\pi {r_{\rm clp}}^3 P_{\rm ex},
\end{equation}
where the internal kinetic energy $U$ and the potential energy $\Omega$ are
\begin{equation} \label{Energy}
 U = \frac{3}{2} M_{\rm clp} {\sigma_{\rm 1D}}^2,\\\Omega = - \frac{3}{5}\alpha\beta\frac{G{M_{\rm clp}}^2}{r_{\rm clp}}.
\end{equation}
The factors $\alpha$ and $\beta$ are related with power-law index of the density profile and clump eccentricity \citep{LiDi13}, respectively. Using $\alpha$ derived from the 870~\micron\ radial profile at the irradiated side and $\beta$ derived from the 870~\micron\ geometry \citep{Urquhart14}, we find that $F<0$ and there is no solution of the virial mass \mvir\ for the quadratic equation $F\left(M_{\rm vir}\right) = 0$. Thus, I18290 is always bound or subvirialized. (2) \textit{Without the external pressure from IBL.} This could be treated as the pre-compression status. With $P_{\rm ex}=0$ and $\alpha$ derived from the 870~\micron\ emission radial profile at the non-irradiated side, we solve $F\left(M_{\rm vir}\right) = 0$. The resulting \mvir~$\simeq2320$~\msun. Consequently, I18290 is unbound if IBL compression is ignored.

Second, the IBL actually only covers a part of the clump surface rather than the whole surface. The factor of $4\pi$ in the pressure item of Eq.~\ref{VrialEquation} could be replaced by an actual solid angle $\theta$ covered by the IBL on the clump surface. The potential energy $\Omega$ is correlated with the distribution of density ($\alpha$) and the shape of clump ($\beta$). The changes of IBL coverage $\theta$ propagates to $\alpha$ and $\beta$ due to the corresponding changes of compression. Therefore, the full form of the $F$ - $\theta$ equation is much complicated. A steeper density profile and a more elongated shape lead to a more bound status for clump \citep{LiDi13}. Here we simply estimate a critical IBL coverage solid angle $\theta_{\rm cri}$, which is the minimum solid angle allowing IBL to bind the clump. By setting $F(M_{\rm clp}) = 2U + \Omega - \theta_{\rm cri} {r_{\rm clp}}^3 P_{\rm ex} = 0$, $\alpha$ derived from non-irradiated sides and $\beta$ derived from a spherical shape (to have a smaller $|\Omega|$), we have an upper limit of 3.5~$rad$ for $\theta_{\rm cri}$. Accounting for this small $\theta_{\rm cri}$ which is around a quarter of the clump surface, it is rational to suggest that the clump is still bound even considering the actual coverage of IBL.  

Recalling our previous virial calculations without external pressure for the candidate high-mass starless clumps which are impacted or non-impacted by ionized regions \citep{Zhang20}, we found a noteworthy difference that 90\% of the non-impacted clumps are bound while only 50\% of the impacted clumps are bound. The lower bound fraction for the impacted high-mass starless clumps could be just a biased result because the ignored IBL compression could contribute significantly in binding clumps.

\section{Sequential star formation within clump-fed scenario} \label{STARFORMATIONCLUMPFED}
The observations targeting RDI in low-mass cases reveal that star formation follows a small-scale ($\lesssim1$~pc) age sequence along the direction of ionizing radiation. The more evolved protostars or cores are closer to the exciting massive stars. This sequential star formation is thought to be a relic of shock propagation. The shock driven by over-pressure IBL propagates into the clump's interior and then triggers clumps/cores to collapse and sequentially form stars \citep{Sugitani95,Fukuda02,Ikeda08, Getman09,Chauhan09, Choudhury10, Fukuda13, Panwar14, Imai17}. Following the shock speed estimation formula and methods in \citet{Urquhart07}:
\begin{equation}
\centering
{\rm v}^2_{\rm shock} = \alpha_{\rm shock} \frac{\left(P_{\rm s}- P_{\rm n}\right)}{\rho_{\rm n}},
\label{ShockVelocity}
\end{equation}
where $P_{\rm s}$ is the shocked gas pressure, and $P_{\rm n}$ and $\rho_{\rm n}$ are the pressure and mass density of the pre-shock gas, respectively. The $\alpha_{\rm shock}$ is a factor about one to two, depending on the detailed properties of the shock \citep{White99}.  Assuming $P_{\rm s} = P_{\rm ex}$, $P_{\rm n} = P_{\rm in}$, and $\rho_{\rm n} = \rho_{\rm in}$, the estimated ${\rm v}_{\rm shock}$ is about 1.5~\kms. Therefore, the squeezing exerted by I18290 IBL may power a shock with a velocity of $\sim1.5$~\kms\ that triggers star formation sequentially via RDI. We should note that the uncertainties of shock speed estimation may be large and the estimated shock speed is meaningful only regarding the order of magnitude. To explore the star formation activities in I18290, we first extract candidate dense cores and YSOs and then compare their evolutionary stages.

\subsection{Dense cores} \label{DENSECORES} 
\begin{landscape} 
\begin{figure}
     \centering
   \includegraphics[height=0.8\textheight]{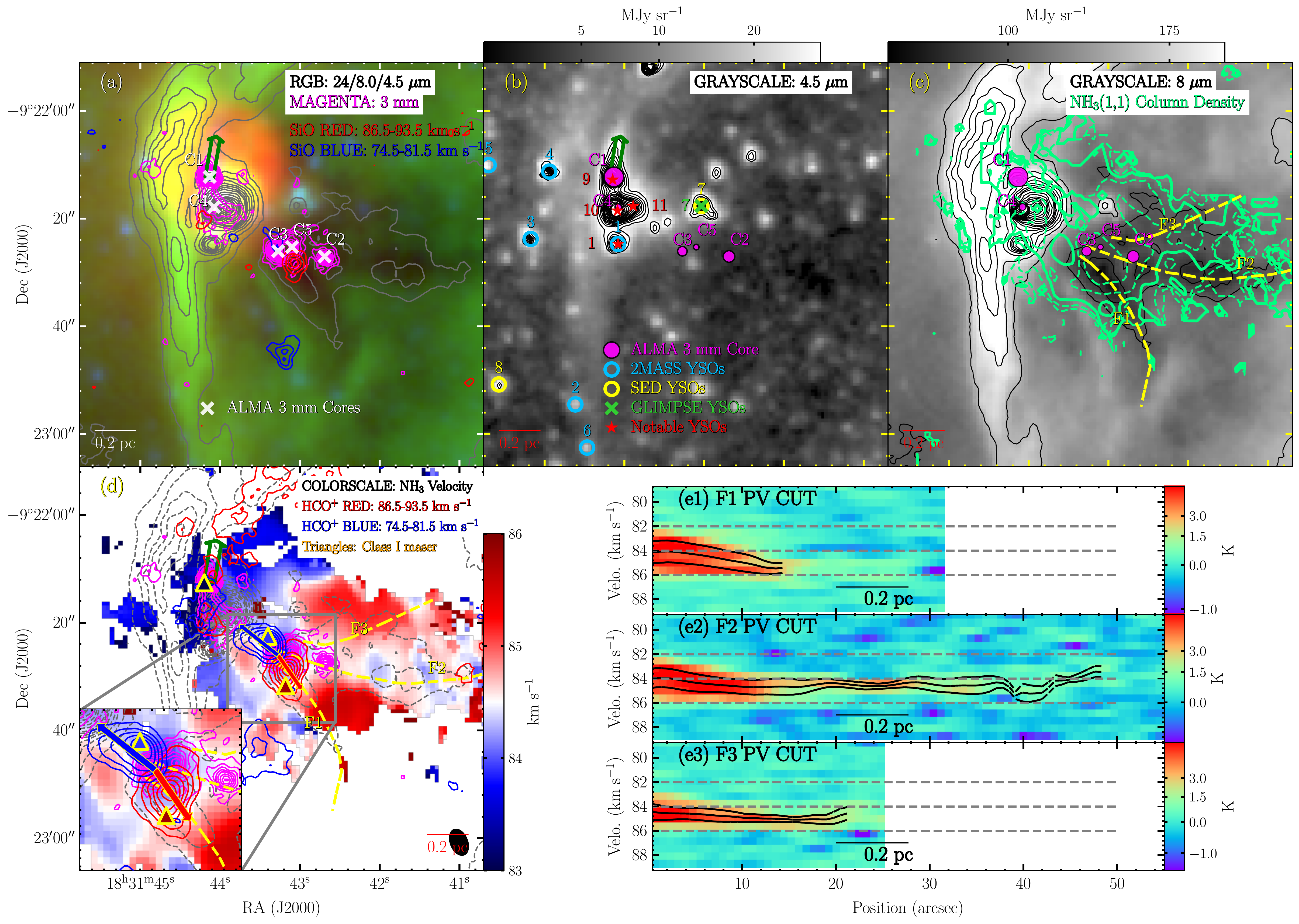}
       \caption{Star formation in the I18290 region. \textit{Panel a}: \textit{Spitzer} 24/8.0/4.5~\micron\ RGB image overlaid with 8~\micron\ emission (the gray contours: [0.8, 0.9, 1, 1.5, 2, 2.5, 3, 3.5, 4, 4.5, 5, 6, 7, 8, 9, 10, 11, 12, 13, 14]~$\times 100$~MJy~sr$^{-1}$), SiO outflow lobes (the red and blue contours: [2, 3, 4, 5, 6, 7, 8, 9, 10]~$\times \sigma_{\rm SiO~lobe}$, where image noise $\sigma_{\rm SiO~lobe} = 1.5$~K~\kms), and 3~mm continuum (the magenta contours: [3, 5, 7, 9, 11, 14, 17, 20, 23]~$\times \sigma_{\rm 3~mm}$, where noise of the 3~mm continuum image $\sigma_{\rm 3~mm} = 0.08$~\mjybeam). \textit{Panel b}: Grayscale image and the black contours (levels: [3, 4, 5, 6, 7, 8, 9, 10, 12, 14, 16, 18, 20, 24, 28, 32]~$\times10$~MJy~sr$^{-1}$) show 4.5~\micron\ emission, overlaid with 3-mm cores and extracted YSOs (see all YSOs in Table~\ref{CandidateYSOtable}). The circles show the 3-mm cores with a size in proportion to core mass. The green arrow indicates EGO extended 4.5~\micron\ emission. \textit{Panel~c}: Grayscale image and the black contours show 8~\micron\ emission, overlaid with the light green contours of \nht\ column density \nnhtcd\ derived from \texttt{fiteach} (levels: [14.5, 14.75, 15, 15.25, 15.5]$\times \rm log(10)~cm^{-2}$) with a highlighted level of 10$^{15}$~\cm\ (solid contour). The yellow dashed lines outline filaments. \textit{Panel d}: \nht\ centroid velocity derived from \texttt{fiteach}. The lobes of \hcop\ outflow are shown in the red and blue contours (levels: [2, 3.5, 5, 6.5, 8, 9.5, 11, 12.5]~$\times \sigma_{\rm HCO^{+}~lobe}$, where the image noise $\sigma_{\rm HCO^{+}~lobe} = 2$~K~\kms). The red and blue arrows indicate outflow directions and lengths. The gray contours show the 8~\micron\ emission. The triangles show \chtoh\ Class I masers found by \citet{Rodrguez17}, with a color indicating maser velocity. The beam size for the observations of \citet{Rodrguez17} is $\sim2$\arcsec. The black ellipse shows the VLA beam of \nht\ image. \textit{Panels e1}, \textit{e2}, and \textit{e3}: \nht\ (1,1) main line position-velocity cuts along filaments. The black lines mark the line centroid and dispersion fitted with \texttt{fiteach}.}
    \label{StarFormationFigure}
\end{figure}
\end{landscape}

 The dust dense cores are extracted from the ATOMS 3~mm continuum image using \texttt{Astrodendrogram} \citep{Rosolowsky08}. A total of five extracted cores (C1 to C5) with masses from 8 to 76~\msun\ are shown and listed in Fig.~\ref{StarFormationFigure} and Table~\ref{ALMACoresPhysical}, respectively (see details of the \texttt{Astrodendrogram} extraction and physical properties estimation in Appendix~\ref{CoreExtraction}). The contamination from free-free emission of the ionized gas on the 3~mm flux is minimal because of non detection of the VLA 20 cm \& 6 cm continuum emission (VLA beam $\sim5$\arcsec, sensitivity $\simeq0.2$ to 0.4~\mjybeam, \citealt{Helfand06}) and the ATOMS \hfourtyalpha\ line emission for these cores. 

\begin{table*}
\caption{\label{ALMACoresPhysical}Physical parameters of cores.} 
\begin{threeparttable}
\setlength{\tabcolsep}{4.pt}
\renewcommand{\arraystretch}{1.0}
\centering
 
\begin{tabular}{clrrcccccrrc}
\hline
\hline
Core & Associated YSO\tnote{\textit{(a)}} &\tkin\tnote{\textit{(b)}} & Min \tkin\tnote{\textit{(b)}} & Max \tkin\tnote{\textit{(b)}} & $M_{\rm core}$\tnote{\textit{(c)}} & $M_{\rm core}^{\rm cold}$\tnote{\textit{(c)}} & $M_{\rm core}^{\rm warm}$\tnote{\textit{(c)}} & $r_{\rm core}$ & $\Sigma_{\rm core}$ & \nhtcd & \nhtnd \\
     &                 & K & K & K & \msun & \msun & \msun & AU & g~cm$^{-2}$ & $10^{23}$~cm$^{-2}$  & $10^6$~cm$^{-3}$ \\
      \hline
C1\tnote{\textit{(d)}} & YSO \#9 & 21.0 & 6.3  & 27.3 & 102.3$\pm$38.9 & 458.4$\pm$174.4 & 76.8$\pm$29.2 & 4656$\pm$465 & 13.3$\pm$4.3 & 28.5$\pm$9.2 & 30.6$\pm$10.4 \\
C2 &           & 17.7 & 17.2 & 17.9 & 36.1$\pm$14.3 & 37.1$\pm$14.7    & 35.5$\pm$14.1 & 4905$\pm$490 & 4.2$\pm$1.4  & 9.0$\pm$3.1  & 9.2$\pm$3.3 \\
C3 &           & 22.0 & 18.4 & 23.4 & 24.0$\pm$9.1  & 29.4$\pm$11.2    & 22.4$\pm$8.5 & 3033$\pm$303  & 7.4$\pm$2.4  & 15.7$\pm$5.1 & 26.0$\pm$8.8 \\
C4 & YSO \#10  & 26.7 & 24.5 & 28.7 & 12.0$\pm$4.7  & 13.1$\pm$5.2     & 11.1$\pm$4.4 & 3949$\pm$394  & 2.2$\pm$0.7  & 4.6$\pm$1.6  & 5.9$\pm$2.1 \\
C5 &           & 18.8 & 17.1 & 22.4 & 8.2$\pm$3.1   & 9.2$\pm$3.5      & 6.8$\pm$2.6 & 1828$\pm$182  & 7.0$\pm$2.3   & 14.8$\pm$4.8 & 40.7$\pm$13.8 \\
\hline
   \end{tabular}
      \begin{tablenotes}
      \item [\textit{(a)}] Candidate YSOs associated with 3-mm cores, see Table~\ref{CandidateYSOtable}.
      \item [\textit{(b)}] VLA beam-averaged \nht\ kinetic temperature \tkin, and the \nht\ max \tkin\ and min \tkin\ in the corresponding core's area, respectively.
      \item [\textit{(c)}] $M_{\rm core}$, $M_{\rm core}^{\rm warm}$, and $M_{\rm core}^{\rm cold}$ are the core masses with the beam-averaged \tkin, the max \tkin, and the min \tkin, respectively.
      \item [\textit{(d)}] C1 is located on the edge where \nht\ intensity decreases dramatically due to effect from IBL. Therefore, the mass of C1 derived from the max pixel \tkin\ ($M_{\rm core}^{\rm warm}$) is probably more accurate than that derived from the beam-averaged \tkin\ ($M_{\rm core}$).
      \end{tablenotes}
      \end{threeparttable}
\end{table*}

The detected cores can be explained by a clump-to-core fragmentation dominated by thermal motions, under a mass sensitivity of 1.1~\msun~beam$^{-1}$ estimated from the rms level (0.08~\mjybeam) of the continuum map assuming $T_{\rm dust}=20$~K. With thermal Jeans fragmentation, the clump is expected to fragment to thermal Jeans cores with a mass and separation around  \mjth\ and \lambdajth:
\begin{equation}  
M_{\rm J}^{\rm th} = \frac{4\pi\rho}{3} \left(\frac{\lambda_{\rm J}^{\rm th}}{2}\right)^3 = \frac{\pi^{5/2}}{6} \frac{{\sigma_{\rm th}}^3}{\sqrt{G^3\rho}},\\
\lambda_{\rm J}^{\rm th} = \sigma_{\rm th} \left( \frac{\pi}{G\rho} \right) ^{1/2},
\label{equ-mj-lambdaj}
\end{equation}
where $\sigma_{\rm th}$ is the thermal velocity dispersion
\begin{equation}  
\sigma_{\rm th} = \left( \frac{k_{\rm B} T}{\mu m_{\rm H}} \right) ^{1/2}.
\label{equ-thermal}
\end{equation}
The mean molecular weight per free particle $\mu = 2.37$ because $\sigma_{\rm th}$ is governed by \hmole\ and He. When taking into account of both thermal and non-thermal motions, the corresponding turbulent Jeans parameters are 
\begin{equation}  
M_{\rm J}^{\rm tot} = \frac{\pi^{5/2}}{6} \frac{{\sigma_{\rm tot}}^3}{\sqrt{G^3\rho}},\\
\lambda_{\rm J}^{\rm tot} = \sigma_{\rm tot} \left(\frac{\pi}{G\rho} \right) ^{1/2},
\label{equ-jeans-tot}
\end{equation}
where the total velocity dispersion $\sigma_{\rm tot} = \left( \sigma^2_{\rm \rm th} + \sigma^2_{\rm nth,^{13}{\rm CO}} \right)^{1/2}$, the \tcoone\ non-thermal dispersion $\sigma_{\rm nth,^{13}{\rm CO}} = \left(\sigma_{\rm 1D} ^2 - \sigma_{\rm th,^{13}{\rm CO}}^2 \right) ^{1/2}$, and  the \tcoone\ thermal dispersion $\sigma_{\rm th,^{13}{\rm CO}} = \left( {k_{\rm B} T}/{\mu_{\rm ^{13}{\rm CO}} m_{\rm H}} \right) ^{1/2}$. The \tco\ molecule weight $\mu_{\rm ^{13}{\rm CO}} = 29$.
In the specific case that the compression of large scale supersonic flow creates density enhancement by a factor of Mach number squared ${\mathcal {M}}^2$, the Jeans parameters are (\citealt{Zhang21} and the references therein)
\begin{equation}  
M_{\rm J}^{\rm com, flow}  = \frac{\pi^{5/2}}{6} \frac{\sigma_{\rm nth, ^{13}{\rm CO}}^3}{\sqrt{G^3\rho_{\rm eff}}},\\
\lambda_{\rm J}^{\rm com, flow} = \sigma_{\rm nth, ^{13}{\rm CO}} \left(\frac{\pi}{G\rho_{\rm eff}} \right) ^{1/2},
\label{equ-jeans-flow}
\end{equation}
where effective density $\rho_{\rm eff} = \rho \mathcal{M} ^2 = \rho \left(\sqrt{3}\sigma_{\rm nth, ^{13}{\rm CO}}/\sigma_{\rm th} \right)^2$.  

We have \mjthclump~$=13\pm5$~\msun, \lambdajth~$=0.46\pm0.1$~pc; \mjflowclump~$=280\pm110$~\msun, \lambdajflowclump~$=0.27\pm0.07$~pc; and \mjtotclump~$=3020\pm1060$~\msun, \lambdajtotclump~$=2.8\pm0.6$~pc. These results suggest that other supportive mechanisms like turbulence and magnetic field only play marginal roles in I18290's fragmentation process. It is consistent with \citet{Liu17} and \citet{Zhang21}, who revealed a thermal fragmentation for the dense cores in massive clumps impacted by ionized regions, but different from the \textit{Spitzer} mid-infrared observations of \citet{Sharma16}, which propose a non-thermally driven fragmentation for the YSOs embedded in BRCs.

\subsection{Young stellar objects}
Candidate YSOs (YSOs hereafter) are extracted from the GLIMPSE and 2MASS point source catalogs using color-color criteria or YSO SED models and then the background/foreground sources are excluded with the Gaia measurements \citep[detailed extraction methods can be found in Appendix~\ref{appendix:yso},][]{Robitaille06, Wang07, Gaia21}. The YSOs with valid Gaia measurements are selected to keep those within the distance of [4.34, 6.34]~kpc (1~kpc different from that of I18290) while the YSOs without Gaia measurements are simply assumed to have the same distance as I18290. With such selection, one of the two GLIMPSE, and six of the nineteen 2MASS color-color selected YSOs, and two of the four SED fitted YSOs remain. The extended emission of the 8-\micron\ rim PDR heavily influences source extraction and photometric measurements of the point sources. Therefore, three point sources bright at GLIMPSE band (YSOs \#9, 10, 11 in \textit{Panel b} of Fig.~\ref{StarFormationFigure} and Table~\ref{CandidateYSOtable}) are additionally included, which are located at the rim but not identified as YSOs by the three mentioned methods due to very bright background contamination in the IR images. All extracted YSOs are shown and listed in \textit{Panel b} of Fig.~\ref{StarFormationFigure} and Table~\ref{CandidateYSOtable}, respectively.

There are a total of four sources are found at the rim, YSOs~\#1, 9, 10, and 11. YSOs~\#9 and 10 are associated with C1 and C4, respectively, whereas YSOs~\#1 and 11 are not associated with any detected core. YSO~\#11 is the brightest 8~\micron\ point source and its 8~\micron\ emission partly overlaps with the 8~\micron\ emission of YSO~\#10 (C4). Gaia measurements suggest that YSO~\#11 is a foreground source located at a distance of $1.22_{-0.38}^{+1.0}$~kpc. YSO~\#1 seems to be a Herbig Ae/Be (HAeBe) star according to its 2MASS color (see Appendix~\ref{appendix:yso}). Whether YSO~\#1 is associated with I18290 is uncertain, no associated dense gas or dust emission is detected in the ATOMS data set. YSO~\#1 may couple with the rim structure based on the morphology that YSO~\#1 8-\micron\ emission is likely located in an emission dip in the zeroth moment maps of a number of the ATOMS molecules tracing the rim, such as \cch, \cs, and \so\ shown in Appendix Fig.~\ref{moleculartracer}. This complementary morphology between the 8~\micron\ emission of YSO~\#1 and the molecular emission of the rim can be explained as the dispersal of circumstellar dense gas by HAeBe star YSO~\#1, akin to the cases shown by \citet{Fuente98,Fuente02}.

\begin{table*}
\caption{\label{CandidateYSOtable}Candidate YSOs.}  
\begin{threeparttable}
\tiny
\centering
\setlength{\tabcolsep}{4.5pt}
\renewcommand{\arraystretch}{1.0}

\begin{tabular}{cccccccccccccc}
\hline
\hline
YSOs & RA     & DEC & D\tnote{\textit{(a)}}   & GLIMPSE & 2MASS & $J$\tnote{\textit{(b)}}   & $H$         & $K_{S}$ &     3.6~\micron &  4.5~\micron   &    5.8~\micron  &  8.0~\micron  &  Method \tnote{\textit{(c)}} \\
     & \degree & \degree & kpc                &         &       & mag &  mag      &   mag   &    mag &   mag &    mag &  mag &                               \\
\hline
 1         & 277.933667 &  -9.37353 &   5.34 & G022.3534+00.0648 & 18314408-0922247 &  15.11             & 14.402 $\pm$ 0.129 & 12.636 $\pm$ 0.064 & 10.713 $\pm$ 0.163 &  9.983 $\pm$ 0.167 &   8.33 $\pm$ 0.085 &                    &            2MASS \\ 
 2         & 277.935895 & -9.381802 &   5.34 &                   & 18314461-0922544 & 17.339             & 16.078             & 13.892 $\pm$ 0.078 &                    &                    &                    &                    &            2MASS \\ 
 3         &  277.93823 & -9.373271 &   5.34 & G022.3556+00.0610 & 18314517-0922237 & 16.248             & 13.384 $\pm$ 0.074 & 11.742 $\pm$ 0.044 & 10.546 $\pm$ 0.055 & 10.533 $\pm$ 0.083 & 10.291 $\pm$ 0.122 &                    &            2MASS \\ 
 4         & 277.937258 & -9.369796 &   5.34 & G022.3582+00.0634 & 18314494-0922112 & 17.164             & 14.434 $\pm$  0.08 &  12.03 $\pm$ 0.031 &                    & 10.045 $\pm$  0.15 &  9.356 $\pm$ 0.268 &                    &            2MASS \\ 
 5         & 277.940428 & -9.369465 &   5.34 & G022.3600+00.0608 & 18314570-0922100 & 17.178             & 14.803             & 13.441 $\pm$ 0.062 & 12.129 $\pm$ 0.084 & 11.868 $\pm$ 0.134 & 11.683 $\pm$ 0.215 &                    &            2MASS \\ 
 6         & 277.935281 & -9.384027 &   5.34 & G022.3448+00.0586 & 18314446-0923024 &  14.32             & 14.268 $\pm$ 0.101 & 13.341 $\pm$ 0.054 & 12.522 $\pm$ 0.128 & 12.448 $\pm$ 0.135 &                    &                    &            2MASS \\ 
 7         & 277.929311 & -9.371574 &   5.34 & G022.3531+00.0696 & 18314301-0922174 & 15.392 $\pm$ 0.265 & 13.053 $\pm$  0.08 & 11.801 $\pm$  0.04 &  10.34 $\pm$ 0.246 & 10.029 $\pm$ 0.118 &    9.2 $\pm$ 0.053 &  8.324 $\pm$ 0.15  &      GLIMPSE+SED \\ 
 8         & 277.939892 & -9.380771 &   5.34 & G022.3498+00.0561 & 18314555-0922506 & 15.841 $\pm$ 0.102 & 13.352 $\pm$ 0.024 & 12.263 $\pm$ 0.024 & 11.461 $\pm$  0.06 & 11.403 $\pm$ 0.075 & 11.243 $\pm$ 0.114 & 11.147 $\pm$ 0.158 &              SED \\ 
 9         & 277.933946 & -9.370218 &   5.34 & G022.3564+00.0662 &                  &                    &                    &                    &                    &  9.225 $\pm$ 0.162 &  8.267 $\pm$ 0.068 &                    &                  \\ 
 10        & 277.933692 & -9.371816 &   5.34 & G022.3549+00.0656 &                  &                    &                    &                    &                    &  8.202 $\pm$ 0.177 &  6.876 $\pm$ 0.122 &                    &                  \\ 
 11        & 277.932878 &  -9.37158 &   1.22 &                   & 18314389-0922176 & 14.069             & 13.939 $\pm$ 0.111 & 12.525 $\pm$ 0.087 &                    &                    &                    &                    &                  \\

\hline
\end{tabular}
      \begin{tablenotes}
      \item [\textit{{\normalsize (a)}}] {\normalsize If there is no Gaia measurement, an assumed distance of 5.34~kpc for candidate YSOs is used.} 
      \item [\textit{{\normalsize (b)}}] {\normalsize Measurements without error are upper limit magnitudes.} 
      \item [\textit{{\normalsize (c)}}] {\normalsize Labels ``2MASS'', ``GLIMPSE'', and ``SED'' mark the YSO extraction methods described in Appendix~\ref{appendix:yso}.}
      \end{tablenotes}
      \end{threeparttable}
\end{table*}

\subsection{Age sequence}
\begin{figure}
  \centering
  \includegraphics[width=.95\linewidth]{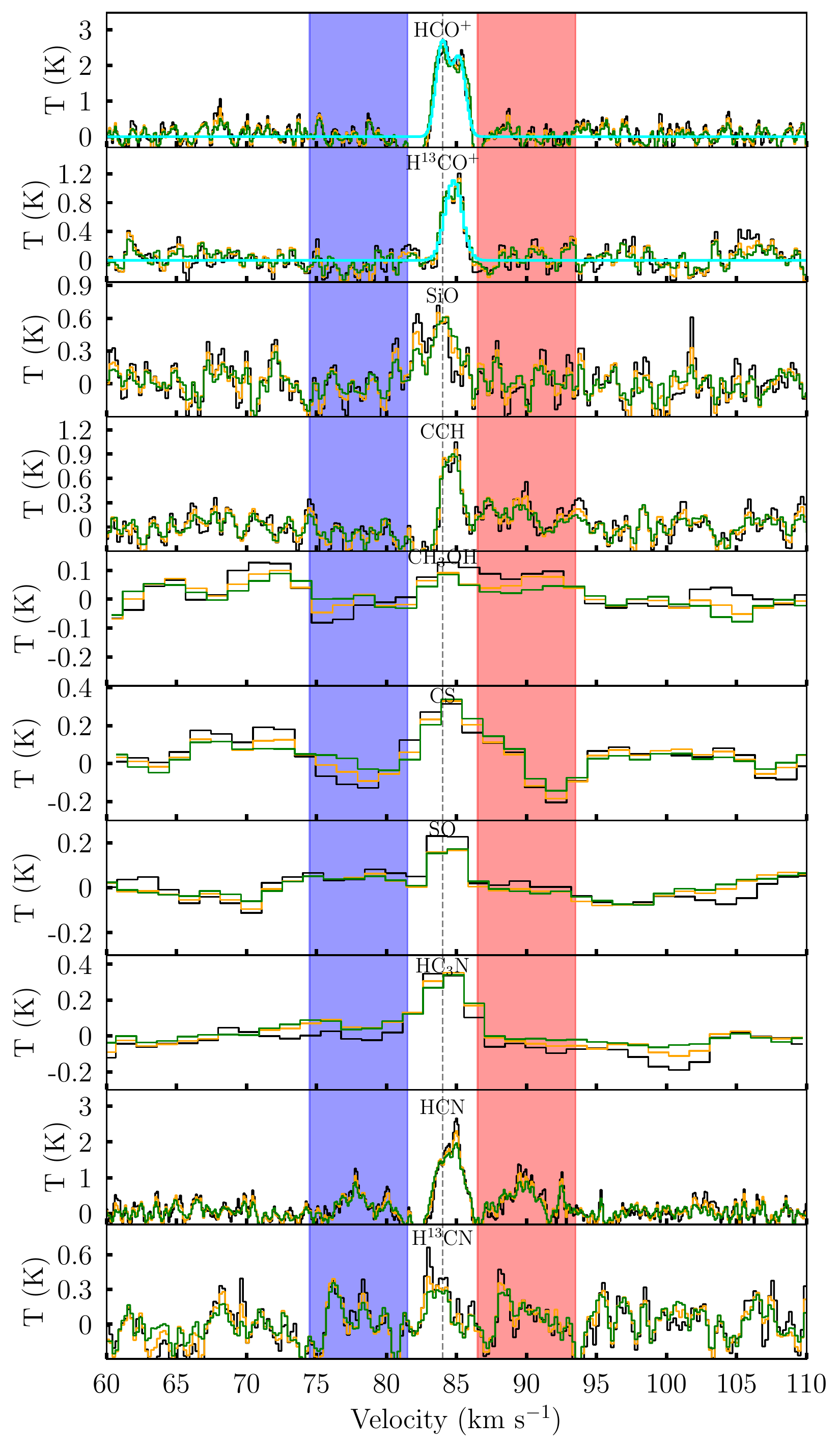}  
  \caption{C2 ATOMS spectra. The spectra extracted from semi-axis, double semi-axes, and triple semi-axes areas are shown in black, orange, and green, respectively. The red and blue shadows indicate the velocity ranges of red and blue wings, respectively. The presented molecules include: \hcop, \htcop, \hcn, \htcn, \sio, \cs, \cch, \chtoh, \so, and \hctn. Corresponding transitions could be found in Table~\ref{SSIMtable} (only $N_{J,F}=1_{3/2,2}-0_{1/2,1}$ is shown for \cch). The gray dashed line shows the clump systemic velocity of 84~\kms. The \texttt{Hill5} infall and Gaussian fitting results are shown with cyan lines in panels of \hcop\ and \htcop, respectively. The spectra of other cores can be found in Fig.~\ref{corespectral3} and Appendix~\ref{App:CoreSpectra145}.}
  \label{corespectral2}
\end{figure}

The cores' kinetic temperature \tkin\ from \nht\ observations of \citet{Lu14} follows a decreasing trend with the distance from the rim, from $\sim27$~K for C1 and C4, then 20~K for C3 and C5, and finally 18~K for C2. The reason for this \tkin\ gradient is the combined heating from the external IBL and the internal protostars at various evolutionary stages. The cores and YSOs in I18290 could be divided into three groups according to their locations and evolutionary stages:
\begin{itemize}
\item \textit{IR-bright protostellar sources closest to or on the top of the rim.} This group contains YSOs~\#9 (C1), 10 (C4), and possibly YSOs~\#1, 3, 4, and 5. The most massive core C1 is identified as an extended green object (EGO) whose 4.5~\micron\ emission is proposed to be dominated by \hmole\ ($v = 0 - 0$, S(9, 10, 11)) from outflow shocked gas \citep{Cyganowski08, Rodrguez17}. The 4.5~\micron\ emission of C1 shows an arc structure extending to north, and this 4.5-\micron\ arc has a morphology parallel to and complementary to the 8~\micron\ rim,  which probably suggests that rim compression is reorienting C1 outflow (see green arrow in \textit{Panels a} and \textit{b} of Fig.~\ref{StarFormationFigure}). C1 is also powering the \chtoh\ 44~GHz Class I \citep{Rodrguez17} and 6.7~GHz periodic Class II masers \citep{Szymczak15}. Class I masers frequently correspond to outflow activities. The protostellar core C4 is probably powering very weak outflows traced by \hcopone\ line wings shown in Appendix Fig.~\ref{corespectral4}. C4 is located on the extended arc which is bright at 4.5~\micron\ and 3~mm and also roughly parallel to the 8~\micron\ rim, indicating a strong rim compression similar to C1 again. IR-bright massive protostellar cores have a probable age of $\lesssim0.3$~Myr according to \citet{Motte18}. The remaining YSOs are likely to be HAeBe (YSOs~\#1, 4) or Class II YSOs (YSOs~\#3, 5) according to their 2MASS colors (see Appendix~\ref{appendix:yso}). Their likely typical age is $\sim1$~Myr \citep{vanDishoeck98, Manoj06}.
\item \textit{IR-quiescent protostellar sources more distant from the rim.} This group contains IR-quiescent cores C3 and C5. The core C3 drives an outflow with a kinematic age of 18~kyr and an outflow mass rate $\dot{M}_{\rm out}$ of 2~\msun~kyr$^{-1}$ (detailed outflow calculations can be found in Appendix~\ref{COLUMNDENSITYANDMASS-OUTFLOW}). C5 is very close to C3 ($\sim0.05$~pc) and may also power an outflow lobe toward the south but it is hard to be distinguished from C3 outflow from channel maps. The probable age of C3 and C5 is $\lesssim0.1$~Myr if we follow the age statistics of IR-quiescent high-mass protostellar cores in \citet{Motte18}.
\item \textit{Prestellar core candidate most distant from the rim.} No outflow is detected towards massive core C2 ($\sim36$~\msun). The narrow line widths of the ATOMS spectra presented in Fig.~\ref{corespectral2} in addition to the non-detection of warm gas tracers in the ATOMS wide spectral windows (97.52-99.39~GHz and 99.46-101.33~GHz, see Appendix Fig.~\ref{widebandspectralcore2}) also support the interpretation that C2 is a candidate high-mass prestellar core with a probable age of 0.01 to 0.07~Myr according to the statistics of \citet{Motte18}.    
\end{itemize}

 On the whole, the three groups of cores/protostars reveal a bona fide age sequence along the direction of ionization flux: the most evolved protostars are closest to the rim PDR and the candidate high-mass prestellar core is the farthest from the rim PDR whereas the IR-quiescent protostellar cores (possibly) powering outflows are located in between. This small-scale age sequence is another piece of evidence for the ongoing RDI process. However, we should note that the age sequence presented in a scale smaller than GMC can be erased by redistribution of the triggered and spontaneously formed stars during evolution of the feedback-driven structure \citep{Dale15, Gonzalez20}. Another potential bias is that the age sequence is observed on 2D sky plane and thus projection may influence our result. A deviation from the observed age sequence is seen for YSO~\#7 which is classified as a Class II YSO with an age of $0.49^{+2.01}_{-0.39}$~Myr and a mass of 1.1~\msun\ given by YSO SED fitting.  C2, C3, and C5 are exactly at the converging end of three filamentary arms (see the next section) whereas YSO~\#7 is offset from this convergent end, which is probably a hint that YSO~\#7 is a spontaneously formed star or redistributed star, or just a background/foreground star.

\subsection{Feeding cores via filaments}
The 8~\micron\ extinction presented in Fig.~\ref{StarFormationFigure} displays a morphology of several filamentary dark lanes converging to the IR-quiescent cores C2, C3, and C5. Visually, we outline tracks of the maximum 8~\micron\ extinction for the three most significant filamentary arms F1, F2, and F3 in \textit{Panel c} of Fig.~\ref{StarFormationFigure}. The convergent morphology indicates that the gas is probably inflowing along filaments onto cores \citep{Ren21}. The zeroth moment maps presented in Appendix Fig.~\ref{moleculartracer} of several molecular line emission (i.e. \hcop, \hcn, \so, and \cs) observed with ATOMS do not show this filamentary morphology. To probe the gas kinematics, we make use of the VLA \nht\ data from \citet{Lu14} which has the spatial and velocity resolutions of $\sim4$\arcsec\ ($\sim0.1$~pc) and 0.6~\kms, respectively. With the \texttt{PySpecKit}\footnote{\url{https://pyspeckit.readthedocs.io/en/latest/index.html}} task \texttt{fiteach} \citep{Ginsburg11}, we fit the \nht\ (1, 1) inversion lines to estimate \nht\ column density \nnhtcd\ and centroid velocity. The \nht\ velocity map presented in \textit{Panel d} of Fig.~\ref{StarFormationFigure} shows that C3, C5, and possibly C2 are immersed in a blue-shifted ``basin'' with the red-shifted surroundings. It favors a possible scenario where the gas is infalling onto the group of IR-quiescent cores \citep{Estalella19, Sepulveda20}. The convergent filaments may represent the mainstreams of inflowing mass and hence play a critical role in transporting gas to the blue-shifted ``basin''. 

In \textit{Panels e1} to \textit{e3} of Fig.~\ref{StarFormationFigure}, we present the position-velocity (PV) cuts with a width of 5\arcsec\ (0.13~pc) along the filaments for \nht\ (1,1) main line. A velocity change of 1 to 2~\kms\ is observed when moving to the inner $\sim0.4$-pc region surrounding the convergent center (velocity gradient $\simeq1.5~{\rm km~s ^{-1}}/0.4~{\rm pc^{-1}}$), in accordance with the blue-shifted ``basin''. This velocity gradient is underestimated because of projection effect. The \nht\ observations can not resolve the filament width seen in the 8~\micron\ image. To estimate the mass inflow rate along filament $\dot{M}_{\rm inflow}$, we assume that the filaments have a width of 0.1~pc (estimated from 8~\micron\ image), a \nht\ column density \nnhtcd\ of 10$^{15}$~\cm\ (shown in solid contour in \textit{Panel c} of Fig.~\ref{StarFormationFigure}), and a [\nht/\hmole] abundance of $3\times10^{-8}$ \citep{Lu14}, respectively. The estimated filament mass inflow rate $\dot{M}_{\rm inflow} \sim M_{\rm fil} \Delta{\rm v}~\tan\left(\alpha_{\rm p} \right)^{-1}$ \citep{Kirk13}, here $M_{\rm fil} \sim30$~\msun, $\Delta{\rm v}$, and $\alpha_{\rm p}$ are the filament mass, velocity gradient, and inclination to the plane of sky, respectively. Assuming $\alpha_{\rm p}\sim20$\degree, the total resulted $\dot{M}_{\rm inflow}$ of three filaments is of the order of 1~\msun~kyr$^{-1}$.

Another interesting observed morphology which possibly relates to filament inflow is that in the core C3, the \hcopone\ outflow red lobe has an edge similar to the \nht\ blue-shifted ``basin'', shown as the zoom-in C3 region in \textit{Panel d} of Fig.~\ref{StarFormationFigure}. Moreover, a Class I \chtoh\ maser is also located on this edge. Given the hypothesis that collisional excitation pumping mechanism of Class I maser makes it able to trace the interface between outflow and surrounding materials \citep{Cyganpwski09, Voronkov14, Gomez16}, the Class I maser here may probe the gas shocked by the encounter of the C3 outflow and the inflowing filament material, similar to the case of SDC335 clump where the Class I masers are explained to be powered by the outflow-filament encounter \citep{Avison21}.

\begin{figure}
  \centering
  \includegraphics[width=.95\linewidth]{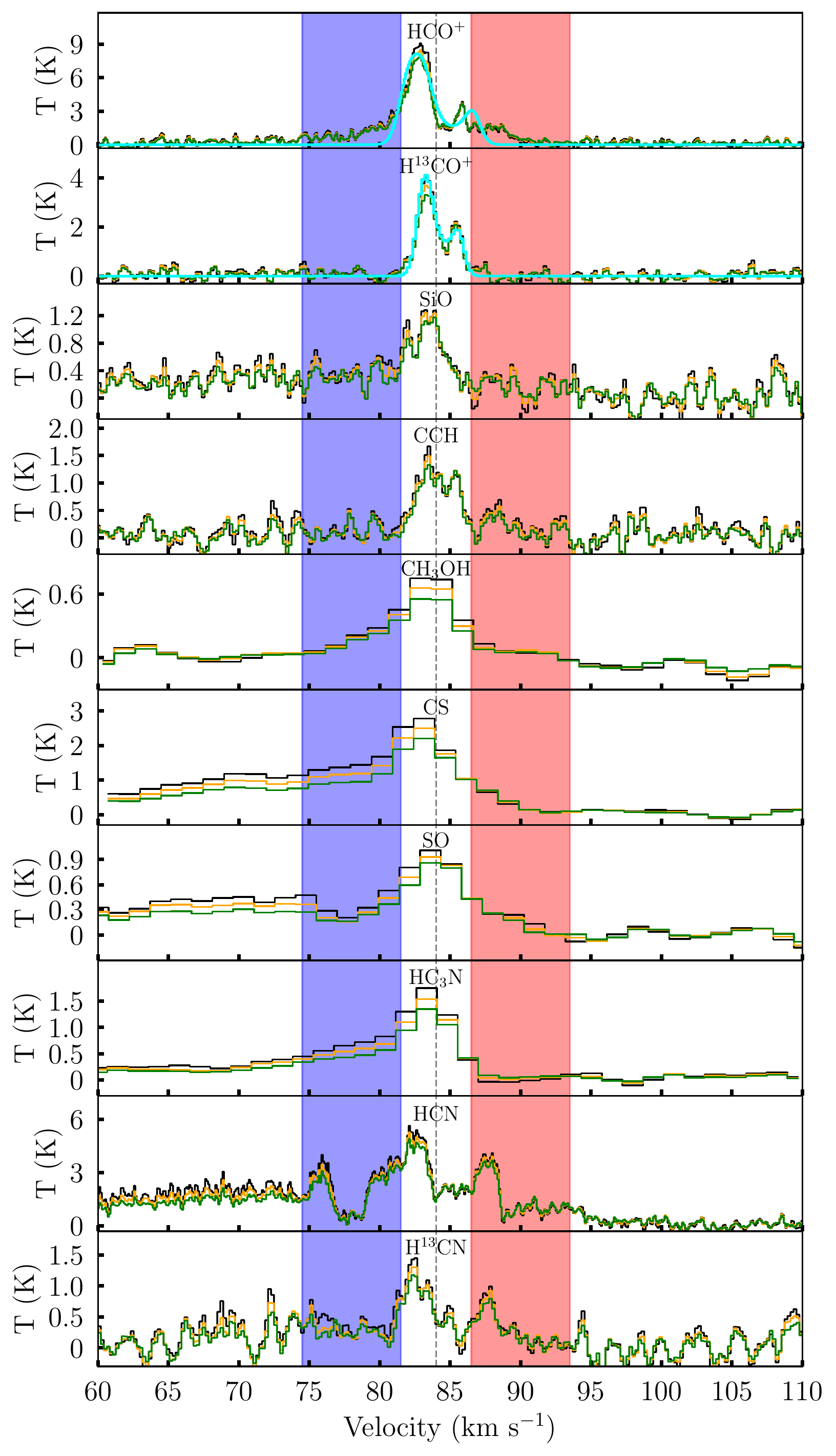}  
  \caption{C3 ATOMS spectra. Similar to Fig.~\ref{corespectral2} except for that cyan lines in panels of \hcop\ and \htcop\ both show the \texttt{Hill5} infall fitting results.}
  \label{corespectral3}
\end{figure}

To further address the inflow nature in the blue-shifted ``basin'', we check the $J = 1 - 0$ spectral profiles of \hcop\ and \htcop\ for C2, C3, and C5 in Figs.~\ref{corespectral2}, \ref{corespectral3}, and Appendix Fig.~\ref{corespectral5}, respectively. The cores C3, C5, and possibly C2 present the blue asymmetries which indicate an infall motion if \tex\ decreases with the distance from core center \citep{Zhou93}. Making use of the \texttt{Hill5} infall model \citep{DeVries05}, we derive core-scale infall velocity \vinfall\ from \htcop\ spectra (details of the \texttt{Hill5} model and fitting procedure are in Appendix~\ref{App:CoreSpectraAndInfall}). In the case that a blue asymmetry of \htcop\ profile is detected, \hcop\ spectra are not preferential because outflow components in \hcop\ profile can heavily bias the model fitting, such as the \texttt{Hill5} modelling for \hcop\ spectra of C3 shown in Fig.~\ref{corespectral3}. Core C2 \vinfall\ is derived from \hcop\ because both self-absorption feature of \htcop\ profile and line wing of \hcop\ are weak. Note that we may overestimate the \vinfall\ of C2 because the double-peak feature of \hcop\ profile is not conspicuous. The modelled \vinfall\ are about 0.6, 0.5, and 0.1~\kms\ for C2, C3, and C5, respectively.  The core-scale mass infall rates estimated via $\dot{M}_{\rm infall} = 4\pi r^2 \rho \rm{v_{infall}}$ \citep{Contreras18} are about 2.7, 2.6, and 0.23~\msun~kyr$^{-1}$ for C2, C3, and C5, respectively. The derived core-scale infall rates are comparable to some similar observations toward HMSF regions, e.g. 3.5~\msun~kyr$^{-1}$ within the innermost 500~AU for \citet{Beuther13} and 1.9~\msun~kyr$^{-1}$ within the innermost 8000~AU for \citet{Contreras18}.\\ 

Although the uncertainties in inclinations, molecule abundances, and calculation methods might be large, the filament inflow rate $\dot{M}_{\rm inflow}$, core-scale infall rate $\dot{M}_{\rm infall}$, and outflow rate $\dot{M}_{\rm out}$ are on the same order of magnitude, which sheds light on a continuous mass-building process from clump to core in I18290 BRC. IR-quiescent massive cores C2, C3, C5 probably gain mass via filamentary arms embedded in the compression-subvirialized BRC, accompanied by ongoing sequential star formation, suggesting that a combined process of clump-fed accretion and RDI is functioning for HMSF in this BRC.

\section{Dynamical PDR with photoevaporative flow} \label{PDRandPEF}
In this section, we further investigate structures and kinematics of the PDR traced by the 8-\micron\ rim. \cch\ is one of the best tracers for the PDR in ATOMS data of I18290. Figure~\ref{PDRDynamics} shows that \cchline\ (the strongest \cch\ hyperfine line in the ATOMS spectral window) zeroth moment has a spatial distribution analogous to the 8~\micron\ emission. \citet{Pety05} proposed that this correlation stems from the fact that PAHs can be the precursors of small hydrocarbons in the PDR. The formation and destruction mechanisms of \cch\ closely correspond to the energetic photons \citep{Huggins84, Teyssier04, Cuadrado15, Nagy15, Buslaeva21, Kirsanova21}, making \cch\ to be a great tracer of PDRs. We first compare the spatial distributions of \cch\ with other molecules and then analyze the rim PDR kinematics with \cch. 

\subsection{Associations of molecules and PDR}
In addition to \cch\ emission, the PDR identified in the 8~\micron\ image is also traced by several other molecular line emission from the ATOMS observations. Their zeroth moment maps are presented in Appendix Fig.~\ref{moleculartracer} where some have poor signal-to-noise (SN) ratio. Using the method of \texttt{structural similarity index measure} \citep[\texttt{SSIM},][]{Wang04} which could compare the similarity of two maps, we categorize the molecules in Table~\ref{SSIMtable} into two types according to their spatial similarity to \cch\ emission (details about \texttt{SSIM} and its calculations are in Appendix~\ref{App:PDRMoleculesAnalysis-SSIM}): (1) PDR tracers whose spectral zeroth moment maps have a morphology similar to \cch. The molecule most similar to \cch\ is \htcop, and then \cs, \hcop, followed by \so\, and \hcn. (2) star formation tracers, which \textit{only} trace the gas directly related with star formation activities such as core dense gas and outflow. They include \sio, \hctn, \chtoh, and \htcn. Note that our classifications of the PDR and star formation tracers are \textit{only} valid for the ATOMS data of this region. Different data sets and star formation regions may lead to different members for these two groups.

\begin{figure*}
     \centering
   \includegraphics[width=0.98\textwidth]{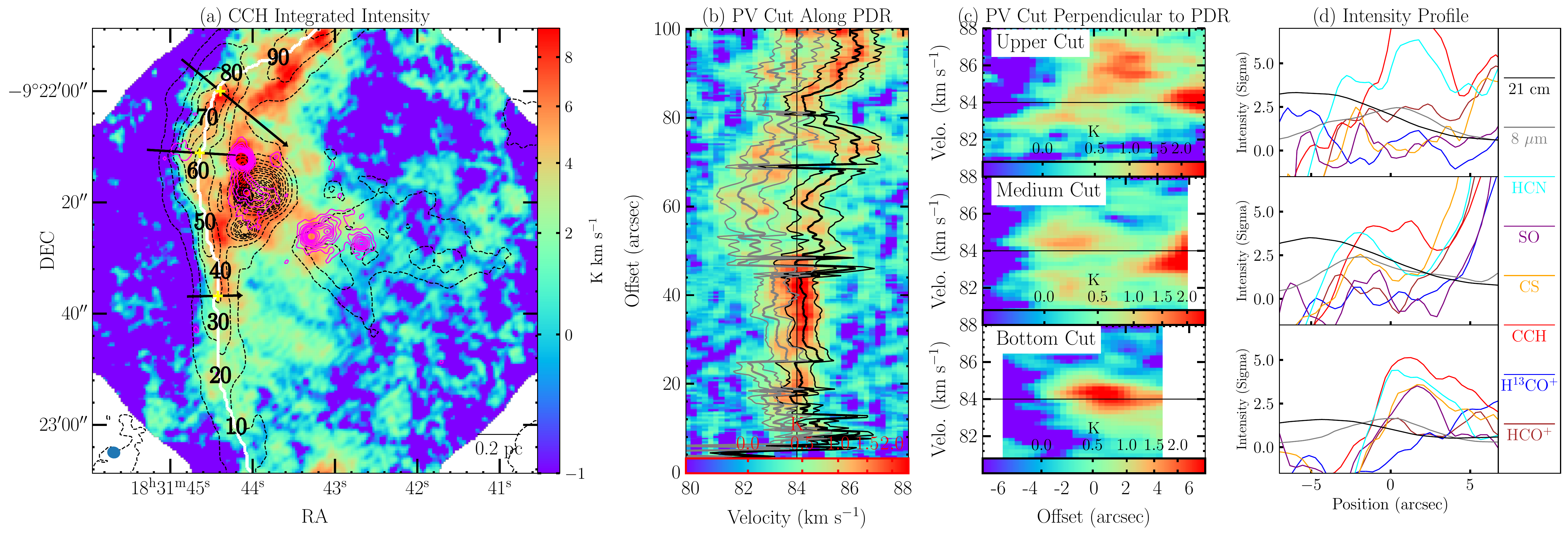}
       \caption{\cchline\ emission. \textit{Panel a:} \cch\ integrated intensity maps overlaid with the contours of 8~\micron\ emission (black) and the 3~mm continuum (magenta) with levels the same as \textit{Panel a} of Fig.~\ref{StarFormationFigure}. The white line indicates the 8-\micron\ rim spine extracted by \texttt{Filfinder} in Sect.~\ref{ENVIRONMENTSANDSTATUS}. The numbers on the spine indicate the offset positions corresponding to \textit{Panel b}. \textit{Panel b:} PV cut along the rim spine. The three black and gray lines show the \texttt{MCMC}-fitted compressed and PeF components with velocity centroid  and dispersion, respectively.  \textit{Panel c:} PV cut (width $=5$\arcsec) along the three arrows shown in \textit{Panel a}. \textit{Panel d:} Intensity profiles of various species along the arrows. In \textit{Panels c} and \textit{d}, the zero offset positions are positions at the spine and a positive offset means a position close to the interior of the clump.}
        \label{PDRDynamics}
    \end{figure*}

PDRs represent transition regions from \hii-dominated to \hmole-dominated with increasing \av. The changes of UV radiation, \hmole\ density, and temperature create physical and chemical stratification from the exterior to the interior of PDRs, leading to a stratified distribution for the emission of various molecular species. \textit{Panel d} of Fig.~\ref{PDRDynamics} shows the zeroth moment profiles of various PDR tracers along three cuts perpendicular to the rim PDR indicated in \textit{Panel a} of Fig.~\ref{PDRDynamics}. No consistent stratification is found in the three cuts and the zeroth moment profiles of various molecular species are peaked at the roughly same position. 

This indistinctive stratification is probably caused by several reasons: (1) The resolution of 0.05~pc along with the noisy nature may be not enough to resolve the PDR stratification. In efforts to resolve the famous Orion Bar PDR, \citet{Goicoechea16} find that the separation between ionization front and dissociation front is only $\sim0.03$~pc, indicating that molecular stratification exists in a narrow space. \citet{Sicilia19} searched for the stratification in BRC 1396A with a spatial resolution of 0.05~pc for a series of molecules, such as \hcn, \hcop, \cs, and \so\ but these authors find that the differences in peaks' position are only one to two beams. (2) The projection effects weaken the observed stratification feature when there is an inclination to the sky plane. (3) The vigorous gas flows in PDRs such as photoevaporative flow (PeF) and rocket effects may mix up the layered distributions. The PeF is off the clump surface and it is driven by the overpressure of the ionization/dissociation region compared to the cold molecular region. Moreover, the conservation of momentum impels the clump to move in the opposite direction. In the next section, we investigate the signatures of PeF in I18290.

\begin{table}
\caption{\label{SSIMtable}Molecular transitions in Sect.~\ref{PDRandPEF}.} 
\begin{threeparttable}
\renewcommand{\arraystretch}{1.0}
\centering

\begin{tabular}{rrrr}
\hline
\hline
Molecular transition                   &       Freq.    &    \texttt{SSIM} peak\tnote{\textit{(a)}}      &    Type\tnote{\textit{(b)}}         \\
                                       &       GHz      &                   &                 \\
\hline
CCH~$N_{J,F}=1_{3/2,2}-0_{1/2,1}$      &       87.3169  &         -         &       PDR       \\   
CCH~$N_{J,F}=1_{3/2,1}-0_{1/2,0}$      &       87.3286  &        0.359      &       PDR       \\   
H$^{13}$CO$^+$~$J=1-0$                 &       86.7543  &         0.256     &        PDR      \\   
CS~$J=2-1$                             &       97.9810  &        0.245      &       PDR       \\   
HCO$^+$~$J=1-0$                        &       89.1885  &          0.229    &         PDR     \\   
SO $v =0, 3(2)-2(1)$                   &       99.2999  &       0.218       &      PDR        \\   
H40$\alpha$                            &       99.0230  &        0.191      &       SF        \\  
SiO~$J=2-1$                            &       86.8470  &        0.188      &       SF        \\   
HC$_3$N~$J=11-10$                      &      100.0764  &        0.185      &       SF        \\  
CH$_3$OH 2(1,1)-1(1,0)A                &       97.5828  &        0.176      &       SF        \\ 
VLA NH$_3$ (1,1)                       &       23.6945  &          -        &       SF\tnote{\textit{(c)}}        \\
HCN~$J=1-0$                            &       88.6318  &          -        &       PDR\tnote{\textit{(d)}}       \\   
H$^{13}$CN~$J=1-0$                     &       86.3399  &          -        &       SF\tnote{\textit{(d)}}        \\   
\hline
   \end{tabular}
      \begin{tablenotes}
      \item [\textit{(a)}] Peak value of the channel-by-channel \texttt{SSIM} calculation figures presented in Appendix Fig.~\ref{SSIM}. A higher peak value indicates a higher similarity between spatial distributions of two molecular emission.
      \item [\textit{(b)}] ``PDR'' and ``SF'' represent the PDR and star formation tracers, respectively.
       \item [\textit{(c)}] VLA NH$_3$ (1,1) is not included in the \texttt{SSIM} calculations because it is not observed by ALMA. No emission is associated with the 8-\micron\ rim PDR in the zeroth moment map and thus NH$_3$ is simply set as SF tracer.
      \item [\textit{(d)}] $J=1-0$ of HCN and H$^{13}$CN are not included in the channel-by-channel \texttt{SSIM} calculations because their hyperfine emission could contaminate each other and then influences channel-by-channel \texttt{SSIM} calculations. A layer aligned with the 8-\micron\ rim is presented in the zeroth moment map of \hcnone\ but not for \htcnone\ and therefore \hcn\ and \htcn\ are classified as PDR and SF tracers, respectively.
      \end{tablenotes}
      \end{threeparttable}
\end{table}

\subsection{Photoevaporative flow traced by \cch?}
Many simulations in BRC clearly present the PeF with a velocity of around 1 to 10~\kms\ and a direction perpendicular to clump surface \citep{Lefloch94, Kessel03, Miao06, Miao09, Haworth13, Nakatani19}. Observationally, however, the velocity fields show the explicit evidence of PeF only in a few BRCs. \citet{Mcleod15} analyzed S~\textsc{ii} spectra PV cut along the direction of ionization radiation for M16 (pillars of creation) and these authors find a blue-shifted dip of $\sim 10$~\kms\ in the PDR which is proposed to be a kinematic feature of PeF. In the PDR of BRC 1396A, \citet{Sicilia19} find a remarkable second component besides the main component tracing the clump body (shift is $\sim2$~\kms) in CN spectra. These authors suggest that the second component is probably from PeF.

Here, we search for PeF with the \cch\ spectra. The core spectra presented in Figs.~\ref{corespectral2}, \ref{corespectral3}, and Appendix Figs.~\ref{corespectral1} to \ref{corespectral5} indicate that the \cch\ emission is relatively unrelated with core outflow motions in our case and therefore the contamination from star formation activities is weak. The \cchline\ PV cuts along and perpendicular to the PDR spine (\textit{Panels b} and \textit{c} of Fig.~\ref{PDRDynamics}) show a complicated velocity mode. For the three cuts perpendicular to the PDR spine, it is obvious that the \cch\ presents at least two velocity components around the spine (offset $\sim0$\arcsec), whereas they merge to one component when moving to the clump interior (offset $>0$\arcsec). The two-velocity structure is not due to self-absorption because stronger \cch\ lines in the region do not show a clear absorption feature.

To check the velocity feature in the entire observed PDR, a \cch\ PV cut with a width of 5\arcsec\ (to minimize the contamination associated with cores) is created along the PDR spine and shown in \textit{Panel b} of Fig.~\ref{PDRDynamics}. The southern parts seem to present a single velocity component with a blue wing. For the strongly irradiated BRC head (40\arcsec\ to 80\arcsec\ position in \textit{Panel b} of Fig.~\ref{PDRDynamics}), the \cch\ emission splits into at least two well-separated components and becomes weaker compared to the southern parts. With a simplified assumption that the \cch\ PV cut along the PDR spine is composed of two Gaussian components at each position, we fit the PV cut position-by-position using the \texttt{Markov Chain Monte Carlo} (\texttt{MCMC}) method (fitting details are in Appendix~\ref{App:PDRMoleculesAnalysis-CCHdecomposition}). The spine PV cut is well modelled by two  Gaussian components with the velocity dispersion of $\sim0.5$~\kms, and the velocity shifts of $\sim1$~\kms\ for the southern part and 2 to 3~\kms\ for the head. The fitting residual $\sim0.17$~K is close to the noise of the \cch\ data cube.

We propose that these two velocity components are from PeF (blue-shifted) and compressed shell (red-shifted), respectively, based on the following observed features in the PV cut along the PDR spine: 
\begin{itemize}
    \item  The red-shifted component is stronger than the blue-shifted component in the modelled \cch\ PV cut, which may stem from the fact that the compressed gas is denser than the PeF gas.
    \item The velocity shift between PeF and compressed components becomes larger at the BRC head, compared to that of the southern part. It is probably driven by more intense ionizing radiation in the BRC head than in the southern part, which powers a stronger compression and photoevaporation in the head to create a larger velocity shift. This enlargement of the velocity shift is also presented in some RDI simulations \citep{Lefloch94, Miao09, Haworth13}.
    \item \cch\ emission is weaker in the BRC head compared to the southern parts, which can be explained as a result of more intense photodestruction of \cch\ in the head.
\end{itemize}

All these observed signatures support the interpretation that the blue- and red-shifted components trace the PeF and compressed gas, respectively. The ATOMS spectral window covers CCH hyperfine lines $N_{J,F}=1_{3/2,2}-0_{1/2,1}$ and $N_{J,F}=1_{3/2,1}-0_{1/2,0}$ which could in principle be used to derive \cch\ optical depths and \tex\ under LTE conditions using hyperfine fitting tools \citep[such as \texttt{HFS} tools in \texttt{CLASS},][]{Kirsanova21}. However, the \cch\ hyperfine line intensity ratio in I18290 PDR is very close to the theoretical value of optically thin emission in LTE, 2:1 ratio. Furthermore, the deviations from the single-Gaussian profile in most pixels and the low SN ratio of the weaker \cchhyperfine\ spectra cause large errors in the hyperfine fitting. Here we only estimate \cch\ column density \ncchcd\ from the Gaussian components extracted by \texttt{MCMC} with the assumption of optically thin emission under LTE. The median \ncchcd\ is about $10^{14}$ to $10^{15}$~\cm\ at a typical \tex\ of 100~K, with the \ncchcd\ ratios of compressed-to-PeF components of 1.1 for the southern parts and 2.7 for the head (see calculations in Appendix~\ref{COLUMNDENSITYANDMASS-CCH}). The abundance of \cch\ in the PDR is highly variable and depends on the detailed physical and chemical status of the PDR, which is beyond the scope of this paper. Further assuming a \cch\ abundance of $10^{-8}$ \citep{Teyssier04,Nagy15, Cuadrado15, Buslaeva21, Kirsanova21}, the column density of the PDR is of the order of $10^{22}$ to $10^{23}$~\cm.

\section{Discussions and conclusions}
The essence of the clump-fed scenario is the central role of inflow motions beyond core scale on mass building of HMSF. Detailed interferometric case studies, e.g. SDC335 \citep{Pereto13}, G22 \citep{Yuan18}, and G34 \citep{LiuHL22ATOMSV}, provide strong evidence for the pc-scale inflow. \citet{Peretto20} proposed that the clump-fed rather than core-fed process is prevalent for the cores hosted in HMSF regions because the mass versus temperature evolutionary track resulted from the clump-fed scenario agrees better with their observations of $\sim200$ cores in 11 massive IRDCs. Most of the studied clump-fed processes are working within pc-scale clumps with a quiescent environment. The necessity of merging external ionizing impact into clump-fed accretion scenarios comes from the prevalence of \hii\ region in HMSF regions \citep{Thompson12, Zhang21}. Mainly existing works involving both ionizing feedback and accretion via filaments are the cases where \hii~regions in the central hub/ridge of hub-filament system (HFS) impact the density structure of the HFS and star formation therein (e.g., example observations by \citealt{Baug18,Watkins19,Kumar20,Dewangan20, LiuXL21}; modellings: \citealt{Gonzalez20, Whitworth21}). The studied interplay between accretion and ionizing feedback in these works has scales much greater than clumps ($\sim1$ pc). 

The HMSF ongoing in BRCs is not brand new in observations considering that various tracers of HMSF embedded in BRCs have been found, such as UC\hii\ regions \citep{Morgan04,Thompson04,Urquhart04, Urquhart06}, masers \citep{Valdettaro05,Valdettaro07,Valdettaro08,Urquhart09}, and massive YSOs \citep{Sharma16}, but a clear case of the filament-mediated accretion within clump accompanied with the signatures of triggered HMSF is not presented by previous \textit{observations} of the clump-scale BRCs. The filament-mediated accretions within RDI-driven objects indeed exhibit in some \textit{simulations}. \citet{Dale14,Dale15} proposed that pillars and globules are the relics of the filaments and accretion flows shaped by ionizing radiations whereas \citet{Bisbas11} found that a filament aligned to the axis of symmetry can form due to the convergence of collapsing material during the evolution of BRC.

Our observations of I18290 BRC offer a pilot picture depicting a scene combined by the clump-fed and the ionizing feedback processes: The pressure exerted by photoionized clump surface constrains I18290 in a subvirial, bound state. Meanwhile, the shock driven by the overpressure of ionized surface propagates into the clump interior, and triggers a sequential star formation with a time scale of 0.1 to 1~Myr and a size scale of $\lesssim1$~pc along the radiation direction. The resulted star formation propagation speed of 1-10~\kms\ is not only on the same order as our estimated shock speed $1.5$~\kms, but also similar to the shock speeds in the low-mass BRC observations and simulations \citep{Marshall19}. These proofs support that induced star formation by RDI is at work in I18290. Compared to the RDI in low-mass case, a remarkable difference for massive I18290 BRC is the inflowing multi-filamentary arms developed in the clump during the RDI process. The filament inflows ($\sim1$~\msun~kyr$^{-1}$) toward the massive core group, along with the strong core-scale infall ($\sim2$~\msun~kyr$^{-1}$), imply that massive cores are gaining their mass from clump-wide environment. 

\begin{figure}
     \centering
   \includegraphics[width=0.48\textwidth]{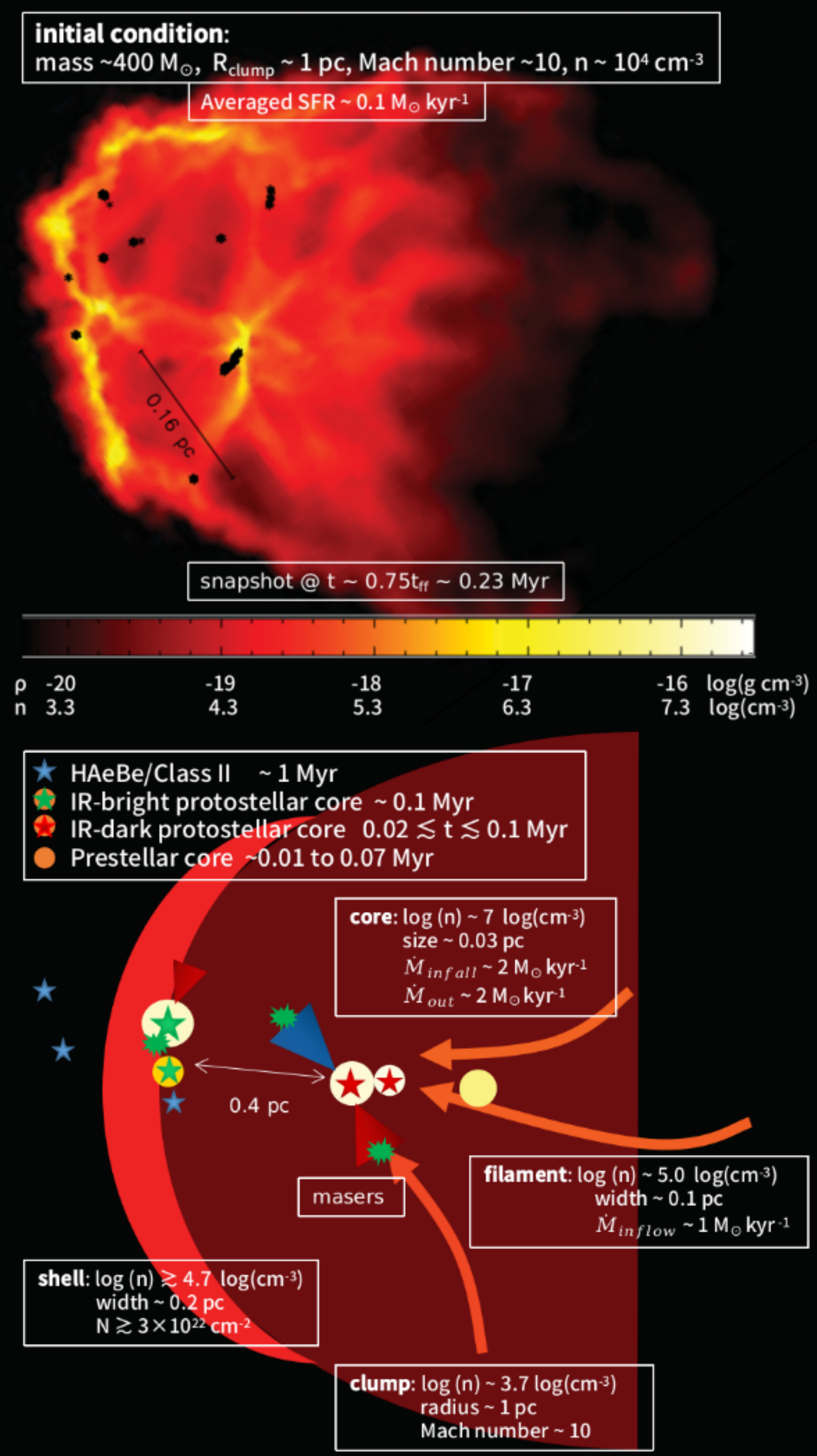}
       \caption{Comparison with model. \textit{Top panel}: Density map of the simulated massive BRC, adapted from \citet{Anathpindika12} with permission. Sink particles are indicated with black asterisks  ``\textasteriskcentered''. \textit{Bottom panel}: Schematic of I18290. The color-coding of various structures (clump, shell, filament, and cores) follows the same colorbar as \textit{top panel}.}
        \label{ModelObservationComparison}
    \end{figure}

The entire picture is very similar to the model of \citet{Anathpindika12} (A\&B2012 hereafter) which is one of the most massive RDI simulations regarding a single BRC. Figure~\ref{ModelObservationComparison} shows a comparison between the schematic of I18290 and the clump modelled by A\&B2012 at a snapshot time of 0.23~Myr. The model initial conditions are quite similar to the current conditions of I18290 estimated in our work, in respects of the clump size ($\sim1$~pc), density ($\sim10^4$~cm$^{-3}$), and turbulence (Mach number $\sim10$). The similar physical conditions make sense to compare between the models and observations.  I18290 presents a morphology strikingly similar to the model. The color-codings of various structures (cores, filaments, PDR shell, and clump) in the schematic of I18290 share the same colorbar of the model panel. Although the observed densities of extended structures (shell and filaments) are one magnitude smaller than in the model owing to beam dilution, a density amplification of one to two orders of magnitude compared to the clump density for these extended structures is akin. I18290 reproduces two main sites of star formation in the model: the dense PDR shell and the central core group with inflowing extended filamentary arms. The modelled averaged star formation rate (SFR) $\sim0.1$~\msun~kyr$^{-1}$ is likely of the same order as in I18290 if we assume a timescale of $\sim0.5$~Myr and a total stellar mass of $\sim50$~\msun\ for I18290. 
    
There are still some notable differences between I18290 and the model: (1) The cores and YSOs identified in I18290 are significantly less than the modelled sink particles that represent forming stars, which could stem from the poor mass sensitivity of the ATOMS data (1.1~\msun~beam$^{-1}$). (2) The corrugated and broken nature of the PDR shell driven by thin-shell instabilities for the model is ambiguous for the observed PDR shell. The possible reason is that the limited resolution and sensitivity dilute the fluctuations at a smaller scale. (3) Sink particles in the modelled shell seem to spread over a larger fraction of shell compared to the cores and YSOs in the shell of I18290. Unfortunately, whether this difference is real is unknown because A\&B2012 did not present the detailed properties of the individual sink particles such as age and mass. 

It will be very interesting to model how the external feedback influences star formation in a scale comparable to or smaller than cores in future work. Outflows near the rim PDR such as C1 outflow may be reoriented due to the compression of ionized surface. Furthermore, the \hcopone\ spectra for the cores (C1 and C4) close to the rim present a red asymmetry indicating expansion. A more striking phenomenon is that the core C1 Class II \chtoh\ masers within a scale of 0.1\arcsec\ (500~AU) seem to align with the large-scale PDR shell (see distribution of Class II masers with a velocity $>81$~\kms\ in Figure~9 of \citealt{Szymczak15}). \citet{Dodson04} proposed that the linear distribution of Class II masers could be caused by an edge-on planar shock propagating into a star-forming core. This scenario may cast light on the potential strong effect of IBL on the interior of core C1.

The outflows sustain a large fraction of I18290 turbulence because the outflow energy rate $\dot{E}_{\rm out}$ is of the same order as the turbulent dissipation rate for I18290 \citep{Baug21}. The gas kinematics are not presented in the A\&B2012 model. It is worthwhile to model and analyze the various gas flows that exist in the irradiated turbulent massive BRC, such as photoevaporation flow and filament inflow \citep{Haworth13}, especially for the questions whether a bound clump status under the assistance of IBL could significantly increase accretion rate of cores in massive clumps \citep{Motoyama07,Maheswar08}.\\

This pilot observational study investigates the triggered star formation via RDI within clump-fed scenario for a massive BRC. Forthcoming ALMAGAL data which involves a few dozens of massive BRCs shown in Fig.~\ref{StatisticsMap} will help us construct a more complete picture, especially for the universality of this combined mechanism in massive BRCs.

\section*{Acknowledgements}
We want to thank the anonymous referee for the constructive comments that helped improve a lot quality of the paper. S.Z. and K.W. acknowledge support from the China Manned Space Project (CMS-CSST-2021-B06, CMS-CSST-2021-A09), the National Science Foundation of China (11973013, 12033005),
the National Key Research and Development Program of China (2019YFA0405100), and the High-performance Computing Platform of Peking University through the instrumental analysis fund of Peking University (0000057511). S.Z. acknowledges the support of the China Postdoctoral Science Foundation through grant No. 2021M700248. T.L. acknowledges support from the National Natural Science Foundation of China (NSFC) through grants No. 12073061 and No. 12122307, the International Partnership Program of the Chinese Academy of Sciences (CAS) through grant No. 114231KYSB20200009, the Shanghai Pujiang Program (20PJ1415500), and science research grants from the China Manned Space Project with no. CMS-CSST-2021-B06. A.Z. thanks the support of the Institut Universitaire de France. M.J. acknowledges support from the Academy of Finland grant No. 348342.  H.-L. Liu is supported by National Natural Science Foundation of China (NSFC) through the grant No.12103045. AS gratefully acknowledges support by the Fondecyt Regular (project code 1220610), and ANID BASAL projects ACE210002 and FB210003. L.B. gratefully acknowledges support by the ANID BASAL projects ACE210002 and FB210003. C.W.L. acknowledges the support by the BasicScience Research Program through the National Research Foundation of Korea (NRF) funded by the Ministry of Education, Science and Technology (NRF-2019R1A2C1010851), and the support by the Korea Astronomy and Space Science Institute grant funded by the Korea government (MSIT) (Project No. 2022-1-840-05). This research was carried out in part at the Jet Propulsion Laboratory, which is operated by the California Institute of Technology under a contract with the National Aeronautics and Space Administration (80NM0018D0004). 

This paper makes use of the following ALMA data: ADS/JAO.ALMA\#2019.1.00685.S. ALMA is a partnership of ESO (representing its member states), NSF (USA), and NINS (Japan), together with NRC (Canada), MOST and ASIAA (Taiwan), and KASI (Republic of Korea), in cooperation with the Republic of Chile. The Joint ALMA Observatory is operated by ESO, AUI/NRAO, and NAOJ

\section*{Data availability}
The data underlying this article will be shared on the request to the corresponding author.



\bibliographystyle{mnras}
\bibliography{example} 




\appendix
\section{Figure information}
\label{App:Figure1Information}
\begin{figure*}
  \centering
  \includegraphics[width=.70\linewidth]{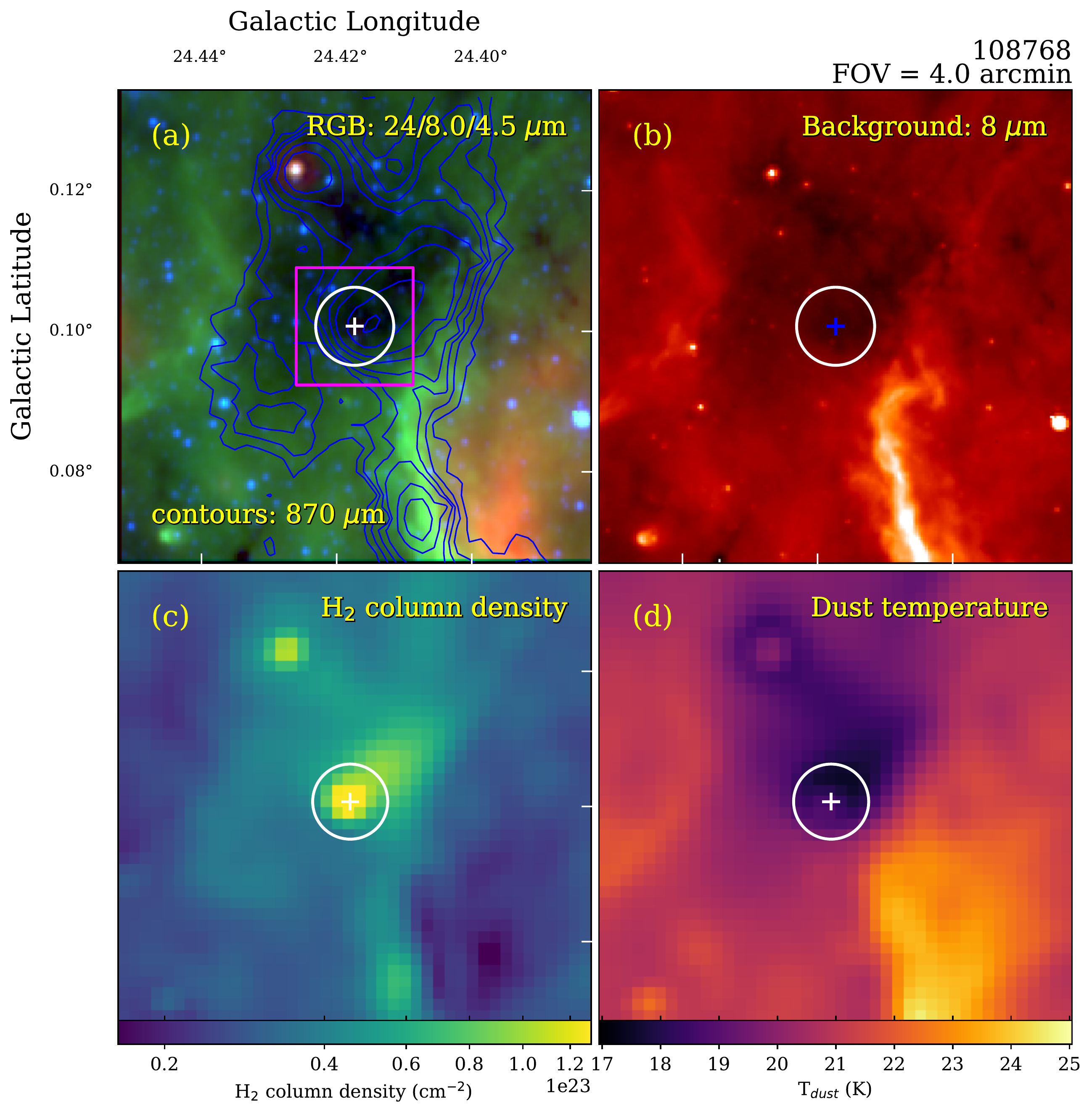}  
  \caption{Example maps for identification of ``ALMAGAL \hii'' source 108768. \textit{Panel a}: \textit{Spitzer} 24/8.0/4.5~\micron\ RGB image overlaid with ATLASGAL 870~\micron\ contours. The cross and circle present the ALMAGAL observation center and field of view, respectively. \textit{Panel b}: 8~\micron\ emission. \textit{Panels c} and \textit{d} shows the Hi-GAL PPMAP column density and dust temperature maps, respectively.}
  \label{108768-fig}
\end{figure*}

\begin{figure*}
  \centering
  \includegraphics[width=.70\linewidth]{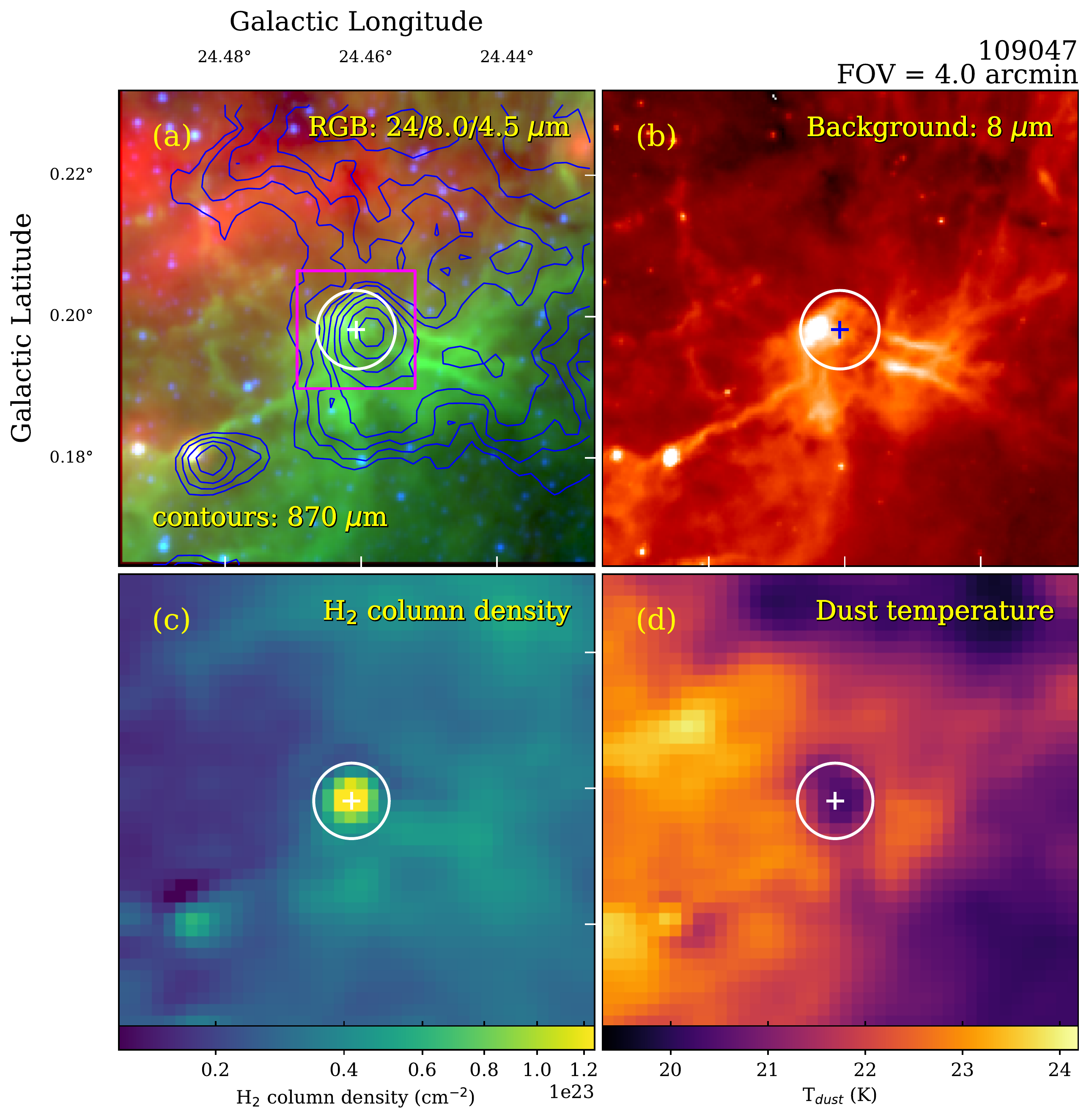}  
  \caption{Example maps for identification of ``ALMAGAL BRC'' source 109047, similar to Fig.~\ref{108768-fig}.}
  \label{109047-fig}
\end{figure*}

We list the detailed source information of Fig.~\ref{StatisticsMap} in Table~\ref{SourceInformation}. Meanwhile, we present two example maps that are used for identification of ``ALMAGAL \hii'' and ``ALMAGAL BRC'' clumps in Figs~\ref{108768-fig} and \ref{109047-fig}, respectively.

\clearpage
\newpage
\onecolumn
\begin{landscape}
\begin{small}
\centering
\begin{ThreePartTable}
\begin{longtable}{llllllll}
\caption{Source information of Figure~\ref{StatisticsMap}.} 
\label{SourceInformation} 
\setlength{\tabcolsep}{1.pt}  
\\  \hline 
    \hline
\thead{Name \\ - }
    & \thead{Catalog \tnote{\textit{(a)}}\\ - }
                        & {\thead{Mass \tnote{\textit{(b)}}\\ \msun}}
                                    & {\thead{Distance \tnote{\textit{(c)}}\\ kpc}}
                                              & {\thead{Spatial Res. \tnote{\textit{(d)}}\\ pc}}  
                                                       & {\thead{Angular Res. \tnote{\textit{(d)}}\\ \arcsec}}  
                                                               & {\thead{Method \tnote{\textit{(b)}}\\ -}} 
                                                                              & {\thead{Note \tnote{\textit{(e)}}\\ -}}   \\
    \hline
\endfirsthead
\caption[]{Source information of Figure~\ref{StatisticsMap}.} \\
    \hline 
    \hline
\thead{Name \\ -}
    & \thead{Catalog \tnote{\textit{(a)}}\\ -}
                        & {\thead{Mass \tnote{\textit{(b)}}\\ \msun}}
                                    & {\thead{Distance \tnote{\textit{(c)}}\\ kpc}}
                                              & {\thead{Spatial Res. \tnote{\textit{(d)}}\\ pc}}  
                                                       & {\thead{Angular Res. \tnote{\textit{(d)}}\\ \arcsec}}  
                                                               & {\thead{Method \tnote{\textit{(b)}}\\ -}} 
                                                                              & {\thead{Note \tnote{\textit{(e)}}\\ -}}  \\
    \hline
\endhead
    \hline 
    \multicolumn{8}{r}{\footnotesize\textit{Continued on next page}} \\ 
\endfoot
    \hline
\endlastfoot

\hline                                     
                   SFO 01 &   SFO         &        2.18 &     0.85 &          0.06 &              14.0 & SD+Continuum    &                                       \citet{Morgan08}      \\
                   SFO 02 &   SFO         &        3.37 &     0.85 &          0.06 &              14.0 & SD+Continuum    &                                       \citet{Morgan08}      \\
                   SFO 03 &   SFO         &        1.15 &     0.85 &          0.06 &              14.0 & SD+Continuum    &                                       \citet{Morgan08}      \\
                   SFO 04 &   SFO         &         2.0 &     0.19 &          0.05 &              54.0 & SD+HCO$^{+}$    &                           \citet{DeVries02,Morgan08}      \\
                   SFO 05 &   SFO         &       25.85 &      1.9 &          0.13 &              14.0 & SD+Continuum    &                            \citet{Morgan08,Fukuda13}      \\
                   SFO 07 &   SFO         &       31.26 &      1.9 &          0.13 &              14.0 & SD+Continuum    &                                       \citet{Morgan08}      \\
                   SFO 09 &   SFO         &        3.03 &      1.9 &          0.13 &              14.0 & SD+Continuum    &                                       \citet{Morgan08}      \\
                   SFO 10 &   SFO         &        2.22 &      1.9 &          0.13 &              14.0 & SD+Continuum    &                                       \citet{Morgan08}      \\
              SFO 11 SMM1 &   SFO         &        20.6 &      1.9 &          0.13 &              14.0 & SD+Continuum    &                            \citet{Thompson2004-414-1017}    \\
            SFO 11NE SMM1 &   SFO         &        13.4 &      1.9 &          0.13 &              14.0 & SD+Continuum    &                            \citet{Thompson2004-414-1017}    \\
             SFO 11E SMM1 &   SFO         &        18.7 &      1.9 &          0.13 &              14.0 & SD+Continuum    &                            \citet{Thompson2004-414-1017}    \\
                   SFO 12 &   SFO         &       16.87 &      1.9 &          0.13 &              14.0 & SD+Continuum    &                                       \citet{Morgan08}      \\
                   SFO 13 &   SFO         &        19.6 &      1.9 &          0.13 &              14.0 & SD+Continuum    &                           \citet{DeVries02,Morgan08}      \\
                   SFO 14 &   SFO         &       51.79 &      1.9 &          0.13 &              14.0 & SD+Continuum    &                                       \citet{Morgan08}      \\
                   SFO 15 &   SFO         &        7.04 &      3.4 &          0.23 &              14.0 & SD+Continuum    &                                       \citet{Morgan08}      \\
                   SFO 16 &   SFO         &        1.25 &      0.4 &          0.03 &              14.0 & SD+Continuum    &                           \citet{DeVries02,Morgan08}      \\
                   SFO 18 &   SFO         &        1.15 &      0.4 &          0.03 &              14.0 & SD+Continuum    &                           \citet{DeVries02,Morgan08}      \\
                   SFO 20 &   SFO         &         6.0 &      0.4 &           0.1 &              54.0 & SD+HCO$^{+}$    &                                      \citet{DeVries02}    \\
                   SFO 22 &   SFO         &        12.0 &     0.46 &          0.03 &              15.0 & SD+$^{13}$CO    &                                     \citet{Motoyama13}    \\
                   SFO 23 &   SFO         &        6.26 &      1.6 &          0.11 &              14.0 & SD+Continuum    &                                       \citet{Morgan08}    \\
                   SFO 25 &   SFO         &        55.0 &     0.78 &           0.2 &              54.0 & SD+HCO$^{+}$    &                                      \citet{DeVries02}    \\
                  SFO 25a &   SFO         &        5.59 &     0.78 &          0.05 &              14.0 & SD+Continuum    &                                       \citet{Morgan08}    \\
                  SFO 25b &   SFO         &        5.69 &     0.78 &          0.05 &              14.0 & SD+Continuum    &                                       \citet{Morgan08}    \\
                   SFO 26 &   SFO         &        1.05 &     1.15 &          0.08 &              14.0 & SD+Continuum    &                                       \citet{Morgan08}    \\
                   SFO 27 &   SFO         &        3.36 &     1.15 &          0.08 &              14.0 & SD+Continuum    &                                       \citet{Morgan08}    \\
                   SFO 29 &   SFO         &        3.57 &     1.15 &          0.08 &              14.0 & SD+Continuum    &                                       \citet{Morgan08}    \\
                   SFO 30 &   SFO         &       29.33 &      2.0 &          0.14 &              14.0 & SD+Continuum    &                                       \citet{Morgan08}    \\
                   SFO 31 &   SFO         &        1.19 &      1.0 &          0.07 &              14.0 & SD+Continuum    &                                       \citet{Morgan08}    \\
                   SFO 33 &   SFO         &        0.45 &     0.75 &          0.05 &              14.0 & SD+Continuum    &                                       \citet{Morgan08}    \\
                   SFO 34 &   SFO         &        1.55 &     0.75 &          0.05 &              14.0 & SD+Continuum    &                                       \citet{Morgan08}    \\
                   SFO 37 &   SFO         &        3.82 &     0.75 &          0.05 &              14.0 & SD+Continuum    &            \citet{Sugitani97, DeVries02, Morgan08}    \\
                  SFO 39a &   SFO         &        2.77 &     0.75 &          0.05 &              14.0 & SD+Continuum    &                                       \citet{Morgan08}    \\
                  SFO 39b &   SFO         &        2.09 &     0.75 &          0.05 &              14.0 & SD+Continuum    &                                       \citet{Morgan08}    \\
                   SFO 40 &   SFO         &        0.45 &     0.75 &          0.05 &              14.0 & SD+Continuum    &                                       \citet{Morgan08}    \\
                   SFO 41 &   SFO         &        0.43 &     0.75 &          0.05 &              14.0 & SD+Continuum    &                                       \citet{Morgan08}    \\
                   SFO 42 &   SFO         &        0.89 &     0.75 &          0.05 &              14.0 & SD+Continuum    &                                       \citet{Morgan08}    \\
                   SFO 43 &   SFO         &       28.34 &      2.4 &          0.16 &              14.0 & SD+Continuum    &                                       \citet{Morgan08}    \\
                   SFO 58 &   SFO         &        10.0 &      0.7 &          0.11 &              32.0 & SD+C$^{18}$O    &                                     \citet{Urquhart06}    \\
                   SFO 68 &   SFO         &        66.0 &      1.7 &          0.26 &              32.0 & SD+C$^{18}$O    &                                     \citet{Urquhart06}    \\
                   SFO 75 &   SFO         &       177.0 &      2.8 &          0.43 &              32.0 & SD+C$^{18}$O    &                        \citet{Urquhart06,Urquhart07}    \\
                   SFO 76 &   SFO         &        48.0 &      1.8 &          0.28 &              32.0 & SD+C$^{18}$O    &                                     \citet{Urquhart06}    \\
                   SFO 79 &   SFO         &        51.0 &     1.35 &          0.21 &              32.0 & SD+C$^{18}$O    &                                     \citet{Urquhart04}    \\
                  SFO 87a &   SFO         &        5.93 &     1.38 &          0.09 &              14.0 & SD+Continuum    &                                       \citet{Morgan08}    \\
                  SFO 87b &   SFO         &        4.45 &     1.38 &          0.09 &              14.0 & SD+Continuum    &                                       \citet{Morgan08}    \\
                   SFO 89 &   SFO         &        5.18 &     1.38 &          0.09 &              14.0 & SD+Continuum    &                                       \citet{Morgan08}    \\
                      CG1 &    CG         &        42.0 &      0.3 &          0.07 &              45.0 &   Extinction    &                          \citet{Maheswar08, Makela13}   \\
                      CG2 &    CG         &        31.0 &      0.3 &          0.07 &              45.0 &   Extinction    &                          \citet{Maheswar08, Makela13}   \\
                      CG3 &    CG         &         2.0 &      0.4 &          0.16 &              84.0 &    SD+NH$_3$    &                          \citet{Bourke95, Maheswar08}   \\
                      CG4 &    CG         &        50.0 &      0.2 &          0.04 &              45.0 & SD+$^{13}$CO    &                \citet{Gonzalez-Alfonso95, Maheswar08}   \\
                      CG5 &    CG         &         0.2 &      0.4 &          0.16 &              84.0 &    SD+NH$_3$    &                          \citet{Bourke95, Maheswar08}   \\
                      CG6 &    CG         &         5.5 &      0.2 &          0.04 &              45.0 & SD+$^{13}$CO    &                \citet{Gonzalez-Alfonso95, Maheswar08}   \\
                     CG7S &    CG         &        85.0 &      1.9 &           0.2 &              22.0 & SD+$^{13}$CO    &                         \citet{Lefloch95, Maheswar08}   \\
                     CG12 &    CG         &       100.0 &     0.63 &          0.07 &              24.0 & SD+C$^{18}$O    &                         \citet{Haikala07, Maheswar08}   \\
                     CG13 &    CG         &        48.0 &     0.45 &          0.39 &             180.0 & SD+$^{13}$CO    &                            \citet{Lohr07, Maheswar08}   \\
                     CG14 &    CG         &         0.6 &      0.4 &          0.16 &              84.0 &    SD+NH$_3$    &                          \citet{Bourke95, Maheswar08}   \\
                     CG15 &    CG         &         0.8 &      0.4 &          0.16 &              84.0 &    SD+NH$_3$    &                          \citet{Bourke95, Maheswar08}   \\
                     CG16 &    CG         &         1.6 &      0.4 &          0.16 &              84.0 &    SD+NH$_3$    &                          \citet{Bourke95, Maheswar08}   \\
                     CG19 &    CG         &         0.9 &      0.4 &          0.16 &              84.0 &    SD+NH$_3$    &                          \citet{Bourke95, Maheswar08}   \\
                     CG20 &    CG         &         1.3 &      0.4 &          0.16 &              84.0 &    SD+NH$_3$    &                          \citet{Bourke95, Maheswar08}   \\
                     CG21 &    CG         &         0.2 &      0.4 &          0.16 &              84.0 &    SD+NH$_3$    &                          \citet{Bourke95, Maheswar08}   \\
                     CG22 &    CG         &       103.0 &     0.45 &          0.39 &             180.0 & SD+$^{13}$CO    &                            \citet{Lohr07, Maheswar08}   \\
                     CG24 &    CG         &         0.2 &      0.4 &          0.16 &              84.0 &    SD+NH$_3$    &                          \citet{Bourke95, Maheswar08}   \\
                     CG26 &    CG         &         1.3 &      0.4 &          0.16 &              84.0 &    SD+NH$_3$    &                          \citet{Bourke95, Maheswar08}   \\
                     CG27 &    CG         &         0.8 &      0.4 &          0.16 &              84.0 &    SD+NH$_3$    &                          \citet{Bourke95, Maheswar08}   \\
                     CG30 &    CG         &        58.0 &     0.45 &          0.39 &             180.0 & SD+$^{13}$CO    &                            \citet{Lohr07, Maheswar08}   \\
                     CG31 &    CG         &        16.5 &     0.45 &          0.09 &              43.0 & SD+$^{13}$CO    &                         \citet{Nielsen98, Maheswar08}   \\
                     CG32 &    CG         &         2.0 &      0.4 &               &                   &                 &                           \citet{Maheswar08, Tobin18}   \\
                     CG34 &    CG         &         2.1 &      0.4 &          0.16 &              84.0 &    SD+NH$_3$    &                          \citet{Bourke95, Maheswar08}   \\
                     CG38 &    CG         &         0.6 &     0.45 &          0.09 &              43.0 & SD+$^{13}$CO    &                         \citet{Nielsen98, Maheswar08}   \\
                     GCD1 &    CG         &         8.7 &      0.4 &          0.16 &              84.0 &    SD+NH$_3$    &                          \citet{Bourke95, Maheswar08}   \\
                     GDC2 &    CG         &         3.0 &      0.4 &          0.16 &              84.0 &    SD+NH$_3$    &                          \citet{Bourke95, Maheswar08}   \\
                     GDC4 &    CG         &         1.1 &      0.4 &          0.16 &              84.0 &    SD+NH$_3$    &                          \citet{Bourke95, Maheswar08}   \\
                     GDC5 &    CG         &         5.8 &      0.4 &          0.16 &              84.0 &    SD+NH$_3$    &                          \citet{Bourke95, Maheswar08}   \\
              Rosette-CG1 &  Individual   &       100.0 &      1.6 &          0.35 &              45.0 & SD+$^{13}$CO    &                                         \citet{Patel93}   \\
              Rosette-CG2 &  Individual   &        50.0 &      1.6 &          0.35 &              45.0 & SD+$^{13}$CO    &                                         \citet{Patel93}   \\
              Rosette-CG3 &  Individual   &       150.0 &      1.6 &          0.35 &              45.0 & SD+$^{13}$CO    &                                         \citet{Patel93}   \\
              Rosette-CG4 &  Individual   &       200.0 &      1.6 &          0.35 &              45.0 & SD+$^{13}$CO    &                                         \citet{Patel93}   \\
              Rosette-CG5 &  Individual   &       300.0 &      1.6 &          0.35 &              45.0 & SD+$^{13}$CO    &                                         \citet{Patel93}   \\
              Rosette-CG6 &  Individual   &       200.0 &      1.6 &          0.35 &              45.0 & SD+$^{13}$CO    &                                         \citet{Patel93}   \\
              Rosette-CG7 &  Individual   &       100.0 &      1.6 &          0.35 &              45.0 & SD+$^{13}$CO    &                                         \citet{Patel93}   \\
              Rosette-CG8 &  Individual   &       100.0 &      1.6 &          0.35 &              45.0 & SD+$^{13}$CO    &                                         \citet{Patel93}   \\
              Rosette-CG9 &  Individual   &        50.0 &      1.6 &          0.35 &              45.0 & SD+$^{13}$CO    &                                         \citet{Patel93}   \\
            W5-E-Pillars1 &  Individual   &        0.33 &      2.0 &          0.35 &              36.0 &     Herschel    &                                     \citet{Deharveng12}   \\
           W5-E-Pillars2a &  Individual   &        1.16 &      2.0 &          0.35 &              36.0 &     Herschel    &                                     \citet{Deharveng12}   \\
           W5-E-Pillars2b &  Individual   &        0.59 &      2.0 &          0.35 &              36.0 &     Herschel    &                                     \citet{Deharveng12}   \\
           W5-E-Pillars2c &  Individual   &        0.83 &      2.0 &          0.35 &              36.0 &     Herschel    &                                     \citet{Deharveng12}   \\
            W5-E-Pillars3 &  Individual   &        0.97 &      2.0 &          0.35 &              36.0 &     Herschel    &                                     \citet{Deharveng12}   \\
            W5-E-Pillars4 &  Individual   &        1.49 &      2.0 &          0.35 &              36.0 &     Herschel    &                                     \citet{Deharveng12}   \\
            W5-E-Pillars5 &  Individual   &        0.47 &      2.0 &          0.35 &              36.0 &     Herschel    &                                     \citet{Deharveng12}   \\
           W5-E-Pillars6a &  Individual   &        1.38 &      2.0 &          0.35 &              36.0 &     Herschel    &                                     \citet{Deharveng12}   \\
           W5-E-Pillars6b &  Individual   &        1.38 &      2.0 &          0.35 &              36.0 &     Herschel    &                                     \citet{Deharveng12}   \\
           W5-E-Pillars7a &  Individual   &        0.96 &      2.0 &          0.35 &              36.0 &     Herschel    &                                     \citet{Deharveng12}   \\
           W5-E-Pillars7b &  Individual   &        1.09 &      2.0 &          0.35 &              36.0 &     Herschel    &                                     \citet{Deharveng12}   \\
            W5-E-Pillars8 &  Individual   &        1.14 &      2.0 &          0.35 &              36.0 &     Herschel    &                                     \citet{Deharveng12}   \\
            W5-E-Pillars9 &  Individual   &        2.46 &      2.0 &          0.35 &              36.0 &     Herschel    &                                     \citet{Deharveng12}   \\
           W5-E-Pillars10 &  Individual   &        0.96 &      2.0 &          0.35 &              36.0 &     Herschel    &                                     \citet{Deharveng12}   \\
           W5-E-Pillars11 &  Individual   &        0.44 &      2.0 &          0.35 &              36.0 &     Herschel    &                                     \citet{Deharveng12}   \\
                W5-E-BRC1 &  Individual   &        55.0 &      2.0 &          0.15 &              15.6 & SD+C$^{18}$O    &                                          \citet{Niwa09}   \\
                W5-E-BRC2 &  Individual   &       550.0 &      2.0 &          0.15 &              15.6 & SD+C$^{18}$O    &                                          \citet{Niwa09}   \\
                W5-E-BRC3 &  Individual   &        67.0 &      2.0 &          0.15 &              15.6 & SD+C$^{18}$O    &                                          \citet{Niwa09}   \\
                W5-E-BRC4 &  Individual   &       170.0 &      2.0 &          0.15 &              15.6 & SD+C$^{18}$O    &                                          \citet{Niwa09}   \\
                W5-E-BRC5 &  Individual   &       110.0 &      2.0 &          0.15 &              15.6 & SD+C$^{18}$O    &                                          \citet{Niwa09}   \\
                W5-E-BRC6 &  Individual   &       230.0 &      2.0 &          0.15 &              15.6 & SD+C$^{18}$O    &                                          \citet{Niwa09}   \\
                W5-E-BRC7 &  Individual   &       740.0 &      2.0 &          0.15 &              15.6 & SD+C$^{18}$O    &                                          \citet{Niwa09}   \\
                W5-E-BRC8 &  Individual   &       180.0 &      2.0 &          0.15 &              15.6 & SD+C$^{18}$O    &                                          \citet{Niwa09}   \\
                W5-E-BRC9 &  Individual   &       130.0 &      2.0 &          0.15 &              15.6 & SD+C$^{18}$O    &                                          \citet{Niwa09}   \\
         CygnusX-Globule1 &  Individual   &      1680.0 &      1.4 &          0.24 &              36.0 &     Herschel    &                                     \citet{Schneider16}   \\
         CygnusX-Globule2 &  Individual   &       282.0 &      1.4 &          0.24 &              36.0 &     Herschel    &                                     \citet{Schneider16}   \\
         CygnusX-Globule3 &  Individual   &        83.0 &      1.4 &          0.24 &              36.0 &     Herschel    &                                     \citet{Schneider16}   \\
         CygnusX-Globule4 &  Individual   &        50.0 &      1.4 &          0.24 &              36.0 &     Herschel    &                                     \citet{Schneider16}   \\
         CygnusX-Globule5 &  Individual   &       156.0 &      1.4 &          0.24 &              36.0 &     Herschel    &                                     \citet{Schneider16}   \\
         CygnusX-Globule6 &  Individual   &      1403.0 &      1.4 &          0.24 &              36.0 &     Herschel    &                                     \citet{Schneider16}   \\
         CygnusX-Globule7 &  Individual   &        86.0 &      1.4 &          0.24 &              36.0 &     Herschel    &                                     \citet{Schneider16}   \\
         CygnusX-Pillars1 &  Individual   &       238.0 &      1.4 &          0.24 &              36.0 &     Herschel    &                                     \citet{Schneider16}   \\
         CygnusX-Pillars2 &  Individual   &       108.0 &      1.4 &          0.24 &              36.0 &     Herschel    &                                     \citet{Schneider16}   \\
         CygnusX-Pillars3 &  Individual   &       180.0 &      1.4 &          0.24 &              36.0 &     Herschel    &                                     \citet{Schneider16}   \\
         CygnusX-Pillars4 &  Individual   &      1356.0 &      1.4 &          0.24 &              36.0 &     Herschel    &                                     \citet{Schneider16}   \\
         LBN 131.54-08.16 & Individual    &       200.0 &      2.0 &               &                   &   Extinction    &                                      \citet{Maheswar08}   \\
                    RNO 6 & Individual    &       190.0 &      2.0 &          0.23 &              24.0 & SD+$^{13}$CO    &                       \citet{Bachiller02, Maheswar08}   \\
                 LDN 1616 & Individual    &       180.0 &      0.4 &          0.12 &              60.0 & SD+$^{13}$CO    &                          \citet{Ramesh95, Maheswar08}   \\
                  Sim 129 & Individual    &       130.0 &      3.4 &               &                   &   Extinction    &                                      \citet{Maheswar08}   \\
                 LDN 1622 & Individual    &       550.0 &     0.45 &               &                   &   Extinction    &                                      \citet{Maheswar08}   \\
                Gal 96-15 & Individual    &       215.0 &          &               &              22.0 & SD+$^{13}$CO    &                           \citet{Olano94, Maheswar08}   \\
               Gal 110-13 & Individual    &        85.0 &     0.44 &          0.12 &              55.0 & SD+$^{12}$CO    &                        \citet{Odenwald92, Maheswar08}   \\
                      CB6 & Individual    &         1.8 &     0.16 &          0.01 &              15.0 & SD+Continuum    &                       \citet{Launhardt10, Maheswar08}   \\
                  ORI-I-2 & Individual    &         3.0 &      0.4 &          0.03 &              17.0 & SD+$^{13}$CO    &                                      \citet{Sugitani89}   \\
                 IC1396-N & Individual    &       150.0 &     0.75 &          0.06 &              17.0 & SD+$^{13}$CO    &                                      \citet{Sugitani89}   \\
                    L1206 & Individual    &        95.0 &     0.91 &          0.08 &              17.0 & SD+$^{13}$CO    &                                      \citet{Sugitani89}   \\
          globule@IC 1396 & Individual    &        20.0 &     0.75 &          0.07 &              20.0 & SD+$^{13}$CO    &                                        \citet{Duvert90}   \\
                 IC 1396E & Individual    &       480.0 &     0.75 &          0.09 &              24.0 &        SD+CS    &                                       \citet{Serabyn93}   \\
              BRC@Sh2-48  & Individual    &       260.0 &      3.8 &          0.61 &              33.0 & SD+Continuum    &                                        \citet{Ortega13}   \\
    Most massive BRC@S233 & Individual    &        90.0 &      2.3 &          0.52 &              47.0 & SD+$^{13}$CO    &                                  \citet{Ladeyschikov15}   \\
                 BRC@CN20 & Individual    &      5200.0 &      5.0 &          0.53 &              22.0 & SD+$^{13}$CO    &                                        \citet{Orgeta16}   \\
         Pillar@G46.5-0.2 & Individual    &        80.0 &      4.0 &           0.7 &              36.0 &     Herschel    &                                         \citet{Paron17}   \\
                  IC1396A & Individual    &       270.0 &    0.945 &          0.16 &              36.0 &     Herschel    &                               \citet{Sicilia19}   \\
               CG@Cep OB3 & Individual    &       350.0 &      0.7 &          0.15 &              45.0 & SD+$^{13}$CO    &                                      \citet{Marshall19}   \\
            Teasure Chest & Individual    &      1000.0 &      2.3 &          0.32 &              29.0 & SD+$^{13}$CO    &                                     \citet{Mookerjea19}   \\
                  BRC@N30 & Individual    &       230.0 &      2.2 &          0.15 &              14.0 & SD+C$^{18}$O    &                                       \citet{Solerno20}   \\
       Massive BRC@RCW120 & Individual    &       200.0 &     1.34 &          0.12 &              19.0 & SD+C$^{18}$O    &                                      \citet{Figueira20}   \\
                   104080 & ALMAGAL \hii   &        2612 &     5.99 &          1.05 &              36.0 & SD+Continuum    &                  \citet{Urquhart18}, AGAL023.389+00.456   \\
                   108768 & ALMAGAL \hii   &        9141 &     7.76 &          1.35 &              36.0 & SD+Continuum    &                  \citet{Urquhart18}, AGAL024.416+00.101   \\
                   109047 & ALMAGAL BRC   &        7798 &     7.76 &          1.35 &              36.0 & SD+Continuum    &                  \citet{Urquhart18}, AGAL024.459+00.197   \\
                   110233 & ALMAGAL \hii   &        5521 &     6.04 &          1.05 &              36.0 & SD+Continuum    &                  \citet{Urquhart18}, AGAL024.673-00.151   \\
                   111140 & ALMAGAL BRC   &         408 &     6.03 &          1.05 &              36.0 & SD+Continuum    &                  \citet{Urquhart18}, AGAL024.849+00.086   \\
                   112942 & ALMAGAL BRC   &         841 &     3.34 &          0.58 &              36.0 & SD+Continuum    &                  \citet{Urquhart18}, AGAL025.224+00.289   \\
                   115679 & ALMAGAL \hii   &        7031 &     8.68 &          1.51 &              36.0 & SD+Continuum    &                  \citet{Urquhart18}, AGAL025.736+00.211   \\
                   115936 & ALMAGAL BRC   &        5058 &     8.68 &          1.51 &              36.0 & SD+Continuum    &                  \citet{Urquhart18}, AGAL025.796+00.242   \\
                   120828 & ALMAGAL \hii   &         243 &     4.42 &          0.77 &              36.0 & SD+Continuum    &                  \citet{Urquhart18}, AGAL026.956-00.076   \\
                   126057 & ALMAGAL BRC   &        2355 &     6.05 &          1.06 &              36.0 & SD+Continuum    &                  \citet{Urquhart18}, AGAL028.321-00.009   \\
                   128280 & ALMAGAL \hii   &         631 &     4.71 &          0.82 &              36.0 & SD+Continuum    &                  \citet{Urquhart18}, AGAL028.834-00.209   \\
                   133651 & ALMAGAL BRC   &        1119 &     5.16 &           0.9 &              36.0 & SD+Continuum    &                  \citet{Urquhart18}, AGAL029.862-00.044   \\
                    59542 & ALMAGAL BRC   &         966 &     2.57 &          0.45 &              36.0 & SD+Continuum    &                  \citet{Urquhart18}, AGAL013.178+00.059   \\
                    62842 & ALMAGAL \hii   &         774 &      3.1 &          0.54 &              36.0 & SD+Continuum    &                  \citet{Urquhart18}, AGAL014.019-00.134   \\
                   704454 & ALMAGAL BRC   &        1340 &      3.8 &          0.66 &              36.0 & SD+Continuum    &                  \citet{Urquhart18}, AGAL305.137+00.069   \\
                   705424 & ALMAGAL \hii   &         883 &      3.8 &          0.66 &              36.0 & SD+Continuum    &                  \citet{Urquhart18}, AGAL305.307-00.039   \\
                   705736 & ALMAGAL BRC   &        1091 &      3.8 &          0.66 &              36.0 & SD+Continuum    &                  \citet{Urquhart18}, AGAL305.362+00.151   \\
                   707313 & ALMAGAL BRC   &         178 &      3.8 &          0.66 &              36.0 & SD+Continuum    &                  \citet{Urquhart18}, AGAL305.667-00.106   \\
                   708358 & ALMAGAL BRC   &         771 &      3.8 &          0.66 &              36.0 & SD+Continuum    &                  \citet{Urquhart18}, AGAL305.887+00.016   \\
                   716353 & ALMAGAL BRC   &         574 &     3.65 &          0.64 &              36.0 & SD+Continuum    &                  \citet{Urquhart18}, AGAL308.646+00.647   \\
                   716713 & ALMAGAL BRC   &         136 &     3.65 &          0.64 &              36.0 & SD+Continuum    &                  \citet{Urquhart18}, AGAL308.731+00.729   \\
                   720116 & ALMAGAL BRC   &        4207 &     6.25 &          1.09 &              36.0 & SD+Continuum    &                  \citet{Urquhart18}, AGAL309.534-00.741   \\
                   730127 & ALMAGAL BRC   &         141 &     3.42 &           0.6 &              36.0 & SD+Continuum    &                  \citet{Urquhart18}, AGAL312.039+00.082   \\
                   737671 & ALMAGAL BRC   &         403 &     4.23 &          0.74 &              36.0 & SD+Continuum    &                  \citet{Urquhart18}, AGAL314.237+00.417   \\
                   737762 & ALMAGAL \hii   &         256 &     4.23 &          0.74 &              36.0 & SD+Continuum    &                  \citet{Urquhart18}, AGAL314.262+00.444   \\
                   744757 & ALMAGAL \hii   &        1239 &     2.52 &          0.44 &              36.0 & SD+Continuum    &                  \citet{Urquhart18}, AGAL316.811-00.059   \\
                   797087 & ALMAGAL BRC   &        1633 &     3.98 &          0.69 &              36.0 & SD+Continuum    &                  \citet{Urquhart18}, AGAL330.674-00.376   \\
                   799034 & ALMAGAL \hii   &        1774 &      5.3 &          0.93 &              36.0 & SD+Continuum    &                  \citet{Urquhart18}, AGAL331.061-00.166   \\
                   800079 & ALMAGAL BRC   &         583 &      5.3 &          0.93 &              36.0 & SD+Continuum    &                  \citet{Urquhart18}, AGAL331.242-00.201   \\
                   800751 & ALMAGAL BRC   &         920 &     3.98 &          0.69 &              36.0 & SD+Continuum    &                  \citet{Urquhart18}, AGAL331.342-00.347   \\
                   807707 & ALMAGAL BRC   &        1268 &     3.57 &          0.62 &              36.0 & SD+Continuum    &                  \citet{Urquhart18}, AGAL332.774-00.584   \\
                   809185 & ALMAGAL \hii   &         705 &     3.57 &          0.62 &              36.0 & SD+Continuum    &                  \citet{Urquhart18}, AGAL333.001-00.436   \\
                   813886 & ALMAGAL \hii   &         676 &     3.57 &          0.62 &              36.0 & SD+Continuum    &                  \citet{Urquhart18}, AGAL333.683-00.256   \\
                   822761 & ALMAGAL BRC   &        1936 &     3.31 &          0.58 &              36.0 & SD+Continuum    &                  \citet{Urquhart18}, AGAL335.789+00.174   \\
                   826927 & ALMAGAL BRC   &         685 &     5.03 &          0.88 &              36.0 & SD+Continuum    &                  \citet{Urquhart18}, AGAL336.573-00.109   \\
                   834656 & ALMAGAL \hii   &         796 &     2.86 &           0.5 &              36.0 & SD+Continuum    &                  \citet{Urquhart18}, AGAL337.934-00.507   \\
                   839208 & ALMAGAL BRC   &         875 &     7.86 &          1.37 &              36.0 & SD+Continuum    &                  \citet{Urquhart18}, AGAL338.836-00.334   \\
                   839678 & ALMAGAL BRC   &        2104 &     4.16 &          0.73 &              36.0 & SD+Continuum    &                  \citet{Urquhart18}, AGAL338.926+00.634   \\
                   876093 & ALMAGAL BRC   &        7362 &     9.84 &          1.72 &              36.0 & SD+Continuum    &                  \citet{Urquhart18}, AGAL347.586+00.216   \\
                   876288 & ALMAGAL BRC   &       13552 &     9.84 &          1.72 &              36.0 & SD+Continuum    &                  \citet{Urquhart18}, AGAL347.627+00.149   \\
        G305.5393+00.3394 & ALMAGAL \hii   &        1531 &      3.8 &          0.66 &              36.0 & SD+Continuum    &                  \citet{Urquhart18}, AGAL305.539+00.339   \\
                       S1 &   Simulation  &           2 &          &         0.015 &                   &                 &            \citet{Sandford82}, resolution is grid size    \\
                       S2 &   Simulation  &          20 &          &         0.008 &                   &                 &                  \citet{Lefloch94}, most massive model    \\
                       S3 &   Simulation  &          85 &          &          0.07 &                   &                 &       \citet{Lefloch95}, CG7S, single dish observation    \\
                       S4 &   Simulation  &          40 &          &           0.1 &                   &                 &                \citet{Kessel03}, SPH simulation    \\
                       S5 &   Simulation  &          30 &          &               &                   &                 &                                     \citet{Motoyama07}    \\
                       S6 &   Simulation  &          96 &          &               &                   &                 &                 \citet{Gritschneder09b}, SPH simulation    \\
                       S7 &   Simulation  &          35 &          &               &                   &                 &     \citet{Miao09}, SPH simulation, most massive model    \\
                       S8 &   Simulation  &          68 &          &               &                   &                 &       \citet{Miao10}, SPH simulation, the Eagle nebula    \\
                       S9 &   Simulation  &           2 &          &               &                   &                 &                   \citet{Miao10}, SPH simulation, IC59    \\
                      S10 &   Simulation  &         666 &          &               &                   &                 &       \citet{Mackey10}, elephant trunks in \hii\ regions    \\
                      S11 &   Simulation  &          15 &          &               &                   &                 &   \citet{Bisbas11}, SPH simulation, most massive model    \\
                      S12 &   Simulation  &         400 &          &               &                   &                 &                 \citet{Anathpindika12}, SPH simulation    \\
                      S13 &   Simulation  &         159 &          &         0.038 &                   &                 &                  \citet{Haworth12}, most massive model    \\
                      S14 &   Simulation  &         185 &          &         0.012 &                   &                 &                  \citet{Haworth13}, most massive model    \\
                      S15 &   Simulation  &          30 &          &               &                   &                 &                      \citet{Kinnear15}, SPH simulation    \\
                      S16 &   Simulation  &         783 &          &               &                   &                 &     \citet{Walch15}, SPH simulation, most massive clump   \\
\end{longtable}
  \begin{tablenotes}
     \item [\textit{(a)}] ~~~~Source catalog. The types of sources described in Sect.~\ref{INTRODUCTION}. ``Simulation'' means the simulation work focusing on the RDI mechanism.
     \item [\textit{(b)}] ~~~Methods of mass calculations. The ``SD'' means single-dish radio observations. The ``Continuum'' means that the molecular mass is derived from mm/submm dust continuum. The ``\textit{Herschel}'' means that the mass is derived from \textit{Herschel} far infrared emissions. The ``Extinction'' means that the mass is derived from optical extinction. HCO$^{+}$, $^{13}$CO, C$^{18}$O, and NH$_{3}$ indicate the molecules that are used to derive the mass with an assumed abundance. 
     \item [\textit{(c)}] ~~~Distance taken from corresponding references.
     \item [\textit{(d)}] ~~~Spatial and angular resolution of the observations. 
     \item [\textit{(e)}] ~~~Corresponding references and some notable information.
  \end{tablenotes}
\end{ThreePartTable}
\end{small}  
\end{landscape}  
\twocolumn
\newpage

\section{ALMA and Other Observational data} \label{App:DataDetails}
\subsection{ALMA data} \label{App:ALMADataDetails}
We made use of ATOMS Band 3 data (ALMA project: 2019.1.00685.S) observed with the main array and 7~m ACA. I18290 was observed on 31st, October, 2019 with the typical precipitable water vapor (PWV) of 3.2 and 5.2~mm for main array and 7~m ACA, respectively. The integration time is three and eight minutes for the main array and 7~m ACA, respectively. 

The data reductions were conducted using \texttt{Common Astronomy Software Applications} 5.6 (\texttt{CASA}, \citealt{Mcmullin07}). A total of eight spectral windows (SPWs) were configured to cover 11 commonly used spectral lines. The narrow spectral windows SPWs 1-6 are at the lower sideband in range of 86.31-89.2~GHz with a spectral resolution of 0.2–0.4~\kms, whereas the wide spectral windows SPWs 7 and 8 are at the upper sideband in the range of 97.52-99.39~GHz and 99.46-101.33~GHz, respectively. The SPWs 7-8 are used for continuum measurements considering that they have a broad bandwidth of 1.87~GHz with a spectral resolution of $\sim1.6$~\kms. 

The main array and 7~m ACA data are combined and then cleaned using  \texttt{CASA} task \texttt{TCLEAN} with a natural weighting and a pixel size ``cell'' of 0.4\arcsec\ (about one fifth of the synthesized beam $\simeq$~2.08\arcsec$\times$1.77\arcsec). The maximum recovering scales are 20\arcsec\ and 76\arcsec\ for the main array and ACA data, respectively.

\subsection{Other observational data} \label{App:SupplementaryDataDetails}
\begin{figure*} 
     \centering
   \includegraphics[width=0.95\textwidth]{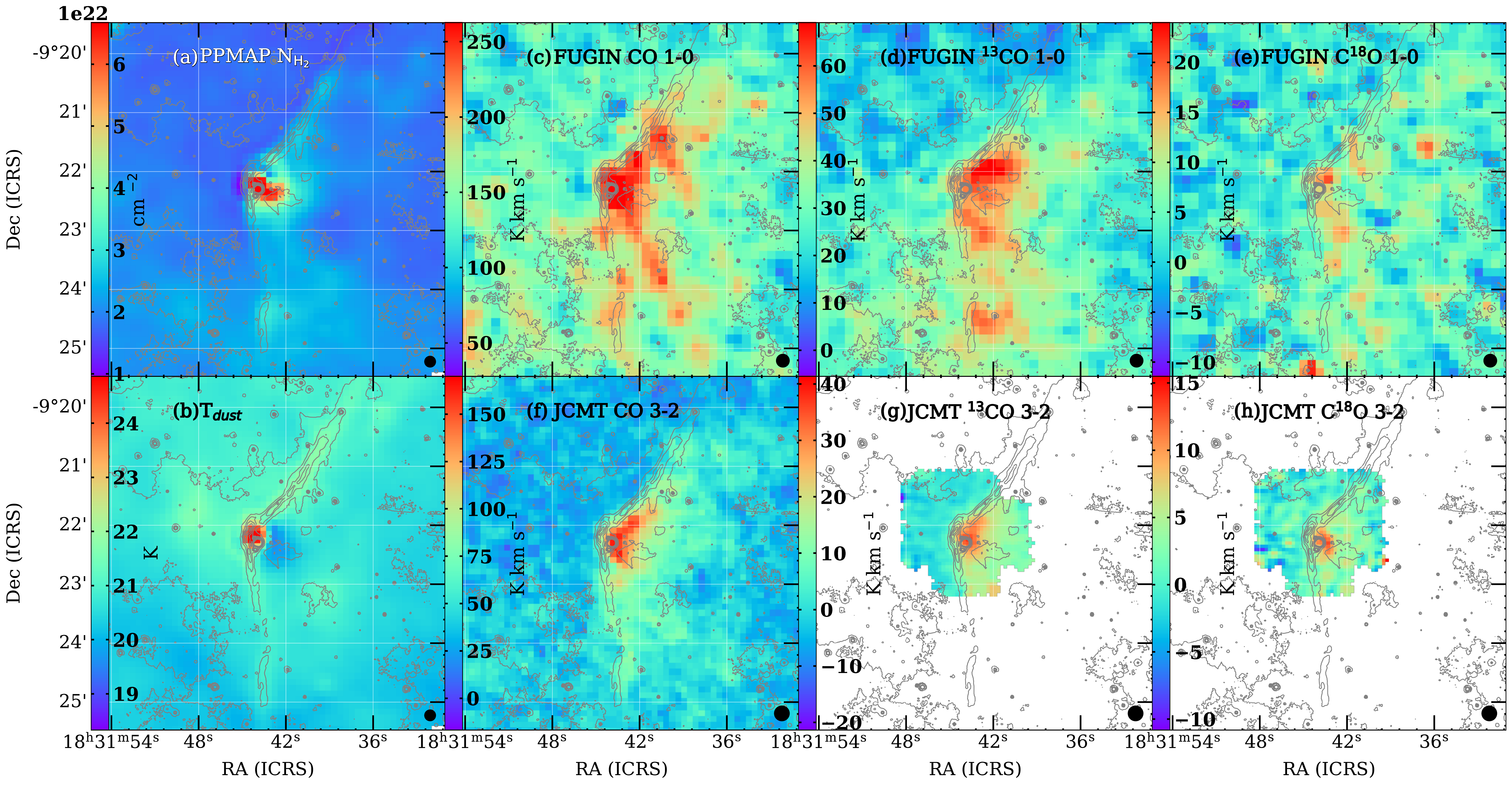}
       \caption{Hi-GAL PPMAP maps, CO and its isotope molecular emission. \textit{Panels a} and \textit{b}: PPMAP column density and dust temperature maps, respectively. \textit{Panels c} to \textit{e}: FUGIN \co, \tco, and \ceo\ $J = 1 - 0$ zeroth moment maps. \textit{Panels f} to \textit{h}: JCMT \co, \tco, and \ceo\ $J = 3 - 2$ zeroth moment maps. The gray contours represent the 8~\micron\ emission with levels of [1, 1.5, 2, 2.5, 3, 3.5, 4, 4.5]~$\times100$~MJy~sr$^{-1}$.}
       \label{PPMAP-CO-FIG} 
    \end{figure*}

\begin{figure}
     \centering
   \includegraphics[width=0.45\textwidth]{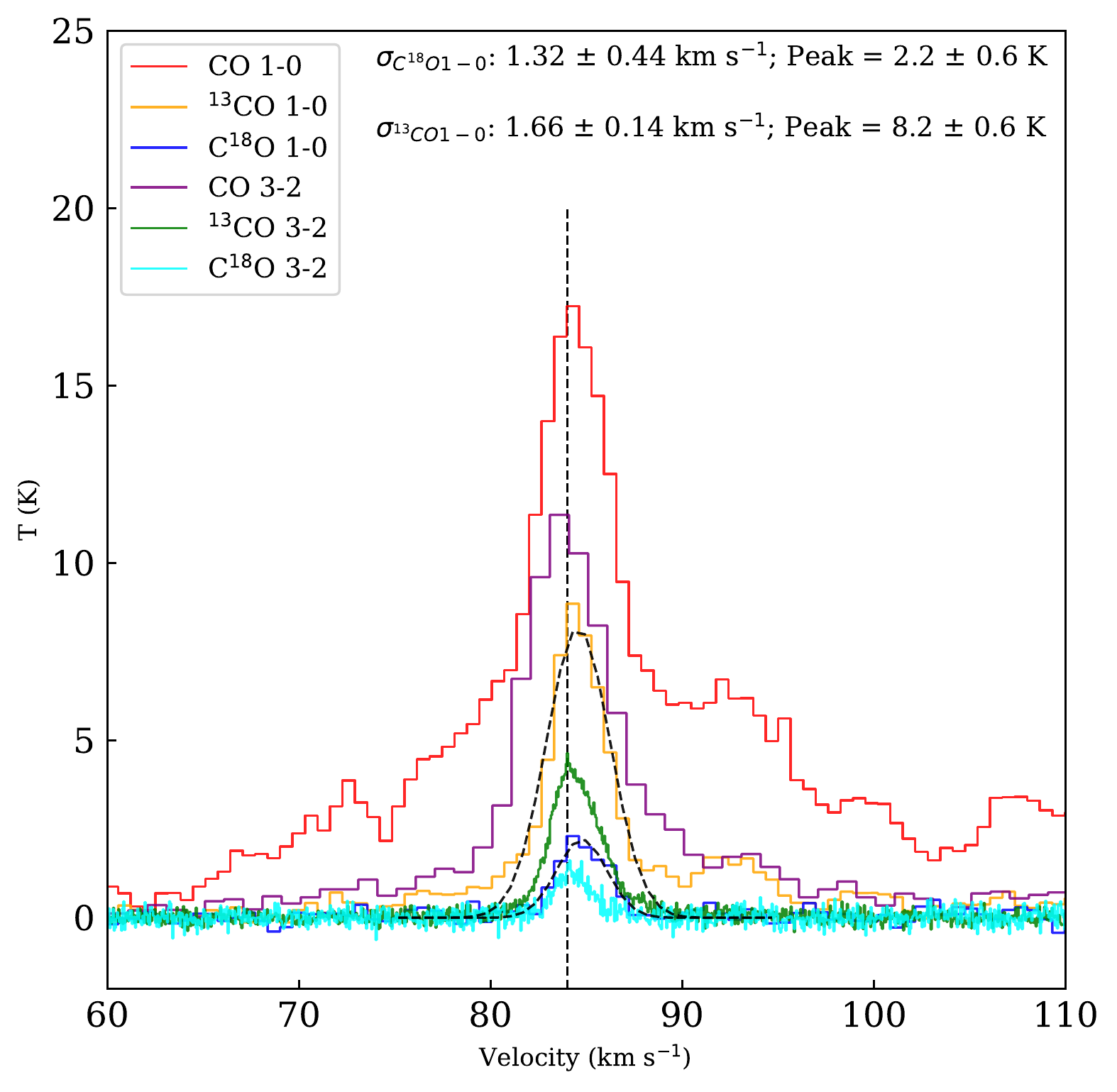}
       \caption{Clump-averaged spectra observed by JCMT and \textit{Nobeyama}. The Gaussian fittings to \tco\ and \ceo\ $J = 1 - 0$ are shown as black dashed lines.}
        \label{SingleDishSpectraFigure}
    \end{figure}

Here we list basic information of the supplementary data used:
 \begin{itemize}
\item \textit{Spitzer} images which include the 4.5~\micron\ and 8.0~\micron\ images taken from GLIMPSE \citep[Galactic Legacy Infrared Mid-Plane Survey Extraordinaire,][]{Churchwell09} and the 24~\micron\ images taken from MIPSGAL \citep[Multiband Imaging Photometer Galactic Plane Survey,][]{Carey09}.
\item \textit{Herschel} PPMAP \nhtcd\ and \dustt\ images. \citet{Marsh17} use the point process mapping (PPMAP) techniques to fit dust emission SED with the \textit{Herschel} infrared Galactic Plane Survey \citep[Hi-GAL,][]{Molinari10} 70, 160, 250, 350, and 500 \micron\ images. The PPMAP results in a resolution of 12\arcsec, much higher than the classical pixel-by-pixel SED fitting. The Hi-GAL PPMAP \nhtcd\ and \dustt\ maps are presented in Fig.~\ref{PPMAP-CO-FIG}.   
\item CO and its isotope molecular spectral cubes. (1) \textit{James Clerk Maxwell} Telescope (JCMT) \co, \tco, and \ceo\ $J = 3 - 2$ cubes. The \cothr\ cube is from CO High-Resolution Survey \citep[COHRS,][]{Dempsey13}, with a spatial and velocity resolution of 16.6\arcsec\ and 1~\kms, respectively. The \tco\ and \ceo\ $J = 3 - 2$ cubes are from \citet{DeVilliers14} observed by the JCMT project M07AU20 (PI: Mark Thompson). Their spatial resolution is similar to \co\ data but with a much higher velocity resolution (0.055~\kms). (2) The \co, \tco, and \ceo\ $J = 1 - 0$ cubes are taken from FUGIN \citep{Umemoto17} which is a Galactic plane survey mainly targeting on the first and third quadrants with a spatial and velocity resolution of 20\arcsec\ and 1.3~\kms, respectively. The zeroth moment maps for these transitions are shown in Fig.~\ref{PPMAP-CO-FIG}. The spectra averaged in the ATLASGAL 870~\micron\ clump region for \co\ and its isotope molecules are shown in Fig.~\ref{SingleDishSpectraFigure}.
\item VLA \nht\ (1, 1) and (2, 2) inversion line cubes from \citet{Lu14} who used VLA D and DnC array configurations to survey 62 high-mass star-forming regions. The spatial and velocity resolutions are around 5.1\arcsec$\times$3.2\arcsec\ and 0.6~\kms, respectively. The zeroth moment map for (1, 1) is shown in Fig.~\ref{moleculartracer}.
\item The twenty-centimeter continuum image combined from New GPS (Galactic Plane Survey) + THOR \citep[The HI/OH/Recombination line survey of the inner Milky Way,][]{Beuther16, Wang20}. New GPS 20 cm image is taken from the project MAGPIS \citep[Multi-Array Galactic Plane Imaging Survey,][]{Helfand06} with a resolution of $\sim$6.2\arcsec $\times$ 5.4\arcsec\ (VLA B configuration). The THOR 20 cm data has a resolution of 25\arcsec\ (VLA C configuration). The THOR data have been combined with the VLA Galactic Plane Survey (VGPS, VLA D + Effelsberg, \citealt{Stil06}) using the \texttt{CASA} task \texttt{feather}. To recover both extended and compact structures, we combined the New GPS data with the THOR data using \texttt{feather}, with $\texttt{sdfactor} = 1$. The flux of the combined 20~cm image is improved about 10\% compared to the original New GPS data.
\end{itemize}

\section{rim and IBL} \label{App:RimAndIBL}
\subsection{Bright rim extracted by Filfinder}
\label{App:BrightRimIdentification}
Spine of the 8~\micron\ rim is extracted using \texttt{Filfinder}\footnote{\url{https://fil-finder.readthedocs.io/en/latest/}} that is based on mathematical morphology \citep{Koch15}. \texttt{Filfinder} first creates a mask using a threshold and then generates spines for the mask region. The bright 8-\micron\ point sources located at the head of BRC bias the extraction of spine by distorting the spine path. We simply remove these bright point sources when extracting the rim mask. 

In the procedure of mask creation (\texttt{Filfinder} task \texttt{Filfinder2D.create$\_$mask}), there are two parameters crucial to determine the shape and size of rim mask: (1) Global threshold (\texttt{glob$\_$thresh}), the minimum intensity of the pixel included in the rim mask, is 129~MJy~sr$^{-1}$ in our case. (2) Adaptive threshold (\texttt{adapt$\_$thresh}), the expected width of the rim, is 0.18~pc in our case.

An overdense and warm layer associated with the 8-\micron\ rim is revealed in \textit{Panels a} and \textit{b} of Fig.~\ref{PPMAP-CO-FIG}. We estimate the rim mass except for the central clump. To limit the rim region, we simply use the 8-\micron\ mask created by \texttt{Filfinder} in the rim extraction process. We got a mass of $\sim1100$~\msun\ for the rim region, by extracting the rim's pixels in the PPMAP \nhtcd\ map. A \nhtcd\ background of $\sim9\times10^{21}$~\cm\ is removed in the integration.

\subsection{IBL calculations} \label{App:IBLCalculations}
The total pressure exerted by IBL (\Pex) is derived from the New GPS + THOR combined 20~cm continuum. Assuming that all ionizing photons are absorbed in IBL, the electron number density $n_{\rm e}$ and photon flux $\Phi$ can be estimated from 20~cm continuum. Following the widely used formulae that are originally from the BRC quantitative analysis of \citet{Lefloch97} and then rearranged by \citet{Thompson04}, 
\begin{equation}
\centering
n_{\rm e} = 122.2\sqrt{\left(\frac{S_{\rm \nu}}{\rm mJy}\right)   \left(\frac{T_{\rm e}}{\rm K}\right)^{0.35} \left(\frac{\nu}{\rm GHz}\right)^{0.1} 
\left(\frac{\eta r_{\rm clp}}{\rm pc}\right)^{-1} \left(\frac{\Theta}{\arcsec}\right)^{-2}}~\rm cm^{-3},
\label{ElectonDensity}
\end{equation}
\begin{equation}
\centering
\Phi = 1.24\times 10^{10} \left(\frac{S_{\rm \nu}}{\rm mJy}\right)   \left(\frac{T_{\rm e}}{\rm K}\right)^{0.35} \left(\frac{\nu}{\rm GHz}\right)^{0.1}  \left(\frac{\Theta}{\arcsec}\right)^{-2}~\rm cm^{-2}~s^{-1},
\label{PhtotFlux}
\end{equation}
where $S_{\rm \nu}$ is the integrated flux at frequency $\nu$, $\Theta$ is the effective angular diameter over which the flux is integrated. There is no $T_{\rm e}$ measurement for our IBL and thus we simply estimate $T_{\rm e}$ using the empirical relation \citep{Tremblin14}:
\begin{equation} \label{TeRGCRelation}
T_{\rm e} =  278~{\rm K}~(\frac{R_{\rm GC}}{\rm kpc}) + 6080~{\rm K}.
\end{equation}

\citet{Bertoldi89} developed an approximate analytical solution for RDI model and found that the thickness of IBL is about $\eta r_{\rm clp}$ with $\eta$ $\sim0.1$ to 0.2 for a wide range of parameter space. I18290 IBL is assumed to be the region enclosed by the 20~cm continuum contour of $\sim2.38$~\mjybeam\ (40\% of the 20~cm continuum peak value), resulting in a width of about 0.38~pc (15\arcsec) which is approximately in line with the analytical model. The resulting photon flux from the exciting OB cluster is $\Phi\simeq1.8 \times 10^{10}$~cm$^{-2}$~s$^{-1}$, corresponding to a stellar photon rate of $\sim4.8\times10^{51}$~s$^{-1}$. Note that the estimated photon rate has large uncertainties because (1) The interstellar absorption is ignored. (2) The initial separation between I18290 and the exciting OB cluster can be smaller than we observe now because of the rocket effect \citep{Saha22} and supernova events. (3) The projection effect leads to an underestimated separation.

Following \citet{Haworth12}, the mass loss rate due to photoevaporation could be estimated with 
\begin{equation}
\centering
\dot{M}_{\rm eva} = 4.4 \times 10^{-3} \left(\frac{\Phi}{\rm cm^{-2} s^{-1}}\right)^{0.5}   \left(\frac{r_{\rm clp}}{\rm pc}\right)^{1.5}~\rm M_\odot~Myr^{-1},
\label{EvaporationMassLossRate}
\end{equation}
We have $\dot{M}_{\rm eva}\sim680$~$\rm M_\odot~Myr^{-1}$. The remaining lifetime of the clump is $M_{\rm clp}/\dot{M}_{\rm eva}\simeq2.1$~Myr.

\section{Density and mass derived from spectra}\label{App:COLUMNDENSITYANDMASS}
Following the derivations in \citet{ZhangYichen16}, the LTE column density ${dN}/{d{\rm v}}$ and the optically-thin LTE column density ${dN}/{d{\rm v}}~|_{\rm thin}$ are correlated by:
\begin{equation}  \label{TauRelation}
 \frac{dN}{d{\rm v}} = \frac{dN}{d{\rm v}} |_{\rm thin} \frac{\tau_{\rm v}}{1 - {\rm exp}\left(-\tau_{\rm v}\right)},
\end{equation}
where $\tau_{\rm v}$ is the optical depth at velocity v. This equation is derived based on the assumptions that the background temperature \tbg~=~2.7~K is much smaller than the excitation temperature \tex\ ($>10$~K) and the studied lines are at a frequency of about 80 to 110~GHz and have a line width of $<100$~\kms. Here, the column densities  
\begin{equation} \label{NoBGColumnDensityEquations}
\frac{dN}{d{\rm v}} \simeq \left(\frac{8\pi k_{\rm B}\nu_{ul}^2}{h c^3 A_{ul} g_u R}\right)
 Q_{\rm rot}(T_{\rm ex}) \exp\left(\frac{E_u}{k_{\rm B}T_{\rm ex}}\right)
\left(\frac{\tau_{\rm v}}{1-\exp(-\tau_{\rm v})}\right) \frac{T_{\rm r}({\rm v})}{f},
\end{equation}
\begin{equation}   \label{OpticalThinColumnDensityEquations}
\frac{dN}{d{\rm v}}|_{\rm thin} \simeq \left(\frac{8\pi k_{\rm B}\nu_{ul}^2}{h c^3 A_{ul} g_u R}\right)
 Q_{\rm rot}(T_{\rm ex}) \exp\left(\frac{E_u}{k_{\rm B}T_{\rm ex}}\right)\frac{T_{\rm r}({\rm v})}{f},
\end{equation}
where $\nu_{ul}$, $A_{ul}$, $R$, $f$, \tr, $Q_{\rm rot}$, $E_u$, and $g_u$, are the transition frequency, Einstein A coefficient, relative intensity of the hyperfine line, beam filling factor, radiation temperature, partition function, energy and degeneracy of the upper level, respectively.

In the following subsections, Eqs.~\ref{NoBGColumnDensityEquations} and \ref{OpticalThinColumnDensityEquations} are used in our calculations of the clump mass, outflow properties, and \cch\ column density \ncchcd.


\subsection{Clump mass} \label{COLUMNDENSITYANDMASS-CLUMPMASS}
The kinematic distances of I18290 estimated by \citet{Urquhart18} and \citet{Mege21} are 5.34~kpc and $4.84\pm0.27$~kpc, respectively. In our work, a distance of 5.34~kpc with 10\% error (0.53~kpc) is adopted. With this distance, \citet{Urquhart18} derived a \mclump\ of $\sim1420$~\msun\ by fitting SED from 8~\micron\ to 870~\micron\ with a two-components (warm and cold) gray-body function. \citet{Lu14} derived a \mclump\ of 1400~\msun\ at this distance, using VLA \nht\ inversion lines with the abundance [\nht/\hmole]~$=3\times10^{-8}$. 

The clump-averaged spectra of \co\ show complicated profiles compared to those of \tco\ and \ceo\ and thus we use $J = 1 - 0$ transitions of \tco\ and \ceo\ to probe the mass with moderate density and temperature.
Assuming two isotopologues I$_{1}$ and I$_{2}$ have the same \tex\ and $f$, rewriting and combing the radiation transfer formulae for I$_{1}$ and I$_{2}$:
\begin{equation} \label{RadiationTransferEquation}
T_{\rm r}({\rm v}) = f\left[J_{\nu}(T_{\rm ex})-J_{\nu}(T_{\rm bg})\right]\left(1-\exp(-\tau_{\rm v})\right),
\end{equation}
\begin{equation}
J_\nu(T)=\frac{h\nu/k_{\rm B}}{\exp\left(\frac{h\nu}{k_{\rm B}T}\right)-1},
\end{equation}
and then we have
\begin{equation} \label{TemperatureRatioEquation}
    \frac{T_{\rm r, I_{1}} \left({\rm v}\right)}{T_{\rm r, I_{2}} \left({\rm v}\right)} = \frac{1 - {\rm exp}\left(-\tau_{\rm v, I_{1}} \right)}{1 - {\rm exp}\left(-\tau_{\rm v, I_{2}} \right)}.
\end{equation}
Further assuming that \ceoone\ is optically thin and $\rm I_{1} =~^{13}CO$, $\rm I_{2} = C^{18}O$, we have 
\begin{equation} \label{TemperatureRatioSimCOEquation}
    \frac{T_{\rm r, ^{13}CO} \left({\rm v}\right)}{T_{\rm r, C^{18}O} \left({\rm v}\right)} = \frac{1 - {\rm exp}\left(-\tau_{\rm v, ^{13}CO} \right)}{\tau_{\rm v, C^{18}O}} \simeq \chi_{\rm ^{13}CO/C^{18}O}\frac{1 - {\rm exp}\left(-\tau_{\rm v, ^{13}CO} \right)}{\tau_{\rm v, ^{13}CO}},
\end{equation}
where abundance ratio $\chi_{\rm ^{13}CO/C^{18}O}$ can be estimated from isotope ratio gradients \citep{Wilson94,Milam05}
\begin{equation} \label{IsotopeRatioEquations}
[^{16}{\rm O}/^{18}{\rm O}] = 58.8R_{\rm GC} + 37.1,\,[^{12}{\rm C}/^{13}{\rm C}] = 7.5R_{\rm GC} + 7.6.
\end{equation}
Therefore, $\tau_{\rm v}$ and associated $N$ can be solved with Eqs.~\ref{TemperatureRatioSimCOEquation} and \ref{NoBGColumnDensityEquations}, respectively.
To reduce noise in the $\tau_{\rm v}$ calculation, we follow the optical depth correction methods usually used in the protostellar outflows \citep[e.g.][]{Dunham14,Feddersen20}. The temperature ratio $T_{\rm r, ^{13}CO} \left({\rm v}\right)/T_{\rm r, C^{18}O} \left({\rm v}\right)$ is fitted with a parabola function for channels with high signal-to-noise (SN) ratio and then $\tau_{\rm r, ^{13}CO}\left({\rm v}\right)$ is derived from this parabola according to Eq.~\ref{TemperatureRatioSimCOEquation}. The maximum $\tau_{\rm v}$ is $\sim1.5$ at 85~\kms. With Eq.~\ref{IsotopeRatioEquations} and \co\ abundance \citep{Fontani06}
 \begin{equation} \label{COAbundanceEquation}
  \chi_{\rm CO} = 9.5\times10^{-5} {\rm exp}\left(1.105-0.13R_{\rm GC}\right),  
 \end{equation}
and adopting $\nu_{ul} = 110.201$~GHz, $g_u = 2\times1+1 = 3$, $A_{ul} = 6.332\times10^{-8}$~s$^{-1}$, $E_{u} = 5.29$~K, $R = 1$, and \tex~=~18~K for \tcoone, we have $M_{\rm clp} \simeq 1450$~\msun\ and \nhtcd$\sim1.72\times10^{22}$~\cm\ for I18290, which are a factor of 1.6 compared to the ones without optical depth correction.

\subsection{Outflow mass and properties} \label{COLUMNDENSITYANDMASS-OUTFLOW}
Among the ATOMS-observed spectra, $J = 1 - 0$ spectra of \hcop\ and \htcop\ are used to probe outflow properties. Following the optical depth corrections in Appendix~\ref{COLUMNDENSITYANDMASS-CLUMPMASS} and assuming that \htcopone\ is optically thin, the optical depth $\tau_{\rm v, HCO^+}$ of outflow \hcopone\ emission could be corrected to some extent. First, a cube of the temperature ratio $T_{\rm r, HCO^{+}} \left({\rm v}\right)/T_{\rm r, H^{13}CO^{+}} \left({\rm v}\right)$ is generated for the voxels with SN ratio larger than two in both \hcop\ and \htcop. The mean temperature ratios for each channel are calculated when that channel has more than five valid voxels. Then, a parabola fitting is applied on the mean temperature ratios of all valid channels. We set $\nu_{ul} = 89.1885247$~GHz, $g_u = 2\times1+1 = 3$, $A_{ul} = 4.187\times10^{-5}$~s$^{-1}$, $E_{u} = 4.28$~K, and $R = 1$ for \hcopone. The \tex\ of \hcopone\  is set as 50~K, similar to the ALMA outflow survey towards HMSF regions carried out by \citet{Baug21}. \citet{Dunham14} propose that a variation of \tex\ from 10 to 50~K could cause to a maximum variation in outflow parameters by a factor of one to three. 

We adopt abundance $\rm [HCO^{+}/H_2] = 5\times10^{-9}$, which is the typical value in the protostellar outflows revealed by \citet{Ospina-Zamudio19}. Note that \hcop\ abundance is enhanced in the outflow region and its variation can reach one magnitude even between different lobes of the same outflow. The outflow mass and its related parameters could be reduced to one fourth of the initial values if we use $\rm [HCO^{+}/H_2] = 2\times10^{-8}$ that is the upper abundance detected by \citet{Ospina-Zamudio19}. Another uncertainty is that \htcopone\ could be optically thick around systematic velocity for dense cores, as the spectra shown in Fig.~\ref{corespectral3}. Therefore, the derived outflow parameters could be underestimated on this point. A series of outflow parameters calculated based on channel-by-channel counting are listed in Table~\ref{OutflowTable}.

\begin{table*}
\caption{\label{OutflowTable}Main outflow properties} 
\begin{threeparttable}
\centering
\footnotesize
\begin{tabular}{cccccccccccc}
\hline
\hline
Lobe & $M_{\rm out}$\tnote{\textit{(a)}} & $P_{\rm out}$\tnote{\textit{(a)}} & $E_{\rm out}$\tnote{\textit{(a)}}     & $\dot{M}_{\rm out}$\tnote{\textit{(b)}}   &  $\dot{P}_{\rm out}$\tnote{\textit{(c)}}    &  $\dot{E}_{\rm out}$\tnote{\textit{(c)}} & $t_{\rm out}$\tnote{\textit{(d)}}   & $l_{\rm out}$\tnote{\textit{(e)}}  & PA\tnote{\textit{(f)}} & OA \tnote{\textit{(g)}} & Velocity Range \tnote{\textit{(h)}} \\

  & $\rm M_\odot$ & $\rm M_\odot km s ^{-1}$    & $\rm M_\odot km ^{2} s^{-2}$  &  $\rm M_\odot kyr^{-1}$  &$\rm M_\odot km s ^{-1} kyr^{-1}$  &   $\rm L_\odot$  & kyr & pc & \degree & \degree       & $\rm km s ^{-1}$    \\
\hline
Red  &  12 & 48    & 112  &  0.66 & 2.6  &   1.0  & 18.4 & 0.20 & $-$145 &  88       & 86.75-95.25  \\
Blue &  10 & 47    & 139 &  0.54  & 2.6  &   1.3  & 17.9 & 0.24 & 52 &  102       & 70.75-81.50    \\
Overall & 22 & 95    & 251 &  1.2 & 5.2 &   2.3 & 18.2 & 0.22 &   &          &   \\
\hline
Thick/Thin\tnote{\textit{(i)}} & 2 & 1.7    & 1.4 &  2.0  & 1.7  &  1.4 &   &   &   &          &  \\
\hline
\end{tabular}
      \begin{tablenotes}
      \item [\textit{(a)}] Total outflow mass, momentum, and energy derived from the sum of outflow mass, momentum, and energy of all included channels, respectively. $M_{\rm out} = \sum M_{\rm v}$, $P_{\rm out} = \sum \left(M_{\rm v}\times {\rm v}\right)$, and $E_{\rm out} = \sum \left(1/2 M_{\rm v}\times {\rm v^2}\right)$. 
      \item [\textit{(b)}] Outflow mass rate $\dot{M}_{\rm out} = M_{\rm out}/t_{\rm out}$, here $t_{\rm out}$ is the kinematic age presented in the eighth column.
      \item [\textit{(c)}] Mechanical force $\dot{P}_{\rm out} = P_{\rm out}/t_{\rm out}$ and mechanical luminosity $\dot{E}_{\rm out} = E_{\rm out}/t_{\rm out}$, respectively.
      \item [\textit{(d)}] Kinematic age $t_{\rm out} = l_{\rm out}/{\rm v}_{\rm max}$, here ${\rm v}_{\rm max}$ and $l_{\rm out}$ are the maximum outflow velocity and length, respectively.
      \item [\textit{(e)}] Outflow length $l_{\rm out}$, derived from the 90th percentile of the outflow maximum length of all included channels. 
      \item [\textit{(f)}] Position angle PA, derived from the median of PA for all included channels. The PA of one channel is the median PA of all outflow voxels in that channel.
      \item [\textit{(g)}] Opening angle OA, derived from median OA of all channels. The OA of one channel is derived from FWHM of OA distribution of all outflow voxels in that channel. 
      \item [\textit{(h)}] Velocity ranges that outflow parameters are integrated and counted.
      \item [\textit{(i)}] Value ratios of optically thick calculations to optically thin calculations.    
      \end{tablenotes}
      \end{threeparttable}
\end{table*}


\subsection{\cch\ column density} \label{COLUMNDENSITYANDMASS-CCH}
Here we derive \cch\ column density \ncchcd\ from \cchline\ following \citet{Sanhueza12} and \citet{Buslaeva21}. The rotational energy levels of \cch\ are
described by rotational quantum number $N$ rather than $J$. The relative intensity of hyperfine line $R = 1.66/4.0$ because we only use one of the six hyperfine lines of $N = 1 - 0$ \citep{Sanhueza12}. The PDR should have a higher temperature compared to the clump body, here we test \tex~=~60, 120, 180, and 240~K for optically thin case using Eq.~\ref{OpticalThinColumnDensityEquations}. Setting $\nu_{ul} = 87.317$~GHz, $g_u = 5$, $A_{ul} = 1.527\times10^{-6}$~s$^{-1}$, $E_{u} = 4.19$~K, we have median \ncchcd\ of 2.8 to $13\times10^{14}$~\cm\ and 1.9 to $8.9\times10^{14}$~\cm\ for the compressed and PeF components, respectively.

\section{Core and YSO candidates} \label{CoreYSOsExtraction}
\subsection{Core extraction by Astrodendro}
\label{CoreExtraction}
The dense cores are extracted in the cleaned main array + 7~m ACA combined continuum image (uncorrected with primary beam to have a uniform noise field) using \texttt{Astrodendrograms}\footnote{\url{https://dendrograms.readthedocs.io/en/stable/}} \citep{Rosolowsky08}. The extracted cores are ``leaves'' which do not have  substructure in the dendrogram. The intensity threshold and the steps to differentiate the leaves are set as 4.5~rms and 0.5~rms, respectively. The minimum size considered as an independent leaf is half of the synthesized beam. Table~\ref{ALMACoresBasics} lists the primary beam-corrected properties for the extracted cores. 

Assuming that dust continuum is optically thin, the core mass is \citep{Hildebrand83}
\begin{equation}
\centering
M_{\rm core} = R_{\rm gd} \frac{S_{\nu}D^{2}}{\kappa_{\nu}{B_{\nu}(T_{\rm dust})}}, 
\label{MassEquation}
\end{equation}
where $S_{\nu}$ is the flux corrected by the primary beam, $B_{\nu}(T_{\rm dust})$ is the Planck function at frequency $\nu$ and \dustt. The gas-to-dust mass ratio $R_{\rm gd}$ is set as 100 in our case. The dust opacity per gram $\kappa_{\nu} = 0.18$~cm$^{2}$~g$^{-1}$, corresponding to the opacity of dust grains with thin ice mantles at a gas density of $\sim10^{6}$~cm$^{-3}$ \citep{Ossenkopf94, LiuT20ATOMSI}. The  $S_{\nu}$ errors are derived by multiplying image rms with core area. The combined errors of ${R_{\rm gd}}/{\kappa_{\nu}}$ are $\sim30$\% (see \citealt{Zhang21} and references therein). 

The \dustt\ is assumed to be equal to \nht\ kinetic temperature \tkin\ \citep{Lu14}. The \tkin\ uncertainty of core is simply set as the value range of the pixels' \tkin\ in one VLA beam. Core masses for the lowest and highest \tkin\ are estimated ($M_{\rm core}^{\rm cold}$ and $M_{\rm core}^{\rm warm}$). The differences between $M_{\rm core}$ and $M_{\rm core}^{\rm cold}$ or $M_{\rm core}^{\rm warm}$ are typically less than 20\%, except for C1. 

We also estimate core mass surface density $\Sigma_{\rm core} = M_{\rm core}/\pi r_{\rm core}^{2}$, number density \nhtnd, and column density \nhtcd\ with the assumptions of a spherical shape and a molecular weight per hydrogen molecule ${\mu}_{\rm H_{2}} = 2.8$ \citep{Kauffmann2008}.

\begin{table*}
\caption{\label{ALMACoresBasics} \texttt{Astrodendrogram} results for combined images.}
\begin{threeparttable}
\centering
\setlength{\tabcolsep}{4.5pt}
\renewcommand{\arraystretch}{1.0}
\begin{tabular}{cccccccccr}
\hline
\hline
Core\tnote{\textit{(a)}} & RA     & DEC     & major\tnote{\textit{(b)}}    & minor\tnote{\textit{(b)}}     & PA     & $r_{\rm core}$\tnote{\textit{(b)}}         & $S$\tnote{\textit{(c)}} & $S_{\rm p}$\tnote{\textit{(c)}} & SN ratio \\
     & \degree & \degree & \arcsec & \arcsec & \degree & \arcsec &  mJy    &     mJy~beam$^{-1}$              &     \\
\hline
C1 & 277.9339 & -9.3701 & 0.96 & 0.79 & 100.4  & 0.87 & 7.76$\pm$0.37 & 6.3$\pm$ 0.11 & 21.1 \\
C2 & 277.9279 & -9.3742 & 1.13 & 0.74 & 150.2  & 0.92 & 2.25$\pm$0.27 & 1.7 $\pm$ 0.08 & 8.5 \\
C3 & 277.9303 & -9.3739 & 0.62 & 0.52 & -179.6 & 0.57 & 1.91$\pm$0.09 & 2.5 $\pm$ 0.08 & 22.3 \\
C4 & 277.9337 & -9.3716 & 1.04 & 0.53 & 128.9  & 0.74 & 1.18$\pm$0.14 & 0.98 $\pm$ 0.10 & 8.6 \\
C5 & 277.9296 & -9.3737 & 0.38 & 0.31 & -178.6 & 0.34 & 0.55$\pm$0.03 & 1.7 $\pm$ 0.08 & 17.9 \\
\hline
\end{tabular}
      \begin{tablenotes}
      \item [\textit{(a)}] Cores are ranked with their mass.
      \item [\textit{(b)}] Major and minor semi-axes, and equivalent radius, respectively.
      \item [\textit{(c)}] $S$ and $S_{\rm p}$ are the primary beam corrected integrated flux and pixel maximum flux, respectively.
      \end{tablenotes}
      \end{threeparttable}
\end{table*}

\subsection{Classification of candidate YSOs} \label{appendix:yso}
\begin{figure}
  \centering
  \includegraphics[width=.95\linewidth]{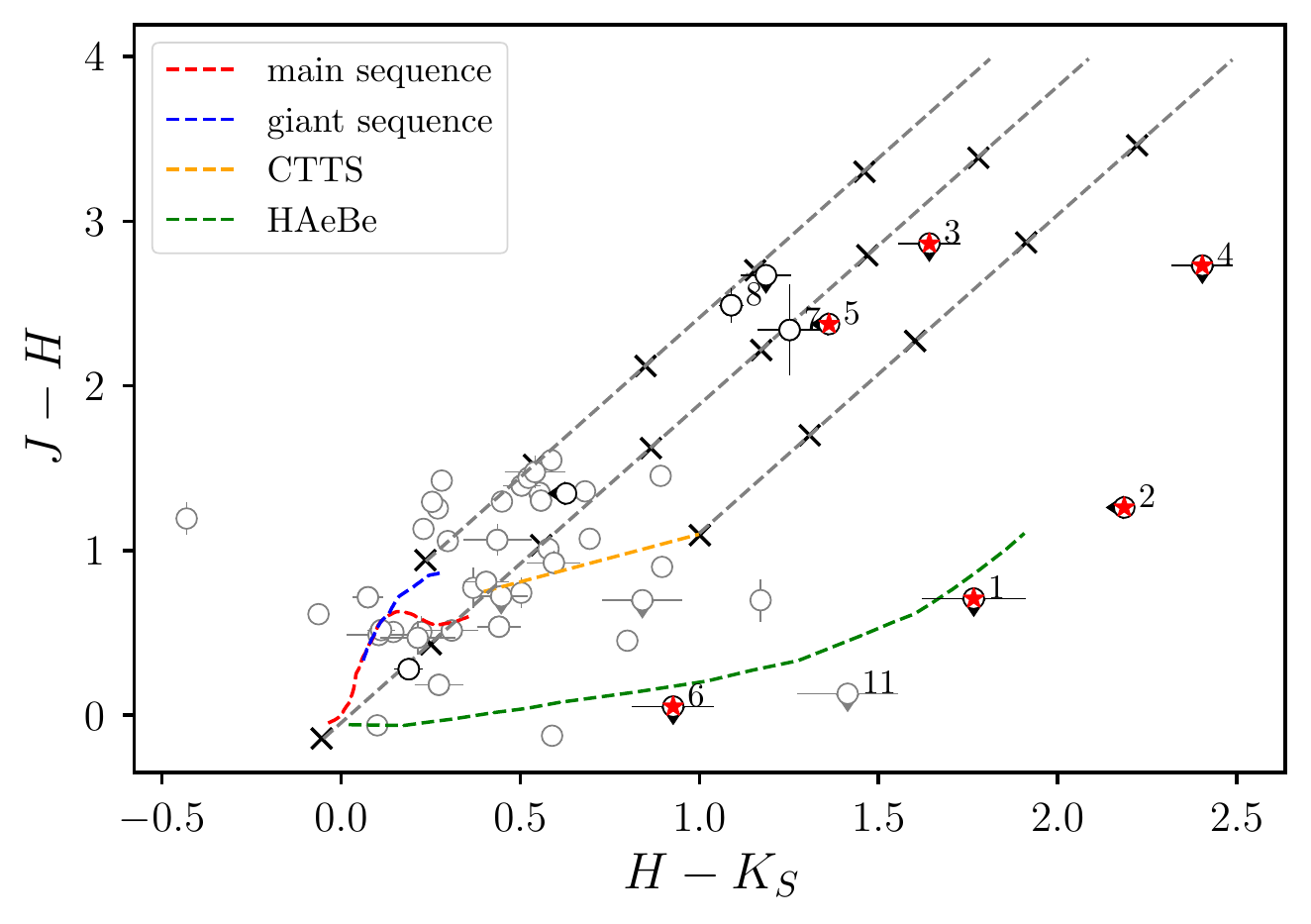}  
  \caption{2MASS color-color diagram. The loci of the main sequence, giants, CTTS, and HAeBe are taken from \citet{Bessell88}, \citet{Lada92}, and \citet{Meyer97}. The gray parallel lines are the reddening vectors from \citet{Rieke85}, with crosses marking the intervals of 5~mag visual extinction. The dots with the gray and black colors represent the 2MASS sources which are projected  and (possibly) associated sources according to the Gaia measurements, respectively. The numbers mark the sources listed in Table~\ref{CandidateYSOtable}. The red stars highlight the 2MASS-selected YSOs.}
  \label{loci-2mass}
\end{figure}

We classify candidate YSOs in a square region centered at RA = 18$^{\rm h}$31$^{\rm m}$43.23$^{\rm s}$, DEC = $-9$\degree$22$\arcmin$28.5$\arcsec\ with a width of 1.3\arcmin, using the 2MASS and GLIMPSE point source catalogs (2MASS All-Sky Point Source Catalog and GLIMPSE I Spring 07 Archive), including bands 2MASS $J$, $H$, $K_{S}$ and IRAC 3.6, 4.5, 5.8, and 8.0~\micron. There are fifty-one 2MASS and sixty-seven GLIMPSE point sources in this field. Gaia measurements are available for 41 and 35 of them, respectively. Most of the derived Gaia 
distances for them are $< 3$~kpc. We make use of three relatively independent methods to extract candidate YSOs from the source catalogs: (1) 2MASS $H - K_{S}$ vs $J - H$ color-color criteria. The 2MASS color-color diagram (CCD) is able to quantify the infrared excess caused by circumstellar grains re-radiation. The candidate YSOs could be classified by comparing colors of the observed sources and a series of loci for different types of stars (main sequence, HAeBe, giant stars, and classical T Tauri (CTTS) stars). Figure~\ref{loci-2mass} shows the $H - K_{S}$ vs $J - H$ CCD and the mentioned loci. The three gray dashed lines in Fig.~\ref{loci-2mass} mark the regimes of Class~I, Class~II, and Class~III (or field) stars from right to left. Several 2MASS-selected YSOs are likely to follow the loci of HAeBe stars, such as YSOs~\#1, 2, 6, and 11. (2) GLIMPSE color-color criteria. We follow the procedures proposed by \citet{Wang07} which extract YSOs based on $J - [3.6]$ vs $K_{S} - [4.5]$ CCD. These authors initially used it to extract YSOs around HAeBe stars. Only one candidate (YSO~\#7) is extracted with this method. (3) YSO SED modelling. By fitting YSO SED models\footnote{\url{https://sedfitter.readthedocs.io/en/stable/}} of \citet{Robitaille07} for the 31 point sources (20 of them have Gaia measurements) with more than three valid photometric measurements in 2MASS and GLIMPSE bands, only two YSOs (YSOs~\#7 and 8) are extracted.

Note that we possibly miss a number of potential YSOs due to bright and diffuse emission of the rim which causes some GLIMPSE sources, such as YSOs~\#9, 10, and 11, to lack valid photometric measurements and leads to a failure in SED fitting or color-color selection. YSO~\#11 is not included in GLIMPSE catalog though its 8~\micron\ emission is the strongest in I18290. The Gaia parallax for YSO~\#11 (Gaia DR3 id: 4155952506159262592) is $0.818\pm0.367$~mas, corresponding to $1.22_{-0.38}^{+1.0}$~kpc, and therefore it is just a foreground source.

\section{Core spectra and infall estimation} \label{App:CoreSpectraAndInfall}
\subsection{Core spectra}
\label{App:CoreSpectra145}
\begin{figure}
  \centering
  \includegraphics[width=.95\linewidth]{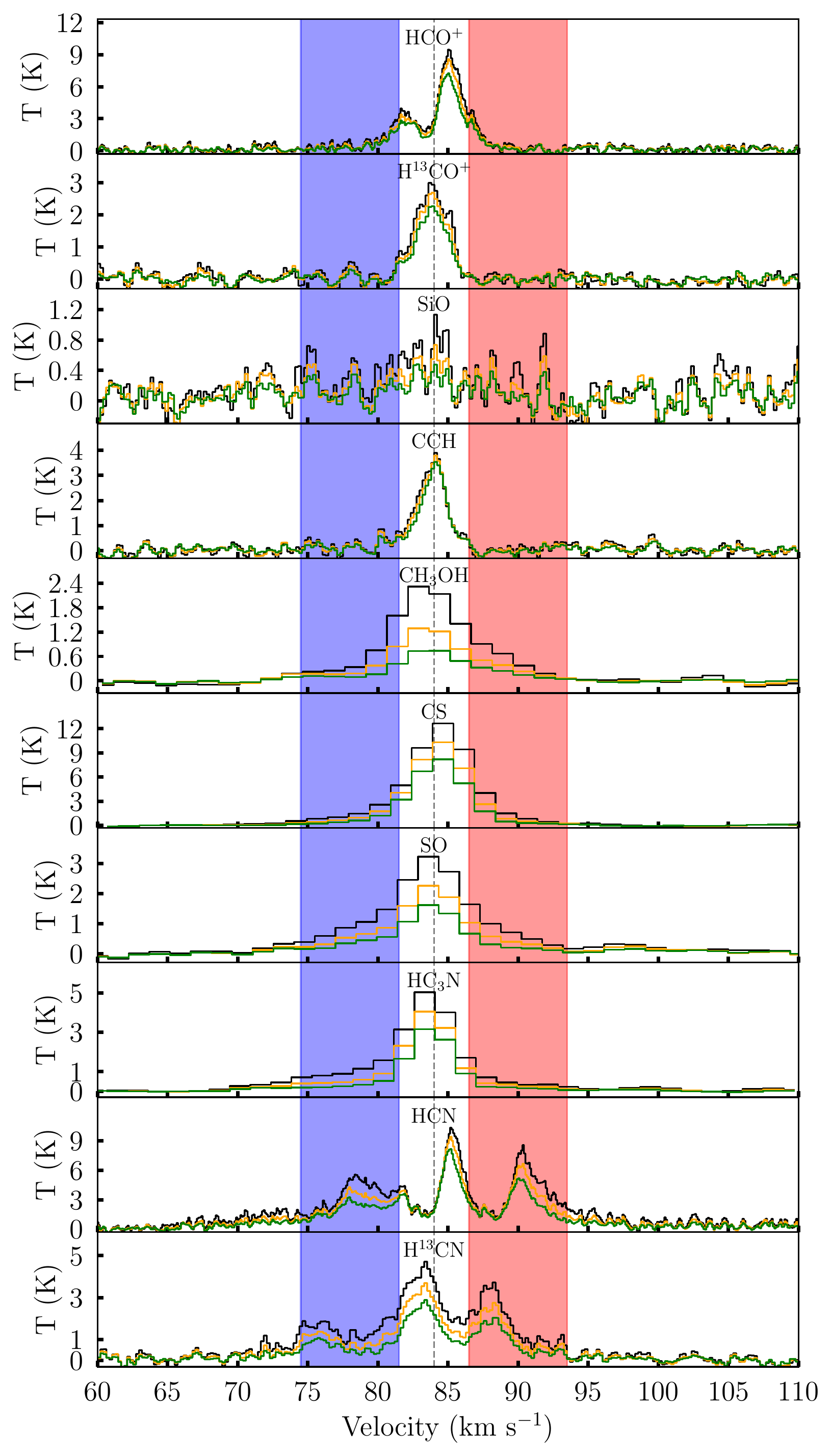}  
  \caption{C1 ATOMS spectra. Similar to Fig.~\ref{corespectral2}.}
  \label{corespectral1}
\end{figure}

\begin{figure}
  \centering
  \includegraphics[width=.95\linewidth]{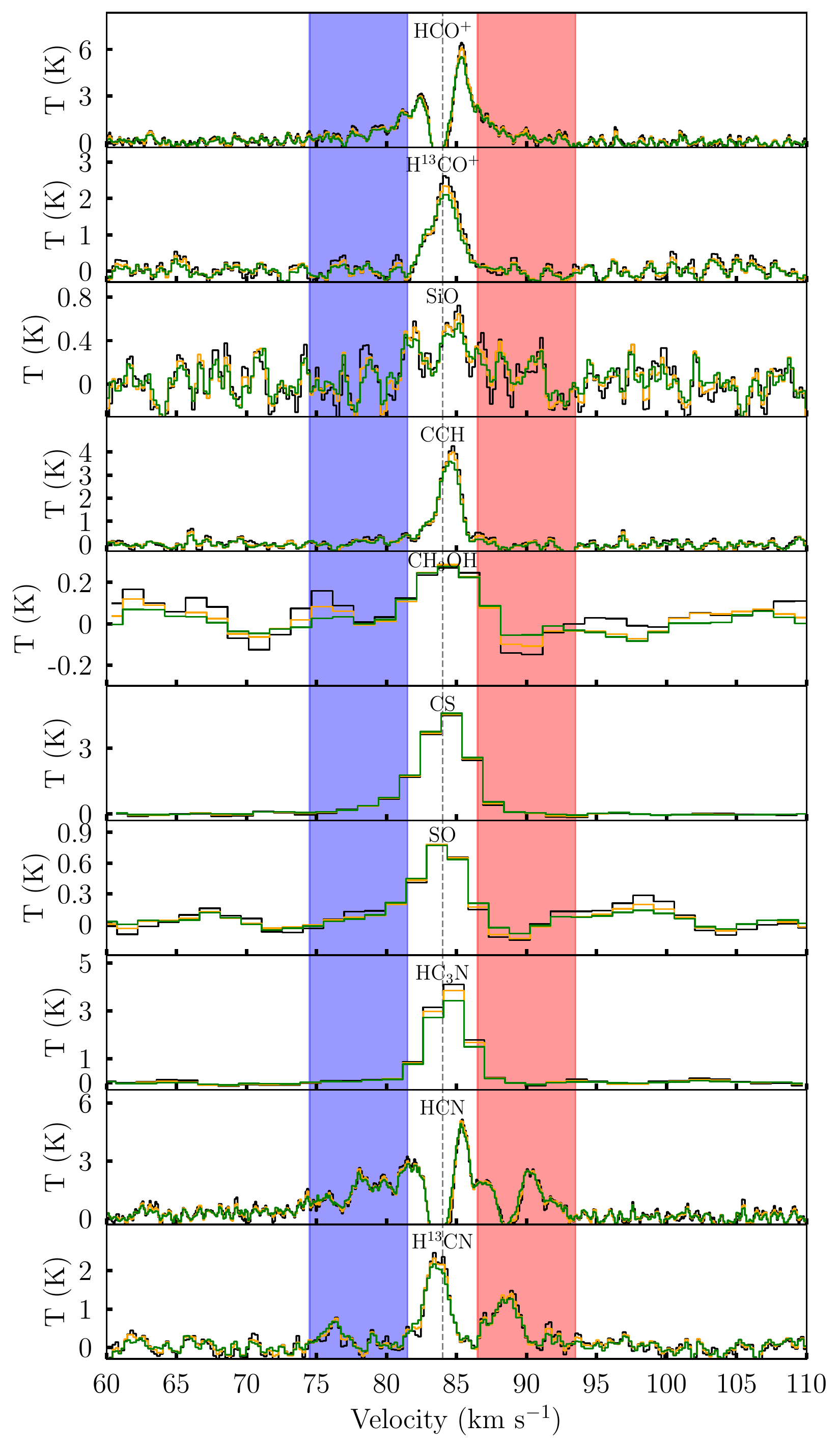}  
  \caption{C4 ATOMS spectra. Similar to Fig.~\ref{corespectral2}.}
  \label{corespectral4}
\end{figure}

\begin{figure}
  \centering
  \includegraphics[width=.95\linewidth]{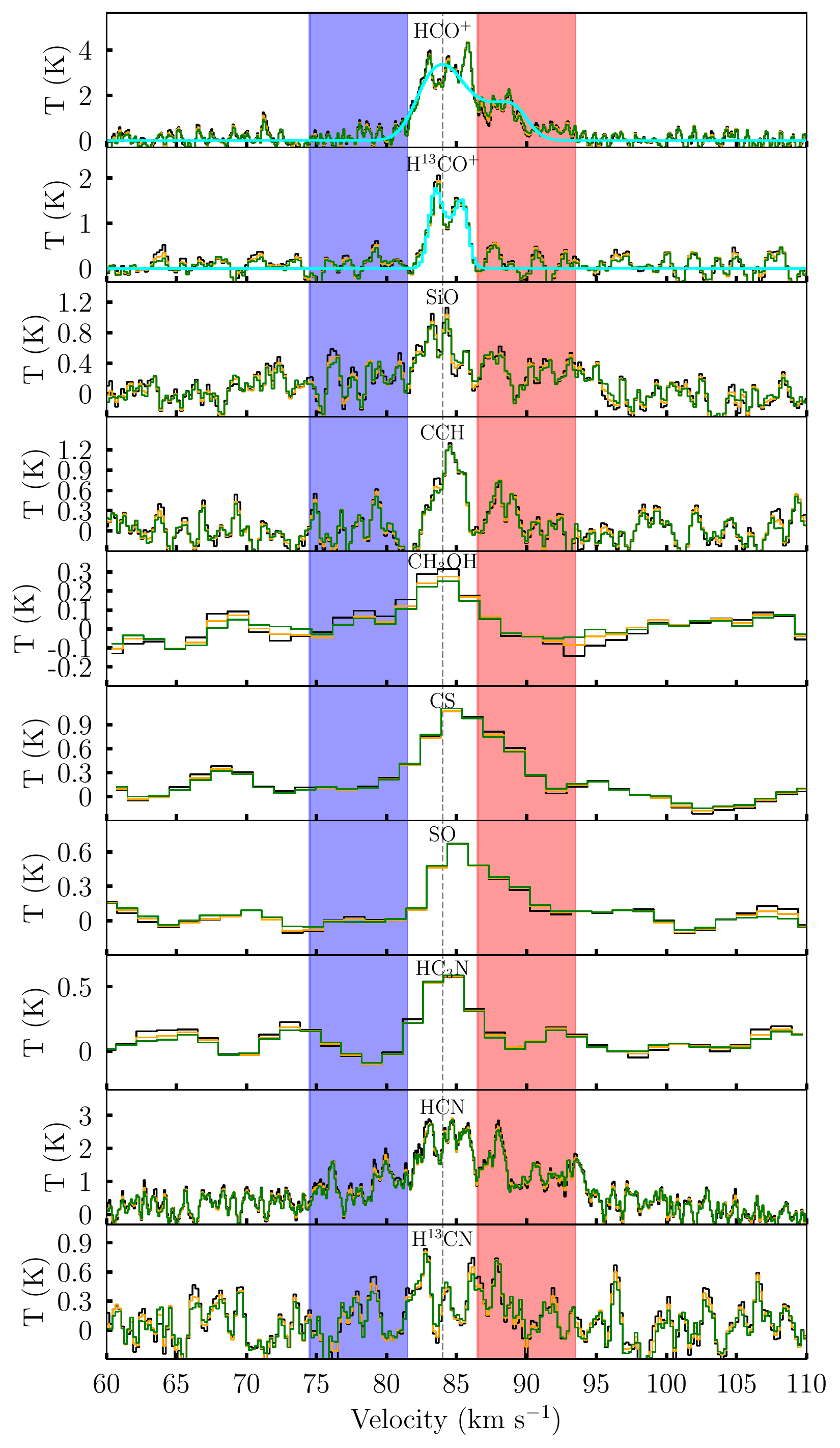}  
  \caption{C5 ATOMS spectra. Similar to Fig.~\ref{corespectral3}.}
  \label{corespectral5}
\end{figure}

Figures~\ref{corespectral1} to \ref{corespectral5} show the ATOMS spectra for cores C1, C4 and C5, respectively. The averaged spectra extracted from semi-axis, double semi-axes, and triple semi-axes areas of core have no significant difference for high velocity resolution spectra such as \hcopone. The $J = 1 - 0$ spectral profiles of \hcop\ and \htcop\ show red asymmetries for C1 and C4 while those of C3, C5, and possible C2 show blue asymmetries. 

\begin{figure*}
  \centering
  \includegraphics[width=.95\linewidth]{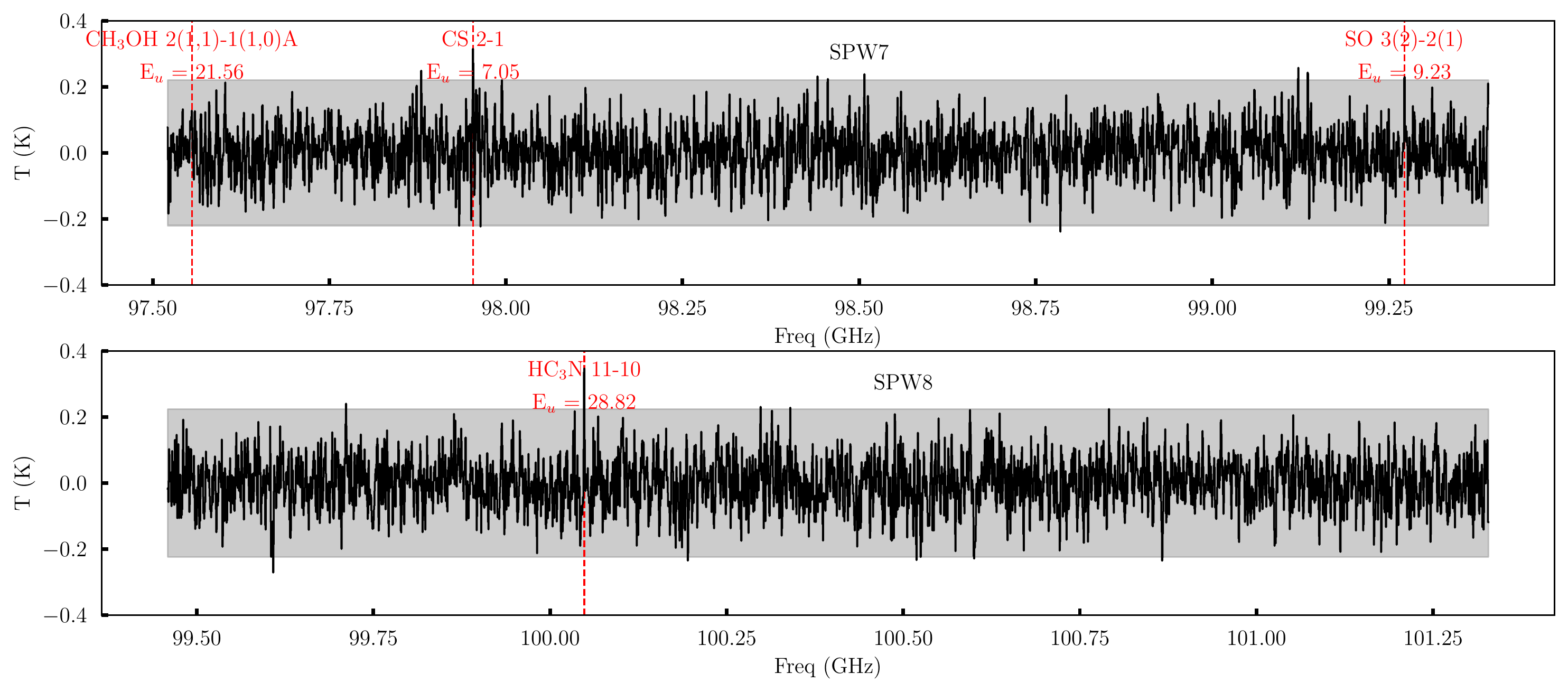}  
  \caption{Core C2 ATOMS wide spectral windows SPW7 and SPW8. The gray shadows highlight the 3$\sigma$ threshold. The red lines mark the spectra presented in Fig.~\ref{corespectral2}.}
  \label{widebandspectralcore2}
\end{figure*}

Figure~\ref{widebandspectralcore2} shows the ATOMS wide spectral windows SPW7 and SPW8 of core C2. The red lines highlight the spectra shown in Fig.~\ref{corespectral2}. Non detection for the warm tracer spectra suggests the cold and probably prestellar nature of core C2. 

\subsection{Infall velocity estimation}
Infall analyses are done using the \texttt{Hill5}\footnote{\url{https://pyspeckit.readthedocs.io/en/latest/hill5infall_model.html}} model in \texttt{pyspeckit}. A total of five free parameters are fitted to the model: line peak $T_{\rm peak}$, velocity dispersion $\sigma$, optical depth $\tau$, systematic velocity \vlsr, and infall velocity \vinfall. The \texttt{Hill5} models perform well when there are two distinct peaks. We first fit \hcopone\ profiles for C2, C3, and C5. The fitted profiles are shown in cyan lines in Figs.~\ref{corespectral2}, \ref{corespectral3}, and \ref{corespectral5}. The secondary peaks at the red-shifted side are poorly fitted for C3 and C5 due to the significant outflow wings. Therefore, we extract \vinfall\ from \htcopone\ for C3 and C5. The fitted \vinfall\ and velocity dispersion are listed in Table~\ref{InfallParameterTable}. The \hcop-derived \vinfall\ is larger than \htcop-derived \vinfall, consistent with the analyses of \citet{DeVries05} which propose that a more optically thin tracer may underestimate actual \vinfall. 

\begin{landscape}
\begin{figure}
  \centering
  \includegraphics[width=0.95\linewidth]{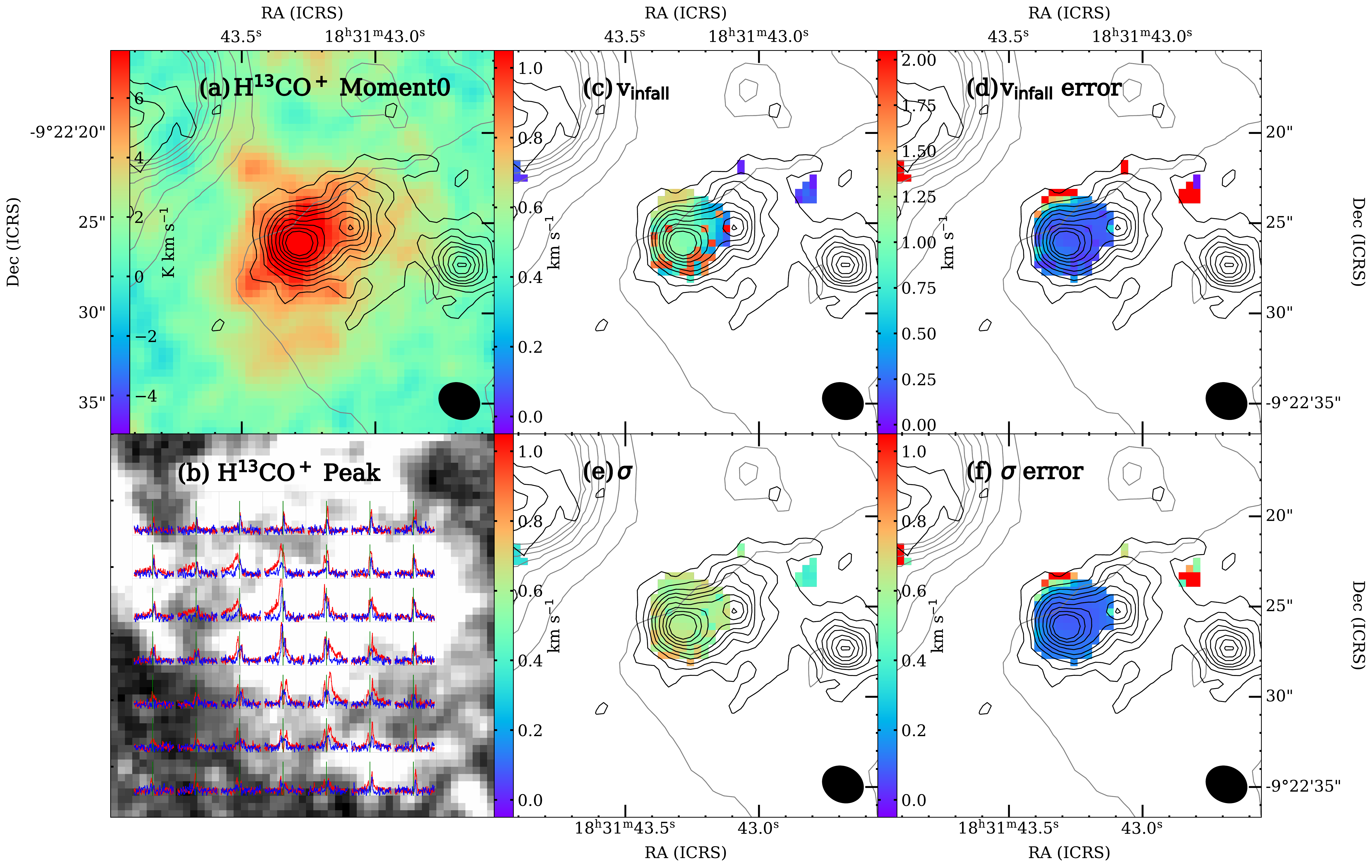}  
  \caption{\htcopone\ infall parameters estimated by the \texttt{Hill5} modelling. \textit{Panel a}: \htcopone\ zeroth moment overlaid with the contours of 8~\micron\ (gray, the levels are the same as Fig.~\ref{PPMAP-CO-FIG}) and 3~mm emission (black, the levels are the same as \textit{Panel a} of Fig.~\ref{StarFormationFigure}). The black ellipse represents the ATOMS beam. \textit{Panel b}: \htcop\ line peak map overlaid with the grid spectra of \hcopone\ (red) and \htcopone\ (blue). \textit{Panels c} to \textit{f}: infall velocity, velocity dispersion and their errors derived from \htcopone\ with the \texttt{Hill5} modelling.}
  \label{hcopinfalltracer}
\end{figure}
\end{landscape}

To further test the stability of \texttt{Hill5} model fitting, we fit \htcopone\ profiles pixel-by-pixel for pixels with \hcopone\ peak $>2$~K. The fitting results are shown in Fig.~\ref{hcopinfalltracer}. Most valid pixels are located in the C3 region and the resulted \vinfall\ and its error are stable at 0.5~\kms\ and 0.2~\kms, respectively. The pixel-by-pixel fitting shows that there is no significant difference for the modelled \vinfall\ between core-averaged profile and pixel profile.

\begin{table}
\caption{\label{InfallParameterTable} \texttt{Hill5} infall parameters.}
\begin{threeparttable}
\scriptsize
\centering
\setlength{\tabcolsep}{4.5pt}
\renewcommand{\arraystretch}{1.0}
\begin{tabular}{ccccccc}
\hline
\hline
Core &  \multicolumn{2}{c}{$\rm v_{infall}$} &  \multicolumn{2}{c}{$\sigma$}  &  \multicolumn{2}{c}{$\dot{M}_{\rm infall}$}    \\
                        &   \hcop    &   \htcop &     \hcop &   \htcop   &  \hcop  & \htcop    \\
                        & \kms & \kms & \kms & \kms & $\rm M_\odot kyr^{-1}$ & $\rm M_\odot kyr^{-1}$    \\
\hline
C2\tnote{\textit{(a)}} & 0.58$\pm$0.02 &   & 0.43$\pm$0.03 &   & 2.7$\pm$1.1  &    \\
C3 & 1.07$\pm$0.05 & 0.51$\pm$0.05   & 0.87$\pm$0.03 & 0.62$\pm$0.02 & 5.4$\pm$2.1 & 2.6$\pm$1.5  \\
C5 & 1.58$\pm$0.71 & 0.08$\pm$0.06 & 1.46$\pm$0.27 & 0.65$\pm$0.03 & 4.5$\pm$2.7 & 0.23$\pm$0.19  \\
\hline
\end{tabular}
      \begin{tablenotes}
      \item [\textit{(a)}] \htcopone\ profile of C2 does not have clearly blue asymmetry and therefore \texttt{Hill5} modelling is not applied.
      \end{tablenotes}
      \end{threeparttable}
\end{table}

\section{PDR molecules analyses} \label{App:PDRMoleculesAnalysis}

\subsection{Structural similarity index measure} \label{App:PDRMoleculesAnalysis-SSIM}
To compare the intensity spatial distributions between \cchline\ and other lines, we make use of the \texttt{SSIM} method\footnote{\url{https://scikit-image.org/docs/stable/api/skimage.metrics.html\#skimage.metrics.structural_similarity}} which is based on three comparison measurements between the samples' luminance, contrast, and structure \citep{Wang04}. The ATOMS data cubes are smoothed to a common velocity resolution of 3~\kms\ (channel width is 1.5~\kms) for the \texttt{SSIM} calculations. \textit{Panels a} to \textit{i} of Fig.~\ref{SSIM} show the channel-by-channel \texttt{SSIM} calculations for the ATOMS spectra, ranked from upper left to bottom right according to the maximum \texttt{SSIM} value of the maps. A larger value in the \texttt{SSIM} map means the emission spatial distributions are more similar. \textit{Panels a} to \textit{e} present a ``bar'' morphology with a slope of one, suggesting that there is a detectable similarity in emission spatial distribution between \cchline\ and the referred lines. We classified these molecules as PDR tracers. \textit{Panels f} to \textit{i} have no significant structure and therefore these molecules do not trace the conspicuous PDR shown in \cch. We classified them as star formation tracers.

\begin{figure}
  \centering
  \includegraphics[width=.95\linewidth]{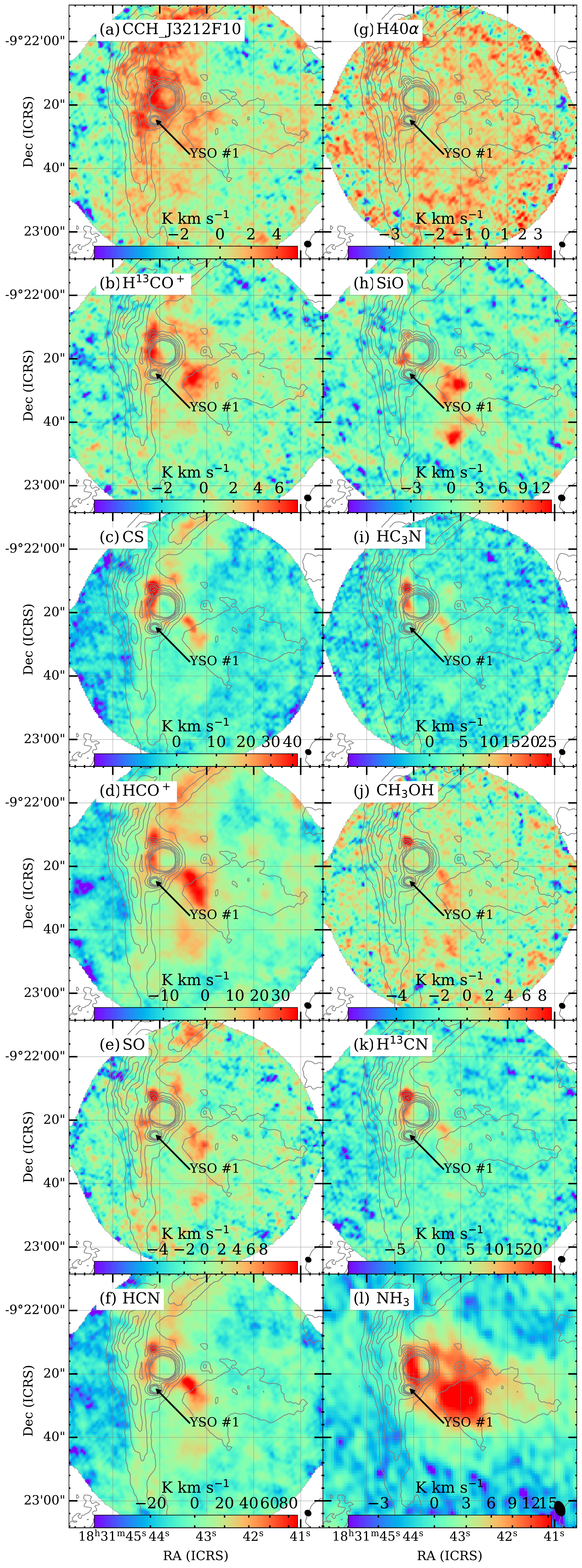}  
  \caption{ATOMS spectra and VLA \nht\ (1,1) zeroth moment maps. The gray contours represent 8~\micron\ emission with levels the same as the gray contours in Fig.~\ref{PPMAP-CO-FIG}. The arrow indicates the position of YSO \#1. The left and right columns show the PDR tracers and star formation tracers, respectively.}
  \label{moleculartracer}
\end{figure}

\begin{figure}
  \centering
  \includegraphics[width=.95\linewidth]{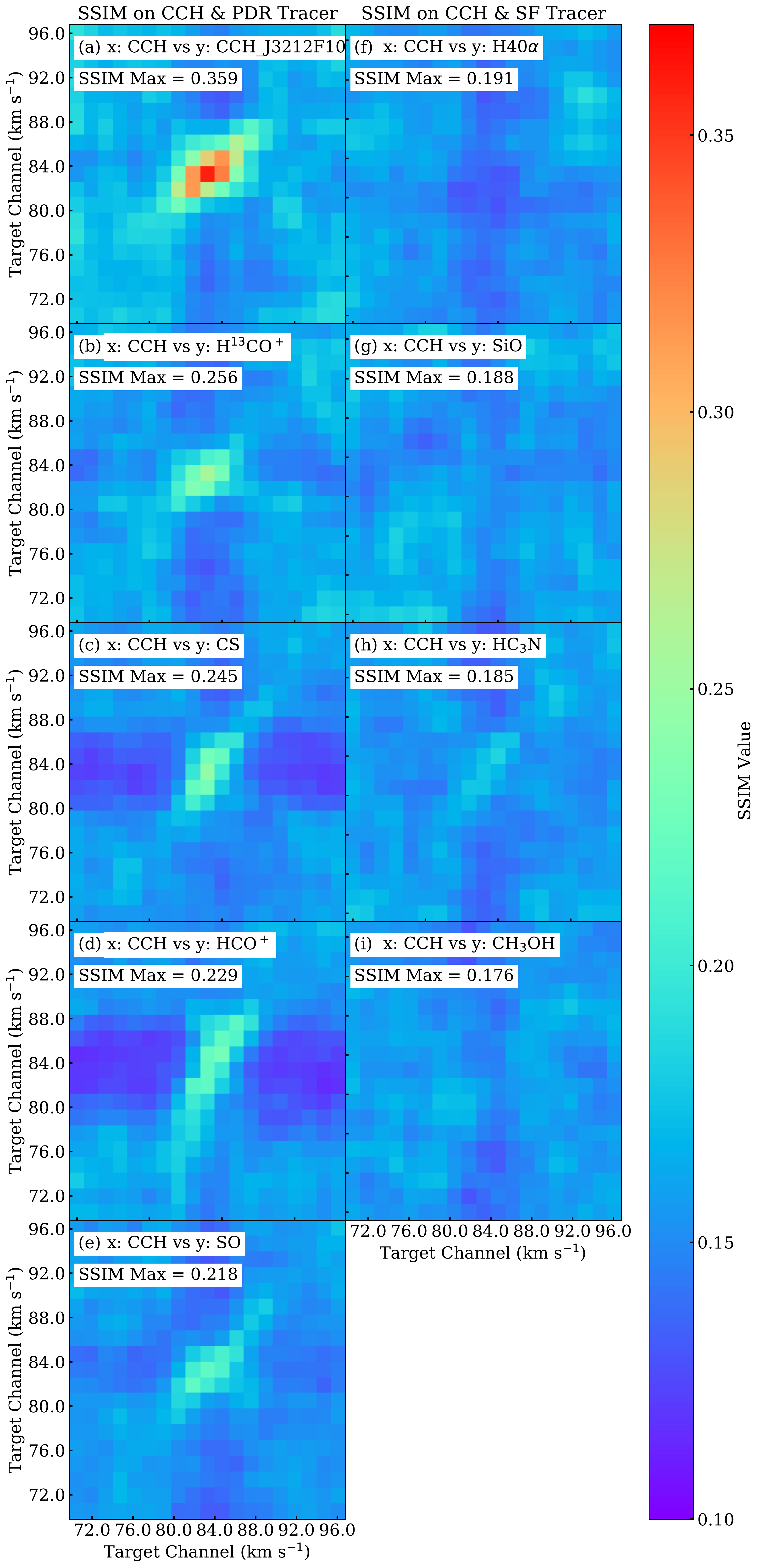}  
  \caption{\texttt{Structural similarity index measure} (\texttt{SSIM}) tests with \cchline\ emission. \textit{Panel a} shows the \texttt{SSIM} test for \cchline\ and \cchhyperfine. The left and right columns show the PDR and star formation tracers, respectively.}
  \label{SSIM}
\end{figure}

The zeroth moment maps for PDR tracers and star formation-tracers are shown in the left and right panels of Fig.~\ref{moleculartracer}, respectively. \nht\ is also classified as the star formation tracer due to the lack of detection in the I18290 PDR.

\subsection{\cch\,velocity decomposition}\label{App:PDRMoleculesAnalysis-CCHdecomposition}
To decompose the evaporative and compressed components, we fit the \cchline\ PDR spine PV cut image position-by-position with double Gaussians using the \texttt{MCMC}. There are five parameters $A_1$, $A_2$, $B_1$, $B_{21} = B_2 - B_1$, $C_1$, and $C_2$ in the double-Gaussian model
\begin{equation}
    model = A_{1}{\rm exp}\left(-\frac{\left({\rm v}-B_1\right)^2}{2C_1^2}\right) + A_{2}{\rm exp}\left(-\frac{\left({\rm v}-B_1 - B_{21}\right)^2}{2C_2^2}\right).
\end{equation}

\begin{figure*}
  \centering
  \includegraphics[width=.99\linewidth]{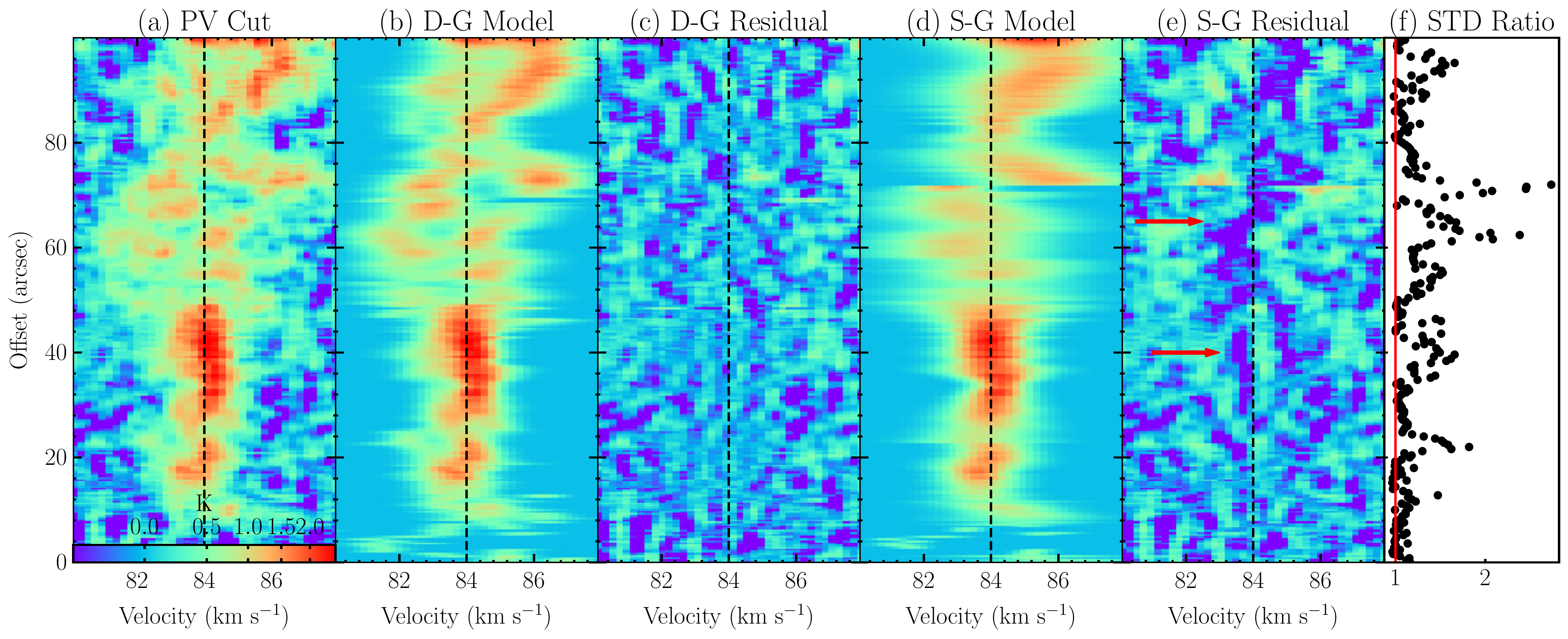}
  \caption{\cchline\ PDR spine PV models. \textit{Panel a}: PV cut along the 8~\micron\ spine. \textit{Panels b} and \textit{c}: Double-Gaussian position-by-position modelling result and its residual, respectively. \textit{Panels d} and \textit{e}: Single-Gaussian position-by-position modelling result and its residual, respectively. \textit{Panel f}: Standard deviation ratio of the single Gaussian model residual-to-double Gaussian model residual at each position.}
  \label{PV-modeling}
\end{figure*}

In the \texttt{MCMC} modelling, we limit parameter space as follows:
\begin{itemize}
    \item $A_1$, $A_2 > 0.4$~K, to ensure SN ratio $>2$.
    \item $81~{\rm km s^{-1}} < B_1 < 87~{\rm km s^{-1}}$ and $1~{\rm km s^{-1}} < \left|B_{21}\right| < 4~{\rm km s^{-1}}$. Choice of the minimum $\left|B_{21}\right|$ is kind of arbitrary but reasonable according to Fig.~\ref{PV-modeling}.
    \item $C_1$ and $C_2$, which represent the velocity dispersion, are required to be more than 0.4~\kms\ (width of two channels) and less than 0.8~\kms. The upper limit of 0.8~\kms, corresponding a FWHM of $\sim1.9$~\kms, is a reasonable value indicated by Fig.~\ref{PV-modeling}.
\end{itemize}

The modelled PV map and its residual are shown in \textit{Panels b} and \textit{c} of Fig.~\ref{PV-modeling}, respectively. The double-Gaussian model reconstructs the raw PV cut map very well, with a residual of $\sim0.17$~K.

Whether the PV cut map could be modelled by the single Gaussian is also tested. The single-Gaussian modelled PV and residual maps are shown in \textit{Panels d} and \textit{e} of Fig.~\ref{PV-modeling}, respectively. Several notable vertical strips indicated by the red arrows in the residual map imply the existence of the second component. The standard deviation ratios of single-Gaussian residual to double-Gaussian residual in each position show that the single-Gaussian model has a larger residual in most positions.


\bsp	
\label{lastpage}
\clearpage
\noindent
$^{1}$Kavli Institute for Astronomy and Astrophysics, Peking University, 5 Yiheyuan Road, Haidian District, Beĳing 100871, China\\
$^{2}$Shanghai Astronomical Observatory, Chinese Academy of Sciences, 80 Nandan Road, Shanghai 200030, China\\
$^{3}$Key Laboratory for Research in Galaxies and Cosmology, Chinese Academy of Sciences, 80 Nandan Road, Shanghai 200030, China\\
$^{4}$Aix Marseille Univ, CNRS, CNES, LAM, Marseille, France\\
$^{5}$Institut Universitaire de France (IUF)\\
$^{6}$Department of Physics, P.O. box 64, FI- 00014, University of
Helsinki, Finland\\
$^{7}$Department of Astronomy, Yunnan University, Kunming, 650091, China\\
$^{8}$Indian Institute of Space Science and Technology, Thiruvananthapuram 695 547, Kerala, India\\
$^{9}$Departamento de Astronom\'{i}a, Universidad de Concepci\'{o}n, Casilla 160-C, Concepci\'{o}n, Chile \\
$^{10}$Max-Planck-Institute for Astronomy, K\"{o}nigstuhl 17, D-69117 Heidelberg, Germany \\
$^{11}$Departamento de Astronomía, Universidad de Chile, Las Condes, 7591245 Santiago, Chile\\
$^{12}$Center for Astrophysics, Harvard \& Smithsonian, Cambridge, MA, USA \\
$^{13}$Jet Propulsion Laboratory, California Institute of Technology, 4800 Oak Grove Drive, Pasadena CA 91109, USA\\
$^{14}$Korea Astronomy and Space Science Institute, 776 Daedeokdaero, Yuseong-gu, Daejeon 34055, Republic of Korea \\
$^{15}$University of Science and Technology, Korea (UST), 217 Gajeong-ro, Yuseong-gu, Daejeon 34113, Republic of Korea \\
$^{16}$Instituto de Radioastronomía y Astrofísica, Universidad Nacional Autónoma de México, Antigua Carretera a Pátzcuaro \# 8701, Ex-Hda. San José de la Huerta, Morelia, Michoacán, México C.P. 58089 \\
$^{17}$Nobeyama Radio Observatory, National Astronomical Observatory of Japan, National Institutes of Natural Sciences, Nobeyama, Minamimaki, Minamisaku, Nagano 384-1305, Japan\\
$^{18}$Department of Astronomical Science, The Graduate University for Advanced Studies, SOKENDAI, 2-21-1 Osawa, Mitaka, Tokyo 181-8588, Japan\\
$^{19}$Department of Astronomy, School of Physics, Peking University, Beijing, 100871, China\\
$^{20}$National Astronomical Observatories, Chinese Academy of Sciences, Beijing 100101, China \\
$^{21}$University of Chinese Academy of Sciences, Beijing 100049, China  \\
\end{document}